\documentstyle[seceq,supplement]{ptptex}
\newcommand{\bra}[1]{\langle {#1} |}     %%
\newcommand{\ket}[1]{| {#1} \rangle}     %%
\newcommand{\bbra}[1]{\langle\langle {#1} |}     %%
\newcommand{\kket}[1]{| {#1} \rangle\rangle}     %%
\newcommand{\rbra}[1]{( {#1} |}     %%
\newcommand{\rket}[1]{| {#1} )}     %%
\newcommand{\rrket}[1]{| {#1} ))}     %%
\newcommand{\dket}[1]{|| {#1} \rangle}     %%
\newcommand{\maru}[1]{\stackrel{\tiny\circ} {#1}} %%
\title{%        %You can use \\ for explicit line-break
Time-Evolution of the Coherent and the Squeezed States
of Many-Body Systems Based on the Basic Idea of\\ the
Boson Mapping and the TDHF Method
\footnote{
To appear in Progress of Theoretical Physics Supplement 
{\it ``Selected Topics in Boson Mapping and Time-Dependent Hartree-Fock 
Methods."}}
}
\author{%       %Use \sc for the family name
Atsushi {\sc Kuriyama}, Jo\~ao da {\sc Provid\^encia}$^*$, 
Yasuhiko {\sc Tsue}$^{**}$ and\\ Masatoshi {\sc Yamamura} 
}

\inst{%         %Affiliation, neglected when [addenda] or [errata]
Faculty of Engineering, Kansai University, Suita 564-8680, Japan\\
$^*$Departamento de Fisica, Universidade de Coimbra, P-3000 Coimbra, 
Portugal\\
$^{**}$Department of Material Science, Kochi University, Kochi 780-8520, 
Japan
}

\abst{%       %this abstract is neglected when [addenda] or [errata]
Various works performed by the present authors in the 1990s 
are reviewed. The topics discussed in this paper are mainly related 
to the time-evolution of the coherent and the squeezed states of 
the systems obeying the $su(2)$- and the $su(1,1)$-algebra. 
The formulations are based on the basic idea of the boson 
mapping and the TDHF theory in canonical form. 
Under the time-dependent variational procedure for trial states 
appropriately chosen, the time-evolution of the systems 
under investigation is described. Further, it is shown that 
this method enables us to obtain the classical counterparts of 
the original quantal systems. 
}

\begin{document}

\maketitle

\section{Introduction and preliminaries}

For the microscopic studies of nuclear collective motions, 
the year 1960 is unforgettable. In this year, a theory was 
proposed by Marumori,\cite{M} also independently by Arvieu 
\& Veneroni and Baranger.\cite{AVB}
This theory is called the quasi-particle random phase approximation 
(QRPA) and, with the help of this theory, microscopic structures 
of collective vibrational states, especially, the first excited 
states of spherical even nuclei, were well described. 
Under the success of the QRPA 
theory for the first excited states, the next was a problem 
how to describe the higher excited states, for example, the 
well-known ($0^+, 2^+, 4^+$) triplet observed in the second 
excited states. In response to the above-mentioned situation, 
in 1962, Belyaev \& Zelevinsky proposed so-called boson 
expansion theory \cite{BZ} and slightly later, in 1964, 
Marumori proposed also boson expansion theory, together with 
Yamamura (one of the present authors) and Tokunaga.\cite{MYT}
The former and the latter are called the BZ and the MYT 
method. Both are suitable for the case of fermion-pairs. 
The case of particle-hole pairs was formulated by 
da Provid\^encia (one of the present authors) \& Weneser and 
Marshalek.\cite{P,Mar} These are called the boson expansion theory 
all together. 
In Ref.\citen{Mar}, the relation between the BZ and the MYT expansion 
was investigated from the side of the former form. 
We can find various further studies concerning 
the boson expansion in the review by Klein \& Marshalek 
in 1991. \cite{KM}

However, we should draw a distinction between the BZ and the MYT 
method. The former is based on the boson realization of 
fermion-pair algebra (Lie algebra) and the latter based on 
the mapping from fermion- to boson-space, which is called 
the boson mapping. The boson mapping is independent from the 
Lie algebra and, if the Lie algebra is treated in the framework 
of the boson mapping, we can arrive at its boson realization. 
In this sense, it should be noted that the boson mapping is 
conceptually wider than the boson realization of the Lie 
algebra. 
However, it is not necessary for the boson mapping 
to restrict ourselves to the 
fermion- and the boson-space. For example, let us consider 
a quantal system which can be described in the framework of 
the orthogonal set $\{ \kket{n} ; n=0, 1, 2, \cdots, N\}$. 
It is not always many-fermion system. Next, we take up 
another set $\{\ket{n}\}$, which is not always composed of 
the boson operators, but, here, following the boson mapping, 
composed of a kind of boson operator $({\hat a}, {\hat a}^*)$ :
\begin{equation}\label{1-1}
\ket{n}=\frac{1}{\sqrt{n!}}({\hat a}^*)^n \ket{0} \ .
\end{equation}
Then, we set up the one-to-one correspondence between the 
two sets :
\begin{equation}\label{1-2}
\kket{n} \sim \ket{n} \ . \quad 
(n=0,1,2,\cdots,N)
\end{equation}
In the boson mapping, the following operator ${\hat U}_N$ 
plays a central role :
\begin{equation}\label{1-3}
{\hat U}_N = \sum_{n=0}^{N} \ket{n}\bbra{n} \ . 
\end{equation}
With the use of ${\hat U}_N$, we have 
\begin{eqnarray}
& &{\hat U}_N \kket{n} = \ket{n} \ ,
\qquad \ \ (n=0,1,2,\cdots,N) \label{1-4}\\
& &{\hat U}_N^{\dagger} \ket{n} = \kket{n} \ , \qquad
\ \ (n=0,1,2,\cdots,N) \label{1-5}\\
& &{\hat U}_N^{\dagger} \ket{N+k}=0 \ . \qquad
(k=1,2,3,\cdots ) \label{1-6}
\end{eqnarray}
Usually, the space spanned by the set 
$\{ \ket{n} ; n=0,1,2,\cdots,N\}$ is called the physical space. 
The operators ${\hat U}_N$ and ${\hat U}_N^{\dagger}$ obey 
\begin{eqnarray}
& &{\hat U}_N^{\dagger} {\hat U}_N = 1 \ , \label{1-7}\\
& &{\hat U}_N {\hat U}_N^{\dagger} = P_N \ , \qquad
P_N=\sum_{n=0}^{N} \ket{n}\bra{n} \ . \label{1-8}
\end{eqnarray}
Here, $P_N$ is a projection operator to the physical space 
which satisfies $P_N^{\dagger}=P_N$ and $P_N^2=P_N$. 
Next, we consider the operator ${\hat O}_\nu$ in the 
original space which satisfies 
\begin{equation}\label{1-9}
\bbra{n+\nu}{\hat O}_\nu \kket{n} 
= \sqrt{\frac{n!}{(n+\nu)!}}O_\nu(n) \ . \qquad
(\nu \ge 0 , n=0,1,2,\cdots, N-\nu)
\end{equation}
Here, $O_\nu(n)$ is a function of $n$ which is obtained by 
calculating the matrix element 
$\bbra{n+\nu}{\hat O}_\nu \kket{n}$. 
Then, in the boson space, ${\hat O}_\nu$ is transformed to 
${\hat U}_N{\hat O}_\nu{\hat U}_N^{\dagger}$ and it is expressed as 
\begin{eqnarray}
& &{\hat U}_N{\hat O}_\nu{\hat U}_N^{\dagger}
=P_N {\widetilde O}_\nu P_N \ , \label{1-10}\\
& &{\widetilde O}_\nu = ({\hat a}^*)^\nu O_\nu ({\hat n}) \ , 
\qquad
{\hat n}={\hat a}^*{\hat a} \ . \label{1-11}
\end{eqnarray}
The above is an outline of the boson mapping and in later section, 
we will use the above-mentioned scheme. 

On the other hand, there exists another powerful approach 
to the microscopic theory of collective motion, which has been 
called the time-dependent Hartree-Fock (TDHF) theory. At the early 
stage, the idea of the TDHF theory was proposed by Nogami \cite{N} 
and slightly later by Marumori \cite{M} for the case of 
the small amplitude vibrational motion. 
The frequency of the small fluctuation around a static HF field 
is calculated in the frames of 
the TDHF theory. At the middle of the 
1970s, the TDHF theory was revived not only for 
the study of anharmonic vibration but also for the studies of 
heavy-ion reaction and nuclear fissions. Many ideas or forms were 
proposed for various problems by many authors.\cite{BV} 
Especially, in 1980, a canonical form for the TDHF theory was proposed 
by Marumori, together with Maskawa, Sakata and Kuriyama (one 
of the present authors).\cite{MMSK} 
The TDHF theory is a possible method for obtaining an approximate 
solution of the time-dependent Schr\"odinger equation. 
First, as a trial state for the time-dependent 
variation, we prepare a Slater determinant 
under a definite form containing 
parameters. Through the variational procedure, we obtain certain 
differential equations with respect to $t$ and, then, by solving 
under an appropriate initial condition, the time-dependence 
of the parameters is determined. Thus, we have an approximate 
solution of the original time-dependent Schr\"odinger equation. 
In the canonical form of the TDHF theory, the parameters are 
required to obey certain relation which the 
present authors call the canonicity condition. Then, the parameters 
can be regarded as canonical variables in classical mechanics 
and the certain differential equation with respect to $t$ 
is reduced to the Hamilton equation of motion in classical 
mechanics, the Hamiltonian of which is given as the expectation 
value of the starting quantal 
Hamiltonian for the trial function.

Introduction of the canonicity condition into the TDHF theory 
suggests us to formulate the variation for the time-dependent 
Schr\"odinger equation in relation to classical mechanics for 
any quantal system. Let ${\hat H}$ denote 
the Hamiltonian under investigation, which is not always 
many-fermion system. For ${\hat H}$, we set up the time-dependent 
Schr\"odinger equation 
\begin{equation}\label{1-12}
i\hbar\frac{\partial}{\partial t}\ket{{\rm exact}}
={\hat H}\ket{{\rm exact}} \ .
\end{equation}
As for an approximate solution of Eq. (\ref{1-12}), we 
prepare a trial state $\ket{p,q}$. Here, $p$ and $q$ 
denote two real parameters obeying the relation 
\begin{eqnarray}\label{1-13}
& &\bra{p,q} i\hbar \frac{\partial}{\partial q}\ket{p,q} 
= p+\frac{\partial S}{\partial q} \ , \nonumber\\
& &\bra{p,q} i\hbar \frac{\partial}{\partial p}\ket{p,q} 
= \frac{\partial S}{\partial p} \ .
\end{eqnarray}
Here, $S$ is arbitrary real function of $p$ and $q$. 
The present authors call the relation (\ref{1-13}) the 
canonicity condition. Then, we have the relation 
\begin{eqnarray}
& &\bra{p,q} i\hbar\frac{\partial}{\partial t}-{\hat H} \ket{p,q} 
=p{\dot q}-H(p,q)+\frac{dS}{dt} \ , \label{1-14}\\
& &\quad H(p,q)=\bra{p,q}{\hat H}\ket{p,q} \ .\label{1-15}
\end{eqnarray}
The relations (\ref{1-14}) and (\ref{1-15}) can be regarded as 
the Lagrangian and Hamiltonian, respectively and the variation 
gives us 
\begin{equation}\label{1-16}
\delta \int \bra{p,q} i\hbar\frac{\partial}{\partial t}
-{\hat H} \ket{p,q} dt 
= \delta \int (p{\dot q}-H(p,q))dt =0 \ . 
\end{equation}
Thus, we have 
\begin{equation}\label{1-17}
{\dot q}=\frac{\partial H}{\partial p} \ , \qquad
{\dot p}=-\frac{\partial H}{\partial q} \ . 
\end{equation}
The differential equation is nothing but the Hamilton equation 
of motion in classical mechanics. Therefore, $(p, q)$ can be 
regarded as canonical variables. By solving Eq. (\ref{1-17}) 
under an appropriate initial condition, $p$ and $q$ can be 
determined as function of $t$. The substitution of the solution 
of Eq. (\ref{1-17}) into $\ket{p,q}$ gives us an approximate 
solution of Eq. (\ref{1-12}). 
If $S$ is chosen as the form $S=pq/2$, the canonicity condition 
is written as 
\begin{eqnarray}
& &\bra{p,q}\partial_X \ket{p,q}=\frac{X^*}{2} \ , \qquad
\bra{p,q}\partial_{X^*} \ket{p,q}=-\frac{X}{2} \ , \label{1-18}\\
& &p=i\sqrt{\frac{\hbar}{2}}(X^*-X) \ , \qquad
q=\sqrt{\frac{\hbar}{2}}(X^*+X) \ .\label{1-19}
\end{eqnarray}
Of course, $(X, X^*)$ denote the complex parameters. 
The above is an outline of the variational method for the 
time-dependent Schr\"odinger equation which is suggested 
from the canonical form of the TDHF theory. In later section, 
we use the above-mentioned scheme.

The above-mentioned time-dependent variational approach to 
quantal system has two characteristic aspects. One is to give an 
approximate description of time-evolution of the quantum state. 
The other is to give a possible classical counterpart for an 
original quantal system under a suitable choice of the trial state. 
Then, it is expected that the original system is reproduced in 
a disguised form under an appropriate requantization 
procedure. This viewpoint was originally proposed by Marshalek 
and Holzwarth for the case of the time-dependent Hartree-Bogoliubov 
theory.\cite{MH} 
They obtained the Holstein-Primakoff boson representation 
for the fermion-pair algebra, which is of the same form as 
that given by the boson mapping.\cite{Mar}

With the aim of formulating the TDHF theory and its extension 
in the canonical form, the present authors (A. K. \& M. Y.) 
have investigated the above-mentioned idea extensively for 
the cases of various many-fermion systems. For example, 
with the aid of the Grassmann variables, odd fermion system 
can be treated in the framework of the above-mentioned scheme. 
These works have been reviewed in the paper by the present 
authors (A. K. \& M. Y.).\cite{YK87}
Further, the above idea was applied to the investigation of 
the behavior of the systems in highly excited states also by 
the present authors (A. K., J. da P. \& M. Y.).\cite{YPKF}
The basic idea for this case is in the description of 
many-fermion systems under the phase space doubling, 
which is fundamental in the thermo field dynamics 
formalism.\cite{TU} 
This description gave us the so-called thermal boson expansion 
and the time-evolution of the thermal states. The above all 
papers are related with many-fermion systems.

Focussing on the $su(2)$- and $su(1,1)$-algebra, in this paper, 
we will review some investigations performed mainly by the 
present authors after 1991. Of course, the topics discussed in this 
paper are related to the title of this paper. 
It is well known that the $su(2)$-algebra has played an important 
and interesting role in the study 
of many-body physics. Even if restricted to nuclear theory, this 
algebra has served us not only for schematic understandings of collective 
dynamics such as the Lipkin model but also for studies of 
realistic phenomena such as superconducting phase 
and rotational motion. We can learn various investigations 
related to the $su(2)$-algebra until 1990 in Refs.\citen{KM} 
and \citen{YK87}. 
A possible boson representation of the $su(2)$-algebra 
was given by Schwinger\cite{Schw} and this form will 
play a central role in this paper. On the other hand, 
as was originally discussed by Celeghini, Rasetti, Tarlini 
and Vitiello,\cite{Vitiello} the $su(1,1)$-algebra is also 
quite interesting. With the aid of this algebra in the 
Schwinger boson representation,\cite{Schw} 
we can describe the ``damped and amplified oscillator" 
quantum mechanically in the conservative form. 
The investigation of the behavior of systems in highly 
excited states, such as nuclear phenomena observed in 
highly excited states, is one of the interesting problems 
in quantum many-body physics. In the theoretical treatment, 
it is necessary to deal with the systems at finite 
temperature. The above means that the $su(1,1)$-algebra presents us 
a possible entrance to the problems related to finite temperature. 
The interest of the present authors in the 1990s has been 
concerned mainly with the time-evolution of many-body 
systems obeying the $su(2)$- and $su(1,1)$-algebra. 
Of course, as is clear from the title of this paper, 
the interest is related with the classical descriptions of 
quantum systems, and the technique of the boson mapping and 
the time-dependent variational method are fully used. 
These investigations are divided into two groups. 
One is related mainly with time-evolution of the coherent 
states in many-boson systems\cite{TKY1}\tocite{PYK} 
and the other with the squeezed states in many-body 
systems,\cite{TFKY}\tocite{TAKY2} 
some of which were published together with 
Fujiwara.\cite{TFKY}\tocite{TF2} 
Our interest is concerned not only with the coherent 
and the squeezed state for a simple system but also 
with those for the systems obeying the $su(2)$- and the 
$su(1,1)$-algebra. Starting from the conventional boson 
coherent state, we formulate various forms which are 
suitable for treating the $su(2)$- and the $su(1,1)$-boson 
models.
Especially, one of the present authors (Y. T.) 
investigated the formalism, which is partially 
presented in this paper, 
in relation to the WKB result and the Maslov phase in the 
limit of $\hbar\rightarrow 0$.\cite{T} 
Modified forms of the conventional Holstein-Primakoff 
boson representation of the $su(1,1)$-algebra and 
$q$-boson realizations of the $su_q(2)$- and the 
$su_q(1,1)$-algebra, which we will not contact with in this 
paper, are discussed in Refs. \citen{KPTY5} and \citen{PYK}, 
respectively.

In \S\S 2 and 3, starting from the Schwinger boson representation 
for the $su(2)$- and the $su(1,1)$-algebra, the Holstein-Primakoff 
boson representations are derived under the use of the basic idea 
of the boson mapping. Further, with the aid of the coherent state, 
the classical counterparts of the both algebras are obtained. 
Section 4 and 5 are devoted to investigating the ``damped 
and amplified oscillation" and the $su(2)$-algebraic model in the 
framework of the $su(1,1)$-algebra and its coherent state. 
The idea is found in the phase space doubling appearing in 
the thermo field dynamics formalism. In particular, the 
thermal effects observed in their systems are discussed. 
In \S\S 6 $\sim$ 9, a boson system interacting 
with the external harmonic oscillator is described. 
Under a certain condition for the interaction, the results 
are almost the same as those obtained in \S\S 4 and 5. 
In \S 10, the time-dependent variational approach with 
the squeezed state is presented and, in \S\S 11 and 12, 
the squeezed state for the $su(2)$-algebraic models of both 
many-boson and many-fermion systems are discussed. Finally, the 
concluding remarks are mentioned in \S 13.

\section{The $su(2)$-algebra---The boson realization and its 
classical counterpart}

First, we will recapitulate the boson representation of the 
$su(2)$-spin system in a form suitable for later discussion. 
For the sake of completeness, various familiar relations will 
be listed. Let us introduce a boson space constructed in terms of 
two kinds of boson operators $({\hat a}, {\hat a}^{*})$ 
and $({\hat b}, {\hat b}^{*})$. 
In this space, we define the following three operators : 
\begin{equation}\label{2-1}
{\hat S}_+=\hbar {\hat a}^*{\hat b} \ , \quad
{\hat S}_-=\hbar {\hat b}^*{\hat a} \ , \quad
{\hat S}_0=\frac{\hbar}{2}({\hat a}^*{\hat a}-{\hat b}^*{\hat b}) 
\ .
\end{equation}
Here, $\hbar$ denotes the Planck constant. The above three 
operators ${\hat S}_{\pm,0}$ form the $su(2)$-algebra, which 
obeys
\begin{eqnarray}
& &{\hat S}_-^*={\hat S}_+ \ , \qquad
{\hat S}_0^*={\hat S}_0 \ , \label{2-2}\\
& &[ {\hat S}_+ , {\hat S}_- ]= +2\hbar {\hat S}_0 \ , \qquad
[{\hat S}_0 , {\hat S}_{\pm} ]=\pm \hbar {\hat S}_{\pm} \ .
\label{2-3}
\end{eqnarray}
The form (\ref{2-1}) is called the Schwinger boson representation 
of the $su(2)$-algebra.\cite{Schw} 
The Casimir operator ${\hat \Gamma}_{su(2)}$ 
can be written as 
\begin{equation}\label{2-4}
{\hat \Gamma}_{su(2)}={\hat S}_0^2+\frac{1}{2}
({\hat S}_-{\hat S}_+ + {\hat S}_+{\hat S}_-) 
={\hat S}({\hat S}+\hbar) \ .
\end{equation}
Here, ${\hat S}$ and its property are given as 
\begin{eqnarray}
& &{\hat S}=+\frac{\hbar}{2}({\hat a}^*{\hat a}
+{\hat b}^*{\hat b}) \ , \label{2-5}\\
& &[{\hat S} , {\hat S}_{\pm, 0} ]=0 \ .\label{2-6}
\end{eqnarray}
We can understand that ${\hat S}$ is an operator denoting 
the magnitude of the $su(2)$-spin.

The eigenstate of ${\hat S}$ and ${\hat S}_0$ with the 
eigenvalues $S$ and $S_0$, respectively, is obtained in the form 
\begin{eqnarray}
& &\ket{S, S_0}=\frac{1}{\sqrt{(s+s_0)!(s-s_0)!}}
({\hat a}^*)^{s+s_0}({\hat b}^*)^{s-s_0}\ket{0} \ , \label{2-7}\\
& &{\hat a}\ket{0}={\hat b}\ket{0}=0 \ . \label{2-8}
\end{eqnarray}
The eigenvalues $S$ and $S_0$ are expressed as 
\begin{eqnarray}
& &S=\hbar s\ , \qquad S_0=\hbar s_0 \ , \label{2-9}\\
& &\ \ s=0, \frac{1}{2}, 1, \frac{3}{2}, \cdots , \qquad
s_0=-s, -s+1, \cdots, s-1, s \ .\label{2-10}
\end{eqnarray}
We call the space spanned by the set $\{ \ket{S, S_0} \}$ 
for fixed value of $S$ the $su(2)$-space with the magnitude 
$S$ and at several occasions, we will use the notation 
$\ket{s, s_0}$ for $\ket{S, S_0}$.

Following Refs.\citen{KPTY1} $\sim$ \citen{KPTY3}, 
let us derive the Holstein-Primakoff representation\cite{HP} 
from the Schwinger representation. 
Basic idea comes from 
the MYT boson mapping method proposed by Marumori, Yamamura 
and Tokunaga. For the above-mentioned aim, we prepare a new 
boson space spanned by the set $\{ \rket{n}=
[\sqrt{n!}]^{-1}\cdot ({\hat X}^*)^n \rket{0} ; 
n=0,1,2,\cdots \}$. Here, $({\hat X} , {\hat X}^{*})$ denotes 
boson operator and $\rket{0}$ is the vacuum for 
$({\hat X}, {\hat X}^*)$. First, following the MYT boson mapping, 
we set up the following correspondence between the original 
and the new boson space : 
\begin{equation}\label{2-11}
\ket{s, s_0} \sim \rket{s+s_0} \ . \qquad
(S=\hbar s \ , S_0=\hbar s_0 )
\end{equation}
The spaces spanned by the set 
$\{ \rket{s+s_0} ; s_0=-s, -s+1, \cdots, s-1, s \}$ and the set 
$\{ \rket{2s+k} ; k=1, 2, 3, \cdots \}$ are physical and 
unphysical space, respectively. Under the correspondence 
(\ref{2-11}), the following mapping operator ${\hat U}$ is 
introduced : 
\begin{equation}\label{2-12}
{\hat U} = \sum_{s_0=-s}^s \rket{s+s_0}\bra{s, s_0} \ .
\end{equation}
The properties of ${\hat U}$ are shown in the form 
\begin{eqnarray}\label{2-13}
& &{\hat U}^{\dagger}{\hat U} 
= \sum_{s_0=-s}^{s}\ket{s, s_0}\bra{s, s_0} = 1 \ , \nonumber\\
& &{\hat U}{\hat U}^{\dagger}
=\sum_{s_0=-s}^{s} \rket{s+s_0}\rbra{s+s_0} = P_{2s} \ .
\end{eqnarray}
Here, $P_{2s}$ denotes a projection operator to the physical 
space :
\begin{equation}\label{2-14}
P_{2s}^* = P_{2s} \ , \qquad
P_{2s}^2=P_{2s} \ .
\end{equation}
From the above conditions, the following relations are derived :
\begin{eqnarray}
& &{\hat U} \ket{s, s_0} = \rket{s+s_0} \ , \label{2-15}\\
& &{\hat U}^{\dagger} \rket{s+s_0} = \ket{s, s_0} \ , \nonumber\\
& &{\hat U}^{\dagger} \rket{2s+k} = 0 \ . \qquad 
(k=1, 2, 3, \cdots )
\label{2-16}
\end{eqnarray}
With the use of ${\hat U}$, the operators ${\hat S}_{\pm, 0}$ 
are transformed to the form
\begin{eqnarray}\label{2-17}
& &{\hat U}{\hat S}_{\pm,0}{\hat U}^{\dagger} 
=P_{2s}\cdot {\widetilde S}_{\pm,0}\cdot P_{2s} \ , \nonumber\\
& &{\widetilde S}_+ = \sqrt{\hbar}{\hat X}^* \cdot
\sqrt{2S-\hbar{\hat X}^*{\hat X}} \ , \nonumber\\
& &{\widetilde S}_- = \sqrt{2S-\hbar{\hat X}^*{\hat X}} \cdot
\sqrt{\hbar}{\hat X} \ , \nonumber\\
& &{\widetilde S}_0 = \hbar {\hat X}^*{\hat X} - S \ .
\end{eqnarray}
The above operators ${\widetilde S}_{\pm,0}$ satisfy 
\begin{eqnarray}
& &{\widetilde S}_-^* = {\widetilde S}_+ \ , \qquad
{\widetilde S}_0^* = {\widetilde S}_0 \ , \label{2-18}\\
& &[ {\widetilde S}_+ , {\widetilde S}_- ] = +2\hbar {\widetilde S}_0 \ , 
\qquad
[ {\widetilde S}_0 , {\widetilde S}_{\pm} ] 
= \pm \hbar {\widetilde S}_{\pm} \ , \label{2-19}\\
& &{\widetilde S}_0^2+\frac{1}{2}
({\widetilde S}_-{\widetilde S}_+ + {\widetilde S}_+{\widetilde S}_-) 
= S(S+\hbar) \ .\label{2-20}
\end{eqnarray}
The relations (\ref{2-18}) $\sim$ (\ref{2-20}) correspond to 
Eqs. (\ref{2-2}) $\sim$ (\ref{2-4}). 
The above is nothing but a refined form given by 
Marshalek\cite{Mar} 
and it is called the Holstein-Primakoff boson representation.

Our third problem is concerned with the coherent 
state $\ket{\gamma}$ for the $su(2)$-spin 
system.\cite{KPTY1}\tocite{KPTY3} 
For the 
construction of the state $\ket{\gamma}$, we note the form 
\begin{equation}\label{2-21}
\ket{s, s_0}=\frac{1}{(2s)!}
\sqrt{\frac{(s-s_0)!}{(s+s_0)!}}\cdot 
\left(\frac{{\hat S}_+}{\hbar}\right)^{s+s_0}\cdot
({\hat b}^*)^{2s} \ket{0} \ . 
\end{equation}
Then, we set up the following correspondence : 
\begin{eqnarray}\label{2-22}
\frac{1}{(2s)!}
\sqrt{\frac{(s-s_0)!}{(s+s_0)!}}\cdot 
\left(\frac{{\hat S}_+}{\hbar}\right)^{s+s_0}
&\sim& \exp\left( \frac{v}{u}\cdot \frac{{\hat S}_+}{\hbar}\right) 
=\exp \left(\frac{v}{u}\cdot {\hat a}^* {\hat b}\right) \ , 
\nonumber\\
({\hat b}^*)^{2s}\ket{0}
&\sim& \exp\left(\sqrt{\frac{2}{\hbar}}wu{\hat b}^*\right)
\ket{0} \ .
\end{eqnarray}
Here, $w$ and $v$ denote complex parameters and $u$ is given as 
\begin{equation}\label{2-23}
u=\sqrt{1-|v|^2} \ .
\end{equation}
Under the correspondence (\ref{2-22}), we define the state 
$\ket{\gamma}$ in the form 
\begin{equation}\label{2-24}
\ket{\gamma}=M_c\cdot \exp\left(
\frac{v}{u}\cdot \frac{{\hat S}_+}{\hbar}\right)\cdot
\exp\left(\sqrt{\frac{2}{\hbar}}wu{\hat b}^*\right) \ket{0} \ .
\end{equation}
Here, $M_c$ denotes the normalization constant given by 
\begin{equation}\label{2-25}
M_c=\exp\left(-\frac{|w|^2}{\hbar}\right) \ .
\end{equation}
We can see that in the state $\ket{\gamma}$ the part 
$\exp(\sqrt{2/\hbar}wu{\hat b}^*)\ket{0}$ 
is a superposition of the states with the maximum weight 
(the operation of ${\hat S}_-$ on this part makes it vanish) 
and the part $\exp[(v/u)\cdot {\hat S}_+/\hbar]$ denotes 
a superposition of the raising operation. The state $\ket{\gamma}$ 
can be rewritten in the form 
\begin{equation}\label{2-26}
\ket{\gamma}=
\exp\left(-\frac{|w|^2}{\hbar}\right)\cdot
\exp\left[\sqrt{\frac{2}{\hbar}}w(v{\hat a}^*+u{\hat b}^*)\right]
\ket{0} \ . 
\end{equation}
It may be clear that $\ket{\gamma}$ is a boson coherent state, 
which satisfies 
\begin{equation}\label{2-27}
{\hat a}\ket{\gamma}=\sqrt{\frac{2}{\hbar}}wv\ket{\gamma} \ , 
\qquad
{\hat b}\ket{\gamma}=\sqrt{\frac{2}{\hbar}}wu\ket{\gamma} \ .
\end{equation}
In other representation, we have 
\begin{eqnarray}\label{2-28}
& &(v^*{\hat a}+u{\hat b})\ket{\gamma} = \sqrt{\frac{2}{\hbar}}
w\ket{\gamma} \ , \nonumber\\
& &(u{\hat a}-v{\hat b})\ket{\gamma} = 0 \ .
\end{eqnarray}
Here, $v^*{\hat a}+u{\hat b}$ and $u{\hat a}-v{\hat b}$ 
represent annihilations of independent boson operators.

The expectation values of ${\hat S}$ and ${\hat S}_{\pm,0}$ 
are calculated in the following form : 
\begin{eqnarray}
& &\bra{\gamma} {\hat S} \ket{\gamma} 
= (S)_c = |w|^2 \ , \label{2-29}\\
& &\bra{\gamma} {\hat S}_+ \ket{\gamma} 
= (S_+)_c = 2|w|^2 v^* u \ , \nonumber\\
& &\bra{\gamma} {\hat S}_- \ket{\gamma} 
= (S_-)_c = 2|w|^2 uv \ , \nonumber\\
& &\bra{\gamma} {\hat S}_0 \ket{\gamma} 
= (S_0)_c = |w|^2 (|v|^2-u^2) \ . \label{2-30}
\end{eqnarray}
Instead of the parameters $(w, w^*)$ and $(v, v^*)$, we introduce 
the parameters $({\mib \psi}, {\mib S})$ and 
$({\mib X} , {\mib X}^*)$, which satisfy the conditions 
\begin{eqnarray}
& &\bra{\gamma} i\hbar \partial_{\mib \psi} \ket{\gamma}
={\mib S} \ , \qquad
\bra{\gamma} i\hbar \partial_{\mib S} \ket{\gamma}
=0 \ , \label{2-31}\\
& &\bra{\gamma} \partial_{\mib X} \ket{\gamma}
=+\frac{{\mib X}^*}{2} \ , \qquad
\bra{\gamma} \partial_{\mib X^*} \ket{\gamma}
=-\frac{{\mib X}}{2} \ . \label{2-32}
\end{eqnarray}
We can learn in the TDHF theory that the parameters 
$({\mib \psi}, {\mib S})$ and $({\mib X}, {\mib X}^*)$ are 
regarded as canonical variables in classical 
mechanics.\cite{MMSK,YK87} 
The variables $({\mib \psi}, {\mib S})$ are called the phase 
angle and the action variable and $({\mib X}, {\mib X}^*)$ 
are classical variables in boson type. For ${\mib S}$, we set up 
the relation 
\begin{equation}\label{2-33}
(S)_c = |w|^2 ={\mib S} \ .
\end{equation}
The conditions (\ref{2-31}) and (\ref{2-32}) give us the following 
relations :
\begin{eqnarray}\label{2-34}
& &w = \sqrt{\mib S} \exp\left(-i\frac{\mib \psi}{2}\right) \ , 
\nonumber\\
& &v=\sqrt{\frac{\hbar}{2{\mib S}}}{\mib X} \ , \qquad
u=\sqrt{1-\frac{\hbar}{2{\mib S}}{\mib X}^*{\mib X}} \ .
\end{eqnarray}
Substituting the relation (\ref{2-34}) into the expectation 
values $(S_{\pm,0})_c$ in Eq. (\ref{2-30}), we have 
\begin{eqnarray}\label{2-35}
& &(S_+)_c = \sqrt{\hbar}{\mib X}^*\cdot 
\sqrt{2{\mib S}-\hbar{\mib X}^*{\mib X}} \ , \nonumber\\
& &(S_-)_c = \sqrt{2{\mib S}-\hbar{\mib X}^*{\mib X}}
\cdot \sqrt{\hbar}{\mib X} \ , \nonumber\\
& &(S_0)_c = \hbar{\mib X}^*{\mib X}-{\mib S} \ . 
\end{eqnarray}
The expectation values $(S_{\pm,0})_c$ shown in the relation 
(\ref{2-35}) satisfy 
\begin{eqnarray}
& &(S_-)_c^* = (S_+)_c \ , \qquad 
(S_0)_c^* = (S_0)_c \ , \label{2-36}\\
& &[ (S_+)_c , (S_-)_c ]_{\rm P} = (-i)[2(S_0)_c] \ , \nonumber\\
& &[ (S_0)_c , (S_{\pm})_c ]_{\rm P} = (-i)[\pm (S_{\pm})_c] \ , 
\label{2-37}\\
& &(S_0)_c^2+\frac{1}{2}((S_-)_c(S_+)_c+(S_+)_c(S_-)_c) 
= {\mib S}^2 \ . \label{2-38}
\end{eqnarray}
Here, $[ A, B ]_{\rm P}$ denotes the Poisson bracket defined by 
\begin{equation}\label{2-39}
[ A, B ]_{\rm P} 
= \left(\frac{\partial A}{\partial {\mib \psi}}
\frac{\partial B}{\partial {\mib S}}
-\frac{\partial B}{\partial {\mib \psi}}
\frac{\partial A}{\partial {\mib S}}\right)
+\frac{1}{i\hbar}
\left(\frac{\partial A}{\partial {\mib X}}
\frac{\partial B}{\partial {\mib X}^*}-
\frac{\partial B}{\partial {\mib X}}
\frac{\partial A}{\partial {\mib X}^*}\right) \ .
\end{equation}
The relations (\ref{2-36}) $\sim$ (\ref{2-38}) correspond 
to Eqs. (\ref{2-18}) $\sim$ (\ref{2-20}) and, in this sense, 
the set $\{(S_{\pm, 0})_c\}$ can be called the classical counterpart 
of the set $\{{\widetilde S}_{\pm, 0} \}$ and, further, 
$\{ {\hat S}_{\pm, 0} \}$. Under the replacements 
$({\mib X}, {\mib X}^*) \rightarrow ({\hat X}, {\hat X}^*)$ 
and ${\mib S}\rightarrow S$, the classical counterpart becomes 
the quantal form.

\section{The $su(1,1)$-algebra---The boson realization 
and its classical counterpart}

Main aim of this section is to recapitulate the boson representation 
of the $su(1,1)$-spin system in a form suitable for later discussion. 
In this case, various relations are not so familiar 
to those in the $su(2)$-spin system. In a manner similar to 
the case of the $su(2)$-spin, we define the following three 
operators :
\begin{equation}\label{3-1}
{\hat T}_+=\hbar {\hat a}^*{\hat b}^* \ , \quad
{\hat T}_-=\hbar {\hat b}{\hat a} \ , \quad
{\hat T}_0=\frac{\hbar}{2}({\hat a}^*{\hat a}+{\hat b}{\hat b}^*) 
\ .
\end{equation}
The operators ${\hat T}_{\pm,0}$ form the $su(1,1)$-algebra, which 
obeys
\begin{eqnarray}
& &{\hat T}_-^*={\hat T}_+ \ , \qquad
{\hat T}_0^*={\hat T}_0 \ , \label{3-2}\\
& &[ {\hat T}_+ , {\hat T}_- ]= -2\hbar {\hat T}_0 \ , \qquad
[{\hat T}_0 , {\hat T}_{\pm} ]=\pm \hbar {\hat T}_{\pm} \ .
\label{3-3}
\end{eqnarray}
The form (\ref{3-1}) is called the Schwinger boson representation 
of the $su(1,1)$-algebra.\cite{Schw} 
The Casimir operator 
${\hat \Gamma}_{su(1,1)}$ 
can be written as 
\begin{equation}\label{3-4}
{\hat \Gamma}_{su(1,1)}={\hat T}_0^2-\frac{1}{2}
({\hat T}_-{\hat T}_+ + {\hat T}_+{\hat T}_-) 
={\hat T}({\hat T}-\hbar) \ .
\end{equation}
Here, ${\hat T}$ and its property are given as 
\begin{eqnarray}
& &{\hat T}=-\frac{\hbar}{2}({\hat a}^*{\hat a}
-{\hat b}{\hat b}^*) \ , \label{3-5}\\
& &[{\hat T} , {\hat T}_{\pm, 0} ]=0 \ .\label{3-6}
\end{eqnarray}
The operator ${\hat T}$ is not positive-definite, but, in a sense 
analogous to the case of the $su(2)$-spin, ${\hat T}$ may be 
called the magnitude of the $su(1,1)$-spin.

The operator ${\hat T}$ is not positive-definite and the 
eigenvalue of ${\hat T}_0$ is larger than $\hbar/2$. Then, 
compared with the case of the $su(2)$-spin system in the 
relations (\ref{2-7}) $\sim$ (\ref{2-10}), the treatment of 
the eigenstate of ${\hat T}$ and ${\hat T}_0$ is rather 
complicated. The eigenstate of ${\hat T}$ and ${\hat T}_0$
with the 
eigenvalues $T$ and $T_0$, respectively, is formally 
obtained in the form 
\begin{eqnarray}
& &\ket{T, T_0}=\frac{1}{\sqrt{(t_0-t)!(t_0+t-1)!}}
({\hat a}^*)^{t_0-t}({\hat b}^*)^{t_0+t-1}\ket{0} \ , \label{3-7}\\
& &T=\hbar t\ , \qquad T_0=\hbar t_0 \ . \label{3-8}
\end{eqnarray}
Since $t_0 \ge t$, $t_0 \ge -t+1$ and $t_0 \ge 1/2$, the state 
$\ket{T, T_0}$ is divided into two cases :
\begin{eqnarray}
& &{\rm (i)} \ \ t=\frac{1}{2}, 1, \frac{3}{2}, 2, \cdots , 
\qquad
t_0=t, t+1, t+2, \cdots \ ,\label{3-9}\\
& &{\rm (ii)} \ t=0, -\frac{1}{2}, -1, -\frac{3}{2}, \cdots , 
\qquad
t_0=-t+1, -t+2, -t+3, \cdots \ .\label{3-10}
\end{eqnarray}
In this paper, we restrict ourselves to case (i), i.e., 
the case $t\ge 1/2$. 
We call the space spanned by the set $\{ \ket{T, T_0} \}$ 
for fixed value of $T\ (\ge \hbar/2)$ the $su(1,1)$-space 
with the magnitude 
$T$ and at several occasions, we will use the notation 
$\ket{t, t_0}$ for $\ket{T, T_0}$.

In the same manner as that in the case of the $su(2)$-spin, let us 
derive the Holstein-Primakoff representation.\cite{TKY1} 
We set up the following correspondence between the original 
and the new boson space : 
\begin{equation}\label{3-11}
\ket{t, t_0} \sim \rket{t_0-t} \ . \qquad
(T=\hbar t \ , \ T_0=\hbar t_0 )
\end{equation}
Since 
$t_0=t, t+1, t+2, \cdots$, the whole space in the new space is 
physical. 
Under the correspondence 
(\ref{3-11}), the following mapping operator ${\hat V}$ is 
introduced : 
\begin{equation}\label{3-12}
{\hat V} = \sum_{t_0=t} \rket{t_0-t}\bra{t, t_0} \ .
\end{equation}
The properties of ${\hat V}$ are shown in the form 
\begin{eqnarray}\label{3-13}
& &{\hat V}^{\dagger}{\hat V} 
= \sum_{t_0=t}\ket{t, t_0}\bra{t, t_0} = 1 \ , \nonumber\\
& &{\hat V}{\hat V}^{\dagger}
=\sum_{t_0=t} \rket{t_0-t}\rbra{t_0-t} = 1 \ .
\end{eqnarray}
In contrast with the relation (\ref{2-13}), there does not 
exist the projection operator such as $P_{2s}$ in the relation 
(\ref{3-13}). From the above conditions, the following relations 
are derived :
\begin{eqnarray}
& &{\hat V} \ket{t, t_0} = \rket{t_0-t} \ , \label{3-14}\\
& &{\hat V}^{\dagger} \rket{t_0-t} = \ket{t, t_0} \ . 
\label{3-15}
\end{eqnarray}
With the use of ${\hat V}$, the operators ${\hat T}_{\pm, 0}$ 
are transformed to the form
\begin{eqnarray}\label{3-16}
& &{\hat V}{\hat T}_{\pm,0}{\hat V}^{\dagger} 
={\widetilde T}_{\pm,0} \ , \nonumber\\
& &{\widetilde T}_+ = \sqrt{\hbar}{\hat X}^* \cdot
\sqrt{2T+\hbar{\hat X}^*{\hat X}} \ , \nonumber\\
& &{\widetilde T}_- = \sqrt{2T+\hbar{\hat X}^*{\hat X}} \cdot
\sqrt{\hbar}{\hat X} \ , \nonumber\\
& &{\widetilde T}_0 = \hbar {\hat X}^*{\hat X} + T \ .
\end{eqnarray}
The above operators ${\widetilde T}_{\pm,0}$ satisfy 
\begin{eqnarray}
& &{\widetilde T}_-^* = {\widetilde T}_+ \ , \qquad
{\widetilde T}_0^* = {\widetilde T}_0 \ , \label{3-17}\\
& &[ {\widetilde T}_+ , {\widetilde T}_- ] = -2\hbar {\widetilde T}_0 \ , 
\qquad
[ {\widetilde T}_0 , {\widetilde T}_{\pm} ] 
= \pm \hbar {\widetilde T}_{\pm} \ , \label{3-18}\\
& &{\widetilde T}_0^2-\frac{1}{2}
({\widetilde T}_-{\widetilde T}_+ + {\widetilde T}_+{\widetilde T}_-) 
= T(T-\hbar) \ .\label{3-19}
\end{eqnarray}
The relations (\ref{3-17}) $\sim$ (\ref{3-19}) correspond to 
Eqs. (\ref{3-2}) $\sim$ (\ref{3-4}). 
Judging from the form, it can be 
called the Holstein-Primakoff boson representation.

In parallel with the case of the $su(2)$-spin system, 
our third problem is concerned with the coherent 
state $\ket{c}$ for the $su(1,1)$-spin system.\cite{TKY1} 
For the 
construction of $\ket{c}$, we note the form 
\begin{equation}\label{3-20}
\ket{t, t_0}=\frac{1}{\sqrt{(t_0-t)!(t_0+t-1)!}}
\left(\frac{{\hat T}_+}{\hbar}\right)^{t_0-t}\cdot
({\hat b}^*)^{2t-1} \ket{0} \ . 
\end{equation}
Then, we set up the following correspondence : 
\begin{eqnarray}\label{3-21}
\frac{1}{\sqrt{(t_0-t)!(t_0+t-1)!}}
\left(\frac{{\hat T}_+}{\hbar}\right)^{t_0-t}
&\sim& \exp\left( \frac{V}{U}\cdot \frac{{\hat T}_+}{\hbar}\right) 
=\exp \left(\frac{V}{U}\cdot {\hat a}^* {\hat b}^*\right) \ , 
\nonumber\\
({\hat b}^*)^{2t-1}\ket{0}
&\sim& \exp\left(\sqrt{\frac{2}{\hbar}}\frac{W}{U}{\hat b}^*\right)
\ket{0} \ .
\end{eqnarray}
Here, $W$ and $V$ denote complex parameters and $U$ is given as 
\begin{equation}\label{3-22}
U=\sqrt{1+|V|^2} \ .
\end{equation}
Under the correspondence (\ref{3-21}), we define the state 
$\ket{c}$ in the form 
\begin{equation}\label{3-23}
\ket{c}=N_c\cdot \exp\left(
\frac{V}{U}\cdot \frac{{\hat T}_+}{\hbar}\right)\cdot
\exp\left(\sqrt{\frac{2}{\hbar}}\frac{W}{U}{\hat b}^*\right) \ket{0} \ .
\end{equation}
Here, $N_c$ denotes the normalization constant given by 
\begin{equation}\label{3-24}
N_c=\frac{1}{U}\exp\left(-\frac{|W|^2}{\hbar}\right) \ .
\end{equation}
In this case, also, 
we can see that in the state $\ket{c}$ the part 
$\exp(\sqrt{2/\hbar}W/U\cdot{\hat b}^*)\ket{0}$ 
is a superposition of the states with the maximum weight 
(the operation of ${\hat T}_-$ on this part makes it vanish) 
and the part $\exp[(V/U)\cdot {\hat T}_+/\hbar]$ denotes 
a superposition of the raising operation. The state $\ket{c}$ 
satisfies 
\begin{eqnarray}\label{3-25}
& &(U{\hat b}-V{\hat a}^*)\ket{c} = \sqrt{\frac{2}{\hbar}}
W\ket{c} \ , \nonumber\\
& &(U{\hat a}-V{\hat b}^*)\ket{c} = 0 \ .
\end{eqnarray}
The operators $U{\hat b}-V{\hat a}^*$ and $U{\hat a}-V{\hat b}^*$ 
represent annihilation of independent boson operators and 
$\ket{c}$ is a coherent state in the sense given in the 
relation (\ref{3-25}). 
As is clear from the form (\ref{2-26}), the state $\ket{\gamma}$ 
is one-mode coherent state and, in the present case, 
$\ket{c}$ is mixed-mode coherent state.

The expectation values of ${\hat T}$ and ${\hat T}_{\pm,0}$ 
are calculated in the following form : 
\begin{eqnarray}
& &\bra{c} {\hat T} \ket{c} 
= (T)_c = |W|^2 + \frac{\hbar}{2}\ , \label{3-26}\\
& &\bra{c} {\hat T}_+ \ket{c} 
= (T_+)_c = 2\left(|W|^2 + \frac{\hbar}{2}\right)V^* U \ , \nonumber\\
& &\bra{c} {\hat T}_- \ket{c} 
= (T_-)_c = 2\left(|W|^2 +\frac{\hbar}{2}\right) UV \ , \nonumber\\
& &\bra{c} {\hat T}_0 \ket{c} 
= (T_0)_c = \left(|W|^2 +\frac{\hbar}{2}\right)(|V|^2+U^2) \ . 
\label{3-27}
\end{eqnarray}
In the same manner as in the case of the $su(2)$-spin system, 
we introduce 
the parameters $({\mib \phi}, {\mib T})$ and 
$({\mib X} , {\mib X}^*)$ for the old $(W, W^*)$ and 
$(V, V^*)$. For these variables, we set up the conditions 
\begin{eqnarray}
& &\bra{c} i\hbar \partial_{\mib \phi} \ket{c}
={\mib T}-\frac{\hbar}{2} \ , \qquad
\bra{c} i\hbar \partial_{\mib T} \ket{c}
=0 \ , \label{3-28}\\
& &\bra{c} \partial_{\mib X} \ket{c}
=+\frac{{\mib X}^*}{2} \ , \qquad
\bra{c} \partial_{\mib X^*} \ket{c}
=-\frac{{\mib X}}{2} \ . \label{3-29}
\end{eqnarray}
The set of the relations (\ref{3-28}) and (\ref{3-29}) 
is called the canonicity condition.\cite{MMSK,YK87} 
For ${\mib T}$, we use the relation 
\begin{equation}\label{3-30}
(T)_c = |W|^2 + \frac{\hbar}{2} = {\mib T} \ .
\end{equation}
The conditions (\ref{3-28}) and (\ref{3-29}) give us the form 
\begin{eqnarray}\label{3-31}
& &W = \sqrt{{\mib T}-\frac{\hbar}{2}} 
\exp\left(-i\frac{\mib \phi}{2}\right) \ , 
\nonumber\\
& &V=\sqrt{\frac{\hbar}{2{\mib T}}}{\mib X} \ , \qquad
U=\sqrt{1+\frac{\hbar}{2{\mib T}}{\mib X}^*{\mib X}} \ .
\end{eqnarray}
Substituting the relation (\ref{3-31}) into the expectation 
values $(T_{\pm,0})_c$ in Eq. (\ref{3-27}), we obtain 
\begin{eqnarray}\label{3-32}
& &(T_+)_c = \sqrt{\hbar}{\mib X}^*\cdot 
\sqrt{2{\mib T}+\hbar{\mib X}^*{\mib X}} \ , \nonumber\\
& &(T_-)_c = \sqrt{2{\mib T}+\hbar{\mib X}^*{\mib X}}
\cdot \sqrt{\hbar}{\mib X} \ , \nonumber\\
& &(T_0)_c = \hbar{\mib X}^*{\mib X}+{\mib T} \ . 
\end{eqnarray}
The expectation values $(T_{\pm,0})_c$ shown in the relation 
(\ref{3-32}) satisfy 
\begin{eqnarray}
& &(T_-)_c^* = (T_+)_c \ , \qquad 
(T_0)_c^* = (T_0)_c \ , \label{3-33}\\
& &[ (T_+)_c , (T_-)_c ]_{\rm P} = (-i)[-2(T_0)_c] \ , \nonumber\\
& &[ (T_0)_c , (T_{\pm})_c ]_{\rm P} = (-i)[\pm (T_{\pm})_c] \ , 
\label{3-34}\\
& &(T_0)_c^2-\frac{1}{2}((T_-)_c(T_+)_c+(T_+)_c(T_-)_c) 
= {\mib T}^2 \ . \label{3-35}
\end{eqnarray}
Here, $[ A, B ]_{\rm P}$ is the same as that shown in 
Eq. (\ref{2-39}) under the replacement (${\mib \psi} \rightarrow 
{\mib \phi}$ and ${\mib S} \rightarrow {\mib T}$). 
The relations (\ref{3-33}) $\sim$ (\ref{3-35}) correspond 
to Eqs. (\ref{3-17}) $\sim$ (\ref{3-19}) and, in this sense, 
the set $\{(T_{\pm, 0})_c\}$ is the classical counterpart 
of the set $\{{\widetilde T}_{\pm, 0} \}$. 
This is in the same situation as the case of the $su(2)$-spin.

\section{The $su(1,1)$-algebra---One-dimensional ``damped and 
amplified oscillator" and the $su(2)$-algebraic model}

Following Ref.\citen{Vitiello}, 
let us consider two-dimensional space, in which we define 
the following Lagrangian : 
\begin{equation}\label{4-1}
L=m{\dot X}{\dot Y}-kXY+m\gamma(X{\dot Y}-{\dot X}Y-\gamma XY) \ .
\end{equation}
Here, $(X, Y)$ denote two kinds of real variables and 
$m$, $k$ and $\gamma$ are positive constants. The Lagrange 
equations of motion lead us to the equations of motion 
\begin{eqnarray}
& &m{\ddot X} = -(k+m\gamma^2)X-2m\gamma{\dot X} \ , \label{4-2}\\
& &m{\ddot Y} = -(k+m\gamma^2)Y+2m\gamma{\dot Y} \ . \label{4-3}
\end{eqnarray}
The solutions of Eqs. (\ref{4-2}) and (\ref{4-3}) are expressed in 
the form 
\begin{eqnarray}
& &X=A^0\exp(-\gamma t)\cos (\omega t+\alpha^0) \ , \label{4-4}\\
& &Y=B^0\exp(+\gamma t)\cos (\omega t+\beta^0) \ , \label{4-5}\\
& &\omega = \sqrt{\frac{k}{m}} \ . \label{4-6}
\end{eqnarray}
The solutions (\ref{4-4}) and (\ref{4-5}) show the 
``damped and amplified oscillation." 
The Lagrangian (\ref{4-1}) can be rewritten as 
\begin{equation}\label{4-7}
L=\left(\frac{1}{2}m{\dot x}^2-\frac{1}{2}kx^2\right)
-\left(\frac{1}{2}{\dot y}^2-\frac{1}{2}ky^2\right)
-m\left[\gamma(x{\dot y}-{\dot x}y)
+\frac{\gamma^2}{2}(x^2-y^2)\right] \ .
\end{equation}
Here, $x$ and $y$ are defined as 
\begin{equation}\label{4-8}
x=\frac{1}{\sqrt{2}}(X+Y) \ , \qquad
y=\frac{1}{\sqrt{2}}(X-Y) \ . 
\end{equation}
The Lagrangian (\ref{4-7}) gives us the following Hamiltonian :
\begin{eqnarray}
& &H=\left(\frac{p_x^2}{2m}+\frac{1}{2}m\omega^2 x^2\right)
-\left(\frac{p_y^2}{2m}+\frac{1}{2}m\omega^2 y^2\right)
-\gamma(xp_y+yp_x) \ , \label{4-9}\\
& & \ \ 
p_x=\frac{\partial L}{\partial{\dot x}}
=m{\dot x}+m\gamma y\ , \qquad
p_y=\frac{\partial L}{\partial{\dot y}}
=-m{\dot y}-m\gamma x\ . \label{4-10}
\end{eqnarray}
After canonical quantization, i.e., $x\rightarrow {\hat x}$, 
$y\rightarrow {\hat y}$, $p_x \rightarrow {\hat p}_x$ and 
$p_y \rightarrow {\hat p}_y$ : 
$[{\hat x} , {\hat p}_x ] = [ {\hat y} , {\hat p}_y ] = i\hbar$ 
and [ other combinations ] =0, we obtain the quantized 
Hamiltonian : 
\begin{eqnarray}\label{4-11}
{\hat H} &=& \hbar\omega{\hat b}^*{\hat b}
-\hbar \omega{\hat a}^*{\hat a}-i\hbar\gamma
({\hat a}^*{\hat b}^*-{\hat b}{\hat a}) \nonumber\\
&=& 2\omega \left({\hat T}-\frac{\hbar}{2}\right)
-i\gamma({\hat T}_+ - {\hat T}_-) \ .
\end{eqnarray}
Here, $({\hat a}, {\hat a}^*)$ and $({\hat b} , {\hat b}^*)$ 
are defined as 
\begin{eqnarray}\label{4-12}
& &{\hat a}=\frac{1}{\sqrt{2\hbar}}\left(\sqrt{m\omega}{\hat y}
+i\frac{1}{\sqrt{m\omega}}{\hat p}_y \right) \ , \quad
{\hat a}^{*}=\frac{1}{\sqrt{2\hbar}}\left(\sqrt{m\omega}{\hat y}
-i\frac{1}{\sqrt{m\omega}}{\hat p}_y \right) \ , \nonumber\\
& &{\hat b}=\frac{1}{\sqrt{2\hbar}}\left(\sqrt{m\omega}{\hat x}
+i\frac{1}{\sqrt{m\omega}}{\hat p}_x \right) \ , \quad
{\hat b}^{*}=\frac{1}{\sqrt{2\hbar}}\left(\sqrt{m\omega}{\hat x}
-i\frac{1}{\sqrt{m\omega}}{\hat p}_x \right) \ . \qquad
\end{eqnarray}
We can see that the Hamiltonian (\ref{4-11}) is expressed 
in terms of ${\hat T}$ and ${\hat T}_{\pm}$. This means that 
the quantized ``damped and amplified oscillator" can be 
described in the framework of the $su(1,1)$-algebra, in other 
word, it is a kind of the $su(1,1)$-spin system.

The above treatment is interpreted as follows : 
In order to treat the system such as the ``damped and amplified 
oscillator" as an isolated system, so-called phase space doubling 
is required. Then, the original intrinsic oscillator expressed in 
terms of the boson $({\hat b} , {\hat b}^*)$ and the 
``external environment" expressed in terms of the boson 
$({\hat a},{\hat a}^*)$ appear and the interaction between 
both system is naturally introduced. This is our interpretation, 
which permits us to generalize the 
Hamiltonian (\ref{4-11}).\cite{TKY3} 
Let the Hamiltonian of the intrinsic system be ${\hat K}$ 
given in the form 
\begin{equation}\label{4-13}
{\hat K}=F(\hbar{\hat b}^*{\hat b}) \ .
\end{equation}
Here, $F$ is a function of $\hbar{\hat b}^*{\hat b}$. 
Then, a possible generalization is performed by setting the 
following Hamiltonian : 
\begin{equation}\label{4-14}
{\hat H}=F(\hbar{\hat b}^*{\hat b})-F(\hbar{\hat a}^*{\hat a}) 
-i\hbar\gamma({\hat a}^*{\hat b}^*-{\hat b}{\hat a}) \ . 
\end{equation}
If $F(\hbar{\hat b}^*{\hat b})=\hbar\omega{\hat b}^*{\hat b}$, 
the Hamiltonian (\ref{4-14}) is reduced to the Hamiltonian 
(\ref{4-11}). The Hamiltonian can be expressed in terms of the 
operators $({\hat T}, {\hat T}_{\pm,0})$ :
\begin{equation}\label{4-15}
{\hat H}=F({\hat T}_0+{\hat T}-\hbar)
-F({\hat T}_0-{\hat T})-i\gamma({\hat T}_+-{\hat T}_-) \ .
\end{equation}
For the convenience of late discussion, we rewrite the 
Hamiltonian (\ref{4-15}) in the form 
\begin{equation}\label{4-16}
{\hat H}=F({\hat T}_z+{\hat T}-\hbar)-F({\hat T}_z-{\hat T}) 
+2\gamma{\hat T}_y \ .
\end{equation}
Here, we adopt the following form : 
\begin{equation}\label{4-17}
{\hat T}_x=\frac{1}{2}({\hat T}_+ + {\hat T}_-) \ , 
\quad
{\hat T}_y=-\frac{i}{2}({\hat T}_+ - {\hat T}_-) \ , 
\quad
{\hat T}_z={\hat T}_0 \ .
\end{equation}

We can find two $su(2)$-algebraic models in nuclear physics. 
One is a model which enables us to describe pairing correlation 
in single-orbit shell model and another is for describing 
particle-hole correlation in two single-particle levels. 
In the first case, the intrinsic Hamiltonian ${\hat K}$ is 
given by 
\begin{equation}\label{4-18}
{\hat K}=2\epsilon({\hat S}_0+S)-G{\hat S}_+{\hat S}_- \ .
\qquad
(S=\hbar(2j+1)/4)
\end{equation}
The quantities $\epsilon$ and $G$ denote the energy of the 
single-particle state and the strength of the pairing interaction, 
respectively. Further, $j$ denotes the magnitude of the angular 
momentum of the orbit. Of course, the set $({\hat S}_{\pm,0})$ 
forms the $su(2)$-algebra and $S$ denotes the magnitude of 
the $su(2)$-spin. Then, if ${\hat S}_{\pm,0}$ are replaced 
with ${\widetilde S}_{\pm,0}$ shown in Eq. (\ref{2-17}), 
${\hat K}$ becomes of the form 
\begin{equation}\label{4-19}
{\hat K}=2\left[\epsilon-G\left(S+\frac{\hbar}{2}\right)\right]
\cdot(\hbar{\hat b}^*{\hat b})+G\cdot(\hbar{\hat b}^*{\hat b})^2 \ . 
\end{equation}
Here, $({\hat X}, {\hat X}^*)$ in Eq. (\ref{2-17}) is replaced 
with $({\hat b}, {\hat b}^*)$. 
Then, the Hamiltonian is given in the form 
\begin{equation}\label{4-20}
{\hat H}=4[\epsilon-G(S+\hbar)]\cdot
\left({\hat T}-\frac{\hbar}{2}\right)
+4G\cdot\left({\hat T}-\frac{\hbar}{2}\right){\hat T}_z + 2\gamma{\hat T}_y \ .
\end{equation}
The other is called the Lipkin model and the intrinsic Hamiltonian 
${\hat K}$ is expressed as 
\begin{equation}\label{4-21}
{\hat K}=2\epsilon({\hat S}+S)-\frac{G}{2}
({\hat S}_+^2+{\hat S}_-^2) \ .
\end{equation}
Here, $S$ denotes the maximum magnitude of the $su(2)$-spin 
that is equal to $\hbar\Omega/2$, where $\Omega$ is the 
degeneracy of the single-particle level. 
And $\epsilon$ and $G \ (>0)$ represent the single-particle 
energy-spacing and the strength of the interaction, respectively. 
The set $({\hat S}_{\pm, 0})$ denotes the $su(2)$-spin. 
In this case, together with the form (\ref{2-17}), 
we adopt the basic assumption of the ATDHF theory and 
the semi-classical approximation. Then, in the 
$T$-conserving framework, the Hamiltonian is given 
in the following : 
\begin{eqnarray}\label{4-22}
{\hat H}&=&\left[2\omega_0-\frac{6\hbar G}{\omega_0^2}
\left(\epsilon+G\left(S+\frac{\hbar}{2}\right)\right)^2\right]
\cdot\left({\hat T}-\frac{\hbar}{2}\right) \nonumber\\
& &+\frac{6G}{\omega_0^2}\left(\epsilon+G\left(S+\frac{\hbar}{2}
\right)\right)^2\cdot
\left({\hat T}-\frac{\hbar}{2}\right){\hat T}_z
+2\gamma{\hat T}_y \ .
\end{eqnarray}
Here, $\omega_0$ is defined by 
\begin{equation}\label{4-23}
\omega_0=2\left( \epsilon + G\left(S+\frac{\hbar}{2}\right)\right) 
\nu_0 \ .
\end{equation}
The quantity $\nu_0$ is a solution of the cubic equation 
\begin{equation}\label{4-24}
4\left(\epsilon + G\left(S+\frac{\hbar}{2}\right)\right)
\nu_0^3 - 4\left(\epsilon-G\left(S+\frac{\hbar}{2}\right)\right)
\nu_0 - 3\hbar G=0 \ .
\end{equation}
The detail of the above derivation has been given in 
Ref.\citen{TKY3}. The Hamiltonians (\ref{4-20}) and (\ref{4-22}) 
depend on $({\hat T}-\hbar/2)$, $({\hat T}-\hbar/2){\hat T}_z$ 
and ${\hat T}_y$ in the same form as each other and if the 
coefficient of the term $({\hat T}-\hbar/2){\hat T}_z$ vanishes, both 
Hamiltonians are reduced to the Hamiltonian (\ref{4-11}). 
Hereafter, we will treat both systems in a unified way : 
\begin{equation}\label{4-25}
{\hat H}=2\omega_T\cdot\left({\hat T}-\frac{\hbar}{2}\right)
+2f_T\cdot\left({\hat T}-\frac{\hbar}{2}\right){\hat T}_z
+2\gamma{\hat T}_y \ .
\end{equation}
The meanings of $\omega_T$ and $f_T$ may be self-evident. 
Hereafter, we will omit the subscript ``$T$" in $\omega_T$ 
and $f_T$.

We are now at stage to discuss the time-evolution of the 
coherent state (\ref{3-23}) for the Hamiltonian (\ref{4-25}) 
under the variational procedure. The expectation value of the 
Hamiltonian (\ref{4-25}) for the state (\ref{3-23}), which 
we denote $H$, is expressed in the form 
\begin{eqnarray}
& &H=2\omega\left({\mib T}-\frac{\hbar}{2}\right)
+2\lambda T_z + 2\gamma T_y \ , \label{4-26}\\
& & \ \ \lambda=\lambda({\mib T})
=f\cdot\left(1+\frac{\hbar}{2{\mib T}}\right)
\left({\mib T}-\frac{\hbar}{2}\right) \ . \label{4-27}
\end{eqnarray}
The expectation values of ${\hat T}_x$, ${\hat T}_y$ and 
${\hat T}_z$ for the state ({\ref{3-23}) are given as follows :
\begin{eqnarray}\label{4-28}
& &T_x = \frac{\sqrt{\hbar}}{2}({\mib X}^* + {\mib X})
\cdot\sqrt{2{\mib T}+\hbar{\mib X}^*{\mib X}} \ , \nonumber\\
& &T_y = -i\frac{\sqrt{\hbar}}{2}({\mib X}^* - {\mib X})
\cdot\sqrt{2{\mib T}+\hbar{\mib X}^*{\mib X}} \ , \nonumber\\
& &T_z = \hbar{\mib X}^*{\mib X}+{\mib T} \ .
\end{eqnarray}
Here, we used the definition (\ref{4-17}) and the form (\ref{3-32}) 
under the omission of the symbol $(\ \ )_c$. 
The set (\ref{4-28}) satisfies the relations 
\begin{eqnarray}
& &[ T_y , T_z]_{\rm P} = T_x \ , \qquad
[ T_z , T_x]_{\rm P} = T_y \ , \qquad
[ T_x , T_y]_{\rm P} = -T_z \ , \label{4-29}\\
& &[ T , T_x]_{\rm P} = [ T , T_y]_{\rm P} 
=[ T , T_z]_{\rm P} = 0 \ , \label{4-30}\\
& &T_z^2-T_x^2-T_y^2={\mib T}^2 \ . \label{4-31}
\end{eqnarray}
Since ${\mib T}$ and ${\mib \phi}$ are mutually canonical, 
the equations of motion are obtained as 
\begin{eqnarray}
& &{\dot {\mib T}} = [ {\mib T} , H ]_{\rm P} = 0 \ , 
\label{4-32}\\
& &{\dot {\mib \phi}} = [ {\mib \phi} , H ]_{\rm P} 
= 2\omega + \frac{\partial}{\partial {\mib T}}
(2\lambda T_z + 2\gamma T_y) \ .
\label{4-33}
\end{eqnarray}
Since ${\mib T}$ is a constant of motion, we set up 
\begin{equation}\label{4-34}
{\mib T}={\mib I} \ .
\end{equation}
The quantity ${\mib I}$ is determined at the initial time. 
Instead of examining the Hamilton equations of motion for 
$({\mib X} , {\mib X}^*)$, let us consider the equations of 
motion for $T_x$, $T_y$ and $T_z$ : 
\begin{eqnarray}
& &{\dot T}_x = [ T_x , H ]_{\rm P}
=-2\lambda T_y - 2\gamma T_z \ , \label{4-35}\\
& &{\dot T}_y = [ T_y , H ]_{\rm P}
=+2\lambda T_x \ , \label{4-36}\\
& &{\dot T}_z = [ T_z , H ]_{\rm P}
=-2\gamma T_x \ , \label{4-37}
\end{eqnarray}
From these equations, we have 
\begin{equation}\label{4-38}
{\ddot T}_x=-4(\lambda^2-\gamma^2)T_x \ . \qquad
(\lambda=\lambda({\mib T}) )
\end{equation}
Clearly, the solutions of Eq. (\ref{4-38}) are classified 
into four cases :
\begin{equation}\label{4-39}
{\rm (i)}\ \lambda^2 > \gamma^2 \ , \quad
{\rm (ii)}\ \lambda^2 < \gamma^2 \ , \quad
{\rm (iii)}\ \lambda = -\gamma \ , \quad
{\rm (iv)}\ \lambda = + \gamma \ .
\end{equation}
The results of the solutions of Eq. (\ref{4-38}) and 
the related points were given in Ref.\citen{TKY4}. 
In this paper, only the results are summarized. 
Including Eqs. (\ref{4-32}) and (\ref{4-33}), we adopt the 
canonical transformation from 
$({\mib \phi}, {\mib T} ; {\mib X}, {\mib X}^*)$ to 
$({\mib \theta}, {\mib I} ; {\mib \chi}, {\mib J})$. 

\vspace{0.2cm}
\noindent
Case (i) : $\lambda^2 > \gamma^2 \ 
(\lambda_0=\sqrt{\lambda^2-\gamma^2})$
\begin{eqnarray}
& &T_x=-\sqrt{{\mib J}^2-{\mib I}^2}\sin{\mib \chi} \ , \nonumber\\
& &T_y=\frac{\lambda}{\lambda_0}\sqrt{{\mib J}^2-{\mib I}^2}
\cos{\mib \chi} - \frac{\gamma}{\lambda_0}{\mib J} \ , \nonumber\\
& &T_z=-\frac{\gamma}{\lambda_0}\sqrt{{\mib J}^2-{\mib I}^2}
\cos{\mib \chi} + \frac{\lambda}{\lambda_0}{\mib J} \ , 
\label{4-40}\\
& &{\mib T}={\mib I} \ , \nonumber\\
& &{\mib \phi}={\mib \theta}+2 \tan^{-1}\left[
\frac{\lambda_0{\mib I}+\lambda {\mib J}
+\gamma\sqrt{{\mib J}^2-{\mib I}^2}}
{\lambda{\mib I}+\lambda_0{\mib J}}
\tan \frac{{\mib \chi}}{2}\right]
+\frac{\gamma}{\lambda_0^2}\frac{d\lambda}{d{\mib I}}\sin{\mib \chi} \ ,
\label{4-41}\\
& &H=2\omega\left({\mib I}-\frac{\hbar}{2}\right)+2\lambda_0{\mib J} \ , 
\label{4-42}\\
& &{\mib \chi}=2\lambda_0 t+\chi_0 \ , \qquad
{\mib \theta}=2\omega t + \theta_0 \ .
\label{4-43}
\end{eqnarray}

\vspace{0.2cm}
\noindent
Case (ii) : $\lambda^2 < \gamma^2 \ 
(\gamma_0=\sqrt{\gamma^2-\lambda^2})$
\begin{eqnarray}
& &T_x=-\sqrt{{\mib J}^2+{\mib I}^2}\sinh{\mib \chi} \ , \nonumber\\
& &T_y=-\frac{\lambda}{\gamma_0}\sqrt{{\mib J}^2+{\mib I}^2}
\cosh{\mib \chi} + \frac{\gamma}{\gamma_0}{\mib J} \ , \nonumber\\
& &T_z=\frac{\gamma}{\gamma_0}\sqrt{{\mib J}^2+{\mib I}^2}
\cosh{\mib \chi} - \frac{\lambda}{\gamma_0}{\mib J} \ , 
\label{4-44}\\
& &{\mib T}={\mib I} \ , \nonumber\\
& &{\mib \phi}={\mib \theta}+2 \tan^{-1}\left[
\frac{\gamma_0{\mib I}-\lambda {\mib J}
-\gamma\sqrt{{\mib J}^2+{\mib I}^2}}
{\lambda{\mib I}+\gamma_0{\mib J}}
\tanh \frac{{\mib \chi}}{2}\right]
+\frac{\gamma}{\gamma_0^2}\frac{d\lambda}{d{\mib I}}\sinh{\mib \chi} \ ,
\label{4-45}\\
& &H=2\omega\left({\mib I}-\frac{\hbar}{2}\right)+2\gamma_0{\mib J} \ , 
\label{4-46}\\
& &{\mib \chi}=2\gamma_0 t+\chi_0 \ , \qquad
{\mib \theta}=2\omega t + \theta_0 \ .
\label{4-47}
\end{eqnarray}

\vspace{0.2cm}
\noindent
Case (iii) : $\lambda = - \gamma$
\begin{eqnarray}
& &T_x={\mib \chi}{\mib J} \ , \nonumber\\
& &T_y=\frac{{\mib \chi}^2}{2}{\mib J}
-\frac{1}{2}\left({\mib J}-\frac{{\mib I}^2}{\mib J}\right) \ , 
\nonumber\\
& &T_z=\frac{{\mib \chi}^2}{2}{\mib J}
+\frac{1}{2}\left({\mib J}+\frac{{\mib I}^2}{\mib J}\right) \ , 
\label{4-48}\\
& &{\mib T}={\mib I} \ , \nonumber\\
& &{\mib \phi}={\mib \theta}+2 \tan^{-1}\left[
\frac{{\mib J}{\mib \chi}}{{\mib I}+{\mib J}}\right] \ ,
\label{4-49}\\
& &H=2\omega\left({\mib I}-\frac{\hbar}{2}\right)+2\lambda{\mib J} \ , 
\label{4-50}\\
& &{\mib \chi}=2\lambda t+\chi_0 \ , \qquad
{\mib \theta}=2\omega t + \theta_0 \ .
\label{4-51}
\end{eqnarray}

\vspace{0.2cm}
\noindent
Case (iv) : $\lambda = + \gamma$
\begin{eqnarray}
& &T_x=-{\mib \chi}{\mib J} \ , \nonumber\\
& &T_y=-\frac{{\mib \chi}^2}{2}{\mib J}
+\frac{1}{2}\left({\mib J}-\frac{{\mib I}^2}{\mib J}\right) \ , 
\nonumber\\
& &T_z=\frac{{\mib \chi}^2}{2}{\mib J}
+\frac{1}{2}\left({\mib J}+\frac{{\mib I}^2}{\mib J}\right) \ , 
\label{4-52}\\
& &{\mib T}={\mib I} \ , \nonumber\\
& &{\mib \phi}={\mib \theta}+2 \tan^{-1}\left[
\frac{{\mib J}{\mib \chi}}{{\mib I}+{\mib J}}\right] \ ,
\label{4-53}\\
& &H=2\omega\left({\mib I}-\frac{\hbar}{2}\right)+2\gamma{\mib J} \ , 
\label{4-54}\\
& &{\mib \chi}=2\gamma t+\chi_0 \ , \qquad
{\mib \theta}=2\omega t + \theta_0 \ .
\label{4-55}
\end{eqnarray}
We can see that ${\mib J}$ is also a constant of motion. 
The variables $({\mib X}, {\mib X}^*)$ are given by 
\begin{equation}\label{4-56}
\sqrt{\hbar}{\mib X}=\frac{T_x-iT_y}{\sqrt{{\mib T}+T_z}} \ , 
\qquad
\sqrt{\hbar}{\mib X}^*=\frac{T_x+iT_y}{\sqrt{{\mib T}+T_z}} \ .
\end{equation}
Then, the time-dependence of various quantities is 
determined as 
functions 
of $t$ for four cases. For example, $|V|^2$ given in Eq. (\ref{3-31}) 
is expressed in the form 
\begin{equation}\label{4-57}
|V|^2=\frac{\hbar{\mib X}^*{\mib X}}{2{\mib T}}
=\frac{T_x^2+T_y^2}{2{\mib T}({\mib T}+T_z)}
=\frac{T_z-{\mib I}}{2{\mib I}} \ .
\end{equation}
Here, the relation (\ref{4-31}) was used. It may be interesting 
to see that there exist four phases.

\section{Description of thermal effects in terms of the 
coherent state for the $su(1,1)$-algebra}

In \S 4, we discussed the $su(1,1)$-algebraic model, which 
permits us to describe influences of the ``external 
environment" on the intrinsic system under investigation 
such as the harmonic oscillator. 
As the result of the interaction between two systems, the 
intrinsic system becomes the ``damped and amplified oscillator." 
Therefore, it may be interesting to investigate such 
influences in the language of the thermal effects. 
We know that, in the thermo field dynamics formalism,\cite{TU} 
statistically mixed state can be described by the conventional 
Schr\"odinger equation with a modified Hamiltonian in a 
space with phase space doubling. In order to specify mixture 
of the pure states, the auxiliary variables are introduced. 
The formalism of the present $su(1,1)$-algebraic model 
resembles the thermo field dynamics formalism. The boson 
operators $({\hat b}, {\hat b}^*)$ describe the intrinsic system 
under investigation and $({\hat a}, {\hat a}^*)$ are 
introduced for the auxiliary variables. The coherent state 
$\ket{c}$ shown in Eq. (\ref{3-23}) contains both boson 
operators and, in the sense of the variational procedure, 
satisfies the Schr\"odinger equation 
%(\ref{4-26}) 
derived 
from the intrinsic Hamiltonian (\ref{4-13}) through 
the form (\ref{4-14}). Therefore, $\ket{c}$ can be regarded 
as the state which describes the mixed state for 
the present system.\cite{TKY2,YKT-p,TKY4,TKY-p}

First, let us expand partially the exponential function 
in the mixed-mode coherent state $\ket{c}$ in Eq. (\ref{3-23}). 
Since the vacuum state $\ket{0}$, which satisfies the relations 
${\hat a}\ket{0}={\hat b}\ket{0}=0$, is represented as 
$\ket{0}=\ket{0}_b \otimes \ket{0}_a \ ({\hat a}\ket{0}_a=
{\hat b}\ket{0}_b=0)$, the state $\ket{c}$ is expressed in the 
following form : 
\begin{equation}\label{5-1}
\ket{c}=\sum_{n=0}^{\infty} \Gamma(n;c)\dket{n(c)}\otimes
\rrket{n} \ .
\end{equation}
Here, $\dket{n(c)}$ and $\rrket{n}$ denote normalized states for 
$b$- and $a$-boson, respectively. 
They are defined together with $\Gamma(n;c)$ as follows :
\begin{eqnarray}
& &\dket{n(c)}=e^{in\sigma/2}N(n;|\delta|^2)^{-\frac{1}{2}}
\exp\left(-\frac{|\delta|^2}{2}\right)
\frac{1}{\sqrt{n!}}({\hat b}^*)^n\exp(\delta{\hat b}^*)
\ket{0}_b \ , \label{5-2}\\
& &\rrket{n}=\frac{1}{\sqrt{n!}}({\hat a}^*)^n\ket{0}_a \ , 
\label{5-3}\\
& &\Gamma(n;c)=|V|^n U^{-(n+1)}
\exp\left(-|V|^2\frac{|\delta|^2}{2}\right)N(n;|\delta|^2)^{\frac{1}{2}} 
\ . 
\label{5-4}
\end{eqnarray}
Here, $V$ is defined as $V=|V|\exp(i\sigma/2)$ and 
$N(n;|\delta|^2)$ and $\delta$ are given by 
\begin{eqnarray}
N(n;|\delta|^2)&=&
\sum_{r=0}^{n}\frac{n!}{(n-r)!r!}(|\delta|^2)^r \nonumber\\
&=&L_n(-|\delta|^2) \ : \ {\rm Laguerre \ polynomial} \ ,
\label{5-5}\\
\delta&=& \sqrt{\frac{2}{\hbar}}\frac{W}{U} \ . \label{5-6}
\end{eqnarray}
We should note that the set of the states $\{ \rrket{n} \}$ 
in Eq. (\ref{5-3}) is an orthonormal one satisfying 
$((m|n))=\delta_{mn}$. 
However, $\dket{n(c)}$ is not orthogonalized but normalized, 
that is, $\langle m(c) || n(c) \rangle \neq \delta_{mn}$ 
for $m\neq n$ but $\langle n(c) || n(c) \rangle =1$. 
Since $\bra{c}c\rangle =1$ and $|n))$ and $\dket{n(c)}$ 
are the normalized states, we have the relation 
\begin{equation}\label{5-7}
\sum_{n=0}^{\infty}\Gamma(n;c)^2=1 \ .
\end{equation}
Since our intrinsic system which we are interested in is given 
only in terms of the $b$-boson, we need to evaluate an 
expectation value for any operator ${\hat O}_b$ consisting of only 
$({\hat b}, {\hat b}^*)$ with respect to $\ket{c}$ :
\begin{equation}\label{5-8}
(O_b)_c = \bra{c} {\hat O}_b \ket{c}
=\sum_{n=0}^{\infty} \Gamma(n;c)^2 
\bra{n(c)}|{\hat O}_b \dket{n(c)} \ .
\end{equation}
This situation is strongly similar to that in the case of 
thermo field dynamics formalism. The relation (\ref{5-8}) 
tells us that the factor $\Gamma(n;c)^2$ denotes the 
weight of the mixture of the pure state $\dket{n(c)}$ in the 
statistically mixed state under consideration and 
the quantum mechanical average gives us the statistical average 
for the physical quantity ${\hat O}_b$ automatically. 
In \S 9, the above will be discussed again.

From the above argument, we learned that $\Gamma(n;c)^2$ 
denotes the statistical weight. Then, it is possible to introduce 
the entropy in the present formalism. First, we note the 
following relation : 
\begin{eqnarray}
& &\ket{c}=\Gamma({\hat n}_a;c)\sum_{n=0}^{\infty}
\dket{n(c)}\otimes \rrket{n} \ , \label{5-9}\\
& &{\hat n}_a = {\hat a}^*{\hat a} \ . \label{5-10}
\end{eqnarray}
Here, $\Gamma({\hat n}_a;c)$ is obtained by replacing $n$ 
in $\Gamma(n;c)$ with the $a$-boson number operator. 
With use of the operator $\Gamma({\hat n}_a;c)$, we define 
the operator ${\hat S}$ in the form 
\begin{equation}\label{5-11}
{\hat S}=-\ln \Gamma({\hat n}_a;c)^2 \ . 
\end{equation}
Then, the expectation value of ${\hat S}$ for the state 
$\ket{c}$ is calculated as follows : 
\begin{equation}\label{5-12}
(S)_c=\bra{c}{\hat S}\ket{c}
=-\sum_{n=0}^{\infty} \Gamma(n;c)^2 \ln \Gamma(n;c)^2 \ .
\end{equation}
The relation (\ref{5-12}) tells that $(S)_c$ is the 
entropy and, then, ${\hat S}$ denotes the entropy operator. 
With the help of $\Gamma(n;c)$ shown in Eq. (\ref{5-4}), 
${\hat S}$ is given as 
\begin{equation}\label{5-13}
{\hat S}=({\hat n}_a+1)\ln U^2 - {\hat n}_a \ln |V|^2 
+|V|^2|\delta|^2-\ln N({\hat n}_a;|\delta|^2) \ .
\end{equation}
The expectation value of ${\hat S}$ for $\ket{c}$ becomes
\begin{eqnarray}\label{5-14}
(S)_c&=&
U^2\ln U^2 - |V|^2\ln |V|^2
+\left(\frac{2}{\hbar}|W|^2\right)
(|V|^2\ln U^2 - |V|^2 \ln |V|^2) \nonumber\\
& &-\left(\frac{2}{\hbar}|W|^2\right)^2\cdot\frac{1}{2}
\frac{|V|^2}{U^2} + \cdots \ . 
\end{eqnarray}
Here, we used the relation 
\begin{equation}\label{5-15}
\ln N({\hat n}_a;|\delta|^2)=|\delta|^2\cdot {\hat n}_a 
-\frac{1}{4}|\delta|^4({\hat n}_a+{\hat n}_a^2)
+\cdots \ . 
\end{equation}
We can show that, compared with the second term on the right-hand 
side of Eq. (\ref{5-14}), the third one is expected to be small 
in a rather wide region of $|V|^2$. Then, hereafter, we 
treat the following form for the entropy : 
\begin{equation}\label{5-16}
(S)_c=U^2\ln U^2-|V|^2 \ln |V|^2 
+\left(\frac{2}{\hbar}|W|^2\right) 
(|V|^2\ln U^2 -|V|^2 \ln |V|^2 ) \ . 
\end{equation}

With the use of the entropy ${\hat S}$ and the intrinsic 
Hamiltonian ${\hat K}$, we define the free energy $F$ 
in the form 
\begin{equation}\label{5-17}
F=\bra{c} {\hat K}-\frac{{\hat S}}{\beta} \ket{c} \ .
\end{equation}
Here, $\beta$ denotes the inverse of the temperature, i.e., 
$\beta=(k_B T)^{-1}$. 
The intrinsic Hamiltonian ${\hat K}$ in the present case 
is expressed as 
\begin{equation}\label{5-18}
{\hat K}=K_0+\hbar e {\hat b}^*{\hat b}
+\hbar^2 g {\hat b}^{*2}{\hat b}^2 \ .
\end{equation}

\vspace{0.3cm}
\noindent
(i) harmonic oscillator
\begin{equation}\label{5-19}
K_0=0 \ , \qquad e=\omega \ , \qquad g=0 \ .
\end{equation}

\vspace{0.3cm}
\noindent
(ii) pairing model
\begin{equation}\label{5-20}
K_0=0 \ , \qquad e=2(\epsilon-GS) \ , \qquad g=G \ .
\end{equation}

\vspace{0.3cm}
\noindent
(iii) Lipkin model
\begin{eqnarray}\label{5-21}
& &K_0=-\hbar \epsilon+\frac{\hbar}{2}\left[
\omega_0-\frac{3\hbar G}{2\omega_0^2}\left(
\epsilon+G\left(S+\frac{\hbar}{2}\right)\right)^2\right] \nonumber\\
& &e=\omega_0 \ , \qquad g=\frac{3G}{2\omega_0^2}
\left(\epsilon+G\left(S+\frac{\hbar}{2}\right)\right)^2 \ .
\end{eqnarray}
The expectation value of ${\hat K}$ for $\ket{c}$ is 
calculated in the form 
\begin{equation}\label{5-22}
(K)_c=K_0+\hbar e \left(|V|^2+\frac{2}{\hbar}|W|^2U^2\right)
+\hbar^2 g \left[2|V|^4+\left(\frac{2}{\hbar}|W|^2\right)\cdot
4U^2|V|^2\right] \ .
\end{equation}
Here, we neglect the term with the order $(2|W|^2/\hbar)^2$. 
Then, combining $(K)_c$ with $(S)_c$ shown in Eq. (\ref{5-16}), 
we can express $F$ as a function of $|V|^2$. 
The variation of $F$ with respect to $|V|^2$ leads us to 
the following relation :
\begin{equation}\label{5-23}
|V|^2\exp[\beta{\cal E}(|V|^2)]=U^2\exp
\left[-\frac{\frac{2}{\hbar}|W|^2}{1+\frac{2}{\hbar}|W|^2}\cdot
\frac{1}{U^2}\right] \ .
\end{equation}
Here, ${\cal E}(|V|^2)$ is defined as 
\begin{equation}\label{5-24}
{\cal E}(|V|^2)=\hbar e + 4\hbar^2 g
\left[|V|^2+\frac{\frac{2}{\hbar}|W|^2}{1+\frac{2}{\hbar}|W|^2}\cdot
\frac{1}{U^2}\right] \ .
\end{equation}
Under the approximation 
$e^{-X}=1-X$ for $X=(2|W|^2/\hbar)/(1+2|W|^2/\hbar)\cdot(1/U^2)$, 
we have 
\begin{equation}\label{5-25}
|V|^2=\frac{1}{1+\frac{2}{\hbar}|W|^2}\cdot
\frac{1}{e^{{\cal E}(|V|^2)\beta}-1} \ .
\end{equation}
The relation (\ref{5-25}) gives us 
\begin{equation}\label{5-26}
\beta=\frac{1}{{\cal E}(|V|^2)}\ln \left(
\frac{U^2+|V|^2\cdot\frac{2}{\hbar}|W|^2}
{|V|^2+|V|^2\cdot\frac{2}{\hbar}|W|^2}
\right) \ .
\end{equation}
The quantities $|W|^2$ and $|V|^2$ are given in Eqs. (\ref{3-30}) 
and (\ref{4-57}), respectively, and $T_z$ is obtained as a 
function of $t$. Therefore, with the aid of the relation 
(\ref{5-26}), $\beta$, i.e., the temperature of the system can be 
determined as a function of $t$.

In order to make further discussion simpler, let us treat the 
case of $\lambda=0$, where $\lambda$ is given in Eq. (\ref{4-27}), 
i.e., the case of harmonic oscillator. As is shown in 
Eq. (\ref{5-19}), $K_0=0$, $e=\omega$ and 
$g=0$, and ${\cal E}(|V|^2)$ in Eq. (\ref{5-24}) becomes 
\begin{equation}\label{5-27}
{\cal E}(|V|^2)=\hbar \omega \ .
\end{equation}
Then, the relation (\ref{5-25}) is reduced to 
\begin{equation}\label{5-28}
|V|^2=\frac{1}{1+\frac{2}{\hbar}|W|^2}\cdot
\frac{1}{e^{\epsilon \beta}-1} \ . \quad
(\epsilon =\hbar \omega )
\end{equation}
With the relations (\ref{5-22}) and (\ref{5-28}), we have 
\begin{eqnarray}\label{5-29}
(K)_c&=&\epsilon\left(
\frac{2}{\hbar}|W|^2+\left(1+\frac{2}{\hbar}|W|^2\right)|V|^2\right)
\nonumber\\
&=&\epsilon\cdot\frac{2}{\hbar}|W|^2+\epsilon\cdot
\frac{1}{e^{\epsilon\beta}-1} \ .
\end{eqnarray}
The quantity $2|W|^2/\hbar$ is a constant of motion and 
at the low temperature limit, the second term on the 
right-hand side of Eq. (\ref{5-29}) vanishes. 
Therefore, the first term represents the energy at the low 
temperature limit and the second the energy coming from 
the thermal fluctuations in the boson distribution.

Next, let us investigate the time-evolution of the system under 
a special initial condition $({\mib I}=I, {\mib J}=0, 
\chi_0=0)$. Then, $|V|^2$ in Eq. (\ref{4-57}) can be expressed as 
\begin{equation}\label{5-30}
|V|^2=[\sinh (\gamma t)]^2 \ .
\end{equation}
Here, we used the relation (\ref{4-44}) with 
$\gamma_0=\gamma$. At the initial time $t=0$, $|V|^2=0$ 
and in this case $\Gamma(n;c)^2$ becomes 
$\Gamma(n=0;c)^2=1$ and $\Gamma(n\neq 0;c)^2=0$. 
This means that the system is in a pure state with $n=0$. 
At the time $t\rightarrow \infty$, $|V|^2\rightarrow \infty$ 
and in this case, all $\Gamma(n;c)^2$ becomes equal to 
one another. This means that the present system is in 
the equal weight mixing. On the basis of the above fact, 
let us investigate the time evolution of the system. 
The energy $(K)_c$ in Eq. (\ref{5-29}) is expressed as 
\begin{equation}\label{5-31}
(K)_c=(2I-\hbar)\omega+\omega\cdot 2I|V|^2 \ (\ =E(t)\ ) \ .
\end{equation}
Since $|V|^2=0$ at $t=0$, $E(t)$ is expressed as 
\begin{equation}\label{5-32}
E(0)=(2I-\hbar)\omega \ . 
\end{equation}
On the other hand, $E(t)$ can be also expressed in the following 
form :
\begin{eqnarray}
& &E(t)=\sum_{n=0}^{\infty} \Gamma(n;c)^2 \bra{n(c)}|{\hat K}_b 
\dket{n(c)} \ , \label{5-33}\\
& &\bra{n(c)}|{\hat K}_b\dket{n(c)}=
\hbar\omega n-\hbar \omega |\delta|^2
\left[\frac{dL_{n+1}(x)}{dx}\right]_{x=-|\delta|^2}
\cdot\frac{1}{L_n(-|\delta|^2)} \ .\label{5-34}
\end{eqnarray}
Then, we have the relations 
\begin{eqnarray}
& &(
{\rm Contribution \ of \ the \ 1st \ term \ in \ Eq. (\ref{5-34})}
)\nonumber\\
& &\ =\sum_{n=0}^{\infty}\Gamma(n;c)^2\cdot \hbar\omega n 
= \omega\cdot 2I|V|^2 \nonumber\\
& &\ =\epsilon\cdot \frac{1}{e^{\epsilon \beta}-1} \ . 
\label{5-35}\\
& &(
{\rm Contribution \ of \ the \ 2nd \ term \ in \ Eq. (\ref{5-34})}
)\nonumber\\
& &\ =\sum_{n=0}^{\infty}\Gamma(n;c)^2 \left\{
-\hbar\omega|\delta|^2\left[
\frac{dL_{n+1}(x)}{dx}\right]_{x=-|\delta|^2} 
\cdot\frac{1}{L_n(-|\delta|^2)}\right\} \nonumber\\
& &\ =\omega(2I-\hbar) \ .
\label{5-36}
\end{eqnarray}
It is interesting to compare the above results with the 
initial value of the energy given in Eq. (\ref{5-32}). 
The ensemble average of the second term shown in Eq. (\ref{5-36}) 
coincides with the initial energy (\ref{5-32}). The initial 
state is in the pure state with $n=0$ and the energy of 
the system concentrates on this state. 
At the time $t\ (> 0)$, the states for $n\neq 0$ exist 
in $\ket{c}$ and the energy is distributed to each state. 
This fact tells us that the energy in the pure state 
with $n=0$ at $t=0$ dissipates to other pure states at $t$ 
under the mixing weight $\Gamma(n;c)^2$. 
However, totally, it conserves. Next, let us investigate 
the contribution of the first term. It is seen from 
the result in Eq. (\ref{5-35}) that this energy originates 
in the thermal effect. Therefore, the energy is supplied from 
the ``external environment" and by this energy 
the temperature of the system increases. More detail informations 
are obtained in Refs.\citen{TKY2}, \citen{TKY4} and \citen{TKY-p}.

\section{The pseudo $su(1,1)$-algebra in the $su(2)$-spin system}

As was discussed in \S 5, the $su(1,1)$-algebraic approach to 
the description of thermal effect gaves us various quite 
interesting features. However, this approach contains an undesirable 
feature. The Hamiltonian for the $su(1,1)$-spin system does not 
represent the energy of the entire system. Thus, it may not be 
justified to assume that the additional degree of freedom 
expressed in terms of the boson $({\hat a}, {\hat a}^*)$ 
can be specified as a physical object characterizing the 
external environment. Rather, it may be natural to regard the 
additional degree of freedom as a theoretical tool for 
describing dynamical motion such as the damped oscillation 
in the frame of isolated system. In this sense, a fictitious 
degree of freedom is introduced in the $su(1,1)$-algebraic 
approach, in spite of giving us interesting results. 
Then, it may be interesting to present, without introducing the 
fictitious degree of freedom, a possible approach which gives us 
the same results as those obtained in the $su(1,1)$-algebraic 
approach. In this section, we will present the basic theoretical 
framework for this problem.\cite{KPTY3} 

Let us prepare a boson space constructed in terms of the 
boson operators $({\hat c}, {\hat c}^*)$ and 
$({\hat d}, {\hat d}^*)$. In this space, we introduce the following 
operators : 
\begin{eqnarray}\label{6-1}
& &\maru{\tau}_+ = \sqrt{\hbar}{\hat c}^*\cdot
\sqrt{2T+\hbar{\hat c}^*{\hat c}}\cdot
\frac{1}{\sqrt{\hbar+\hbar{\hat d}^*{\hat d}}}\cdot \sqrt{\hbar}
{\hat d} \ , \nonumber\\
& &\maru{\tau}_- = \sqrt{\hbar}{\hat d}^*\cdot
\frac{1}{\sqrt{\hbar+\hbar{\hat d}^*{\hat d}}}\cdot 
\sqrt{2T+\hbar{\hat c}^*{\hat c}}\cdot
\sqrt{\hbar}{\hat c} \ , \nonumber\\
& &\maru{\tau}_0 = T+\hbar{\hat c}^*{\hat c} \ .
\end{eqnarray}
Here, $T$ is a real parameter which is assumed to be 
\begin{equation}\label{6-2}
T=\hbar t \ , \quad t=\frac{1}{2} \ , \ 1 \ , \ \frac{3}{2}
\ ,  \ 2 \ , \ \cdots \ .
\end{equation}
The above three operators are rewritten in the form 
\begin{eqnarray}\label{6-3}
& &\maru{\tau}_+ = \frac{1}{\sqrt{\hbar+\maru{S}-\maru{S}_0}}
\maru{S}_+ \sqrt{2T+\maru{S}+\maru{S}_0} \ , \nonumber\\
& &\maru{\tau}_- = \sqrt{2T+\maru{S}+\maru{S}_0}\cdot
\maru{S}_- \frac{1}{\sqrt{\hbar+\maru{S}-\maru{S}_0}}
  \ , \nonumber\\
& &\maru{\tau}_0 = T + \maru{S}+\maru{S}_0 \ .
\end{eqnarray}
The operators $(\maru{S}_{\pm, 0})$ compose the $su(2)$-algebra 
and are defined as 
\begin{eqnarray}
& &\maru{S}_+ = \hbar{\hat c}^* {\hat d} \ , \qquad
\maru{S}_- = \hbar{\hat d}^* {\hat c} \ , \qquad
\maru{S}_0 = \frac{\hbar}{2}
({\hat c}^* {\hat c}-{\hat d}^*{\hat d}) \ , \label{6-4}\\
& &\maru{S} = \frac{\hbar}{2}({\hat c}^*{\hat c}+{\hat d}^*{\hat d}) \ .
\label{6-5}
\end{eqnarray}
It is interesting to see that $(\maru{\tau}_{\pm, 0})$ 
can be expressed in terms of $(\maru{S}_{\pm, 0}, \maru{S})$. 
Further, we can prove the following relations : 
\begin{eqnarray}
& &{\maru{\tau}_-}^*=\maru{\tau}_+ \ , \qquad
{\maru{\tau}_0}^*=\maru{\tau}_0 \ , \label{6-6}\\
& &[ \maru{\tau}_+ , \maru{\tau}_- ] = -2\hbar\maru{\tau}_0
+[(2T+\hbar{\hat c}^*{\hat c})
(\hbar+\hbar{\hat c}^*{\hat c}\maru{D})
+(1-\epsilon)\hbar\cdot\hbar{\hat c}^*{\hat c}]\maru{D} \ , 
\nonumber\\
& &[\maru{\tau}_0 , \maru{\tau}_{\pm} ]
=
\pm\hbar \maru{\tau}_{\pm} \ , \label{6-7}\\
& &\maru{\tau}_0^2-\frac{1}{2}(\maru{\tau}_-\maru{\tau}_+
+\maru{\tau}_+\maru{\tau}_-) \nonumber\\
& &\ \ = T(T-\epsilon\hbar)\nonumber\\
& &\ \ \ +\frac{1}{2}[
(2T+\hbar{\hat c}^*{\hat c})(\hbar+\hbar{\hat c}^*{\hat c})
-(1-\epsilon)(2T+\hbar{\hat c}^*{\hat c})
(\hbar-\hbar{\hat c}^*{\hat c})]\maru{D} \ .
\label{6-8}
\end{eqnarray}
Here, $\epsilon$ and $\maru{D}$ are given as 
\begin{eqnarray}
& &\epsilon = 1 \ , \label{6-9}\\
& &\maru{D}=1-{\hat d}^*\frac{1}{1+{\hat d}^*{\hat d}}{\hat d} \ ,
\qquad
{\maru{D}}^2=\maru{D}\ , \qquad
{\maru{D}}^*=\maru{D} \ .
\label{6-10}
\end{eqnarray}
The relations (\ref{6-6}) $\sim$ (\ref{6-8}) should be compared  
with the relations (\ref{3-2}) $\sim$ (\ref{3-4}). 
If the terms related with the projection operator $\maru{D}$ are 
omitted, all the relations are reduced to those governed 
by the $su(1,1)$-algebra in the $su(1,1)$-space with the 
magnitude $T$. The operator $\maru{D}$ plays a role of the 
projection operator, i.e., 
$\maru{D}\cdot(1/\sqrt{n!})\cdot({\hat d}^*)^n \kket{0}_d 
=\delta_{n0}\kket{0}_d$ for ${\hat d}\kket{0}_d=0$. 
This means that the behavior of the set $(\maru{\tau}_{\pm,0}$) 
is quite similar to that of the $su(1,1)$-spin with the magnitude 
$T$. In this sense, it may be possible to regard the set 
$(\maru{\tau}_{\pm,0})$ as composing the pseudo $su(1,1)$-algebra. 
It is interesting to see that it is constructed in the $su(2)$-spin.

Our next problem is to introduce a possible form of coherent 
state which leads us to the classical counterpart of the set 
$(\maru{\tau}_{\pm,0})$. For the preparation, let us consider 
a state $\rket{\alpha}$ which satisfies 
\begin{equation}\label{6-11}
{\hat \alpha}\rket{\alpha}=\alpha\rket{\alpha} \ .
\end{equation}
Here, the operators ${\hat \alpha}$ and ${\hat \alpha}^*$ 
are defined by 
\begin{equation}\label{6-12}
{\hat \alpha}=f({\hat a}^*{\hat a})^{-1}{\hat a} \ ,
\qquad
{\hat \alpha}^*={\hat a}^* f({\hat a}^*{\hat a})^{-1} \ .
\end{equation}
The operators ${\hat a}$ and ${\hat a}^*$ denote boson annihilation 
and creation operators, respectively. 
Of course, $f({\hat a}^*{\hat a})$ and its inverse 
$f({\hat a}^*{\hat a})^{-1}$ should be defined. 
The eigenvalue $\alpha$ is, in general, complex. We can see that 
if $f({\hat a}^*{\hat a})=1$, the solution of Eq. (\ref{6-11}) 
is the well-known boson coherent state. In association with 
${\hat \alpha}$ and ${\hat \alpha}^*$, we introduce the 
operators ${\hat A}$ and ${\hat A}^*$ in the form 
\begin{equation}\label{6-13}
{\hat A}=f({\hat a}^*{\hat a}){\hat a} \ , \qquad
{\hat A}^*={\hat a}^* f({\hat a}^*{\hat a}) \ .
\end{equation}
The commutation relation between ${\hat \alpha}$ and 
${\hat A}^*$ is given as 
\begin{equation}\label{6-14}
[{\hat \alpha} , {\hat A}^* ] = 1 \ .
\end{equation}
Then, the state $\rket{\alpha}$ is obtained in the following 
form : 
\begin{eqnarray}
& &\rket{\alpha}=M_c\exp(\alpha{\hat A}^*)\rket{0} \ , 
\qquad
{\hat a}\rket{0}=0 \ , \quad (\alpha|\alpha)=1 \ .
\label{6-15}\\
& &(M_c)^{-2}=1+\sum_{n=1}^{\infty}
\frac{1}{n!}\prod_{k=0}^{n-1}f(k)^2\cdot(|\alpha|^2)^n \ .
\label{6-16}
\end{eqnarray}
For the complex variable $z$, we can derive the relation 
\begin{eqnarray}\label{6-17}
\rbra{\alpha}\partial_z \rket{\alpha}
&=&\frac{1}{2}(\alpha^*\partial_z\alpha-\alpha\partial_z\alpha^*)
\rbra{\alpha}f({\hat a}^*{\hat a})\rket{\alpha} \nonumber\\
&=&\frac{1}{2}(a^*\partial_z a-a\partial_z a^*) \ .
\end{eqnarray}
Here, $a$ is defined by 
\begin{equation}\label{6-18}
a=\alpha\sqrt{\rbra{\alpha}f({\hat a}^*{\hat a})\rket{\alpha}}
\ .
\end{equation}
The relation (\ref{6-17}) supports that the parameters $a$ 
and $a^*$ can be regarded as canonical variables in classical 
mechanics. The expectation value of ${\hat a}^*{\hat a}$ 
for $\rket{\alpha}$ is given as 
\begin{eqnarray}\label{6-19}
\rbra{\alpha}{\hat a}^*{\hat a}\rket{\alpha}
&=& \rbra{\alpha}{\hat \alpha}^* f({\hat a}^*{\hat a})^2
{\hat \alpha}\rket{\alpha} \nonumber\\
&=&|\alpha|^2\rbra{\alpha}f({\hat a}^*{\hat a})^2\rket{\alpha}
=a^*a \ .
\end{eqnarray}

Under the above preparation, let us investigate a possible form 
of coherent state which gives us the classical counterpart 
of the set $(\maru{\tau}_{\pm,0})$. 
For this aim, we set up the following wave packet : 
\begin{eqnarray}
& &\kket{c}=M_c\exp\left(\frac{\gamma}{\delta}
\cdot\frac{\maru{\tau}_+}{\hbar}\right)\kket{\delta} \ , 
\label{6-20}\\
& &\kket{\delta}=\sqrt{1-|\delta|^2}\exp\left(
\delta\maru{\Delta}^*\right)
\kket{0} \ , \label{6-21}\\
& &\maru{\Delta}^*={\hat d}^*\sqrt{1+{\hat d}^*{\hat d}} \ .
\label{6-22}
\end{eqnarray}
Here, $\gamma$ and $\delta$ denote the complex parameters. Clearly, 
the state $\kket{\delta}$ is nothing but a simple example of 
$\rket{\alpha}$ shown in Eq. (\ref{6-15}) and satisfies 
\begin{eqnarray}
& &{\hat \delta}\kket{\delta}=\delta\kket{\delta} \ , 
\qquad
{\hat \delta}^*={\hat d}^*
\left(\sqrt{1+{\hat d}^*{\hat d}}\right)^{-1} \ , 
\label{6-23}\\
& &\maru{\tau}_- \kket{\delta} = 0 \ .
\label{6-24}
\end{eqnarray}
Therefore, we can see that the state $\kket{c}$ consists of 
a superposition of successive operation of $\maru{\tau}_+$ 
on the state $\kket{\delta}$ satisfying the relation (\ref{6-24}). 
With the use of the relation (\ref{6-23}), together with the 
definition of $\maru{\tau}_+$ shown in Eq. (\ref{6-1}), 
the state $\kket{c}$ can be rewritten in the form 
\begin{equation}\label{6-25}
\kket{c}
=(1-|\gamma|^2)^t (1-|\delta|^2)^{\frac{1}{2}}
\cdot
\exp\left(\gamma{\hat c}^*\sqrt{2t+{\hat c}^*{\hat c}}\right)
\exp\left(\delta{\hat d}^*\sqrt{1+{\hat d}^*{\hat d}}\right)\kket{0} \ .
\end{equation}
The relation (\ref{6-18}) with $f({\hat c}^*{\hat c})
=2t+{\hat c}^*{\hat c}$ and 
$f({\hat d}^*{\hat d})=1+{\hat d}^*{\hat d}$ tells us that 
$\gamma$ and $\delta$ can be expressed in terms of the canonical 
variables $c$ and $d$ :
\begin{equation}\label{6-26}
\gamma=\frac{c}{\sqrt{2t+c^*c}} \ , \qquad
\delta=\frac{d}{\sqrt{1+d^*d}} \ .
\end{equation}
The relation (\ref{6-19}) gives us 
\begin{equation}\label{6-27}
\bbra{c}{\hat c}^*{\hat c}\kket{c}=c^*c \ , \qquad
\bbra{c}{\hat d}^*{\hat d}\kket{c}=d^*d \ .
\end{equation}
Then, the expectation value of $\maru{S}$ in Eq. (\ref{6-5}) 
is calculated as 
\begin{equation}\label{6-28}
(S)_c=\bbra{c}\maru{S}\kket{c}
=\frac{\hbar}{2}(c^*c+d^*d)
=S \ .
\end{equation}
Instead of the variables $(c, c^*)$ and $(d, d^*)$, we use 
the variables $({\mib \psi}, {\mib S})$ and $({\mib X} , {\mib X}^*)$ 
used in \S 2 :
\begin{eqnarray}\label{6-29}
& &c={\mib X}\exp\left(-i\frac{{\mib \psi}}{2}\right) \ , \qquad
c^*={\mib X}^*\exp\left(+i\frac{{\mib \psi}}{2}\right) \ , 
\nonumber\\
& &d=\frac{1}{\sqrt{\hbar}}\sqrt{2{\mib S}-\hbar{\mib X}^*{\mib X}}
\exp\left(-i\frac{\mib \psi}{2}\right) \ , 
\qquad
%\nonumber\\
%& &
d^*=\frac{1}{\sqrt{\hbar}}\sqrt{2{\mib S}-\hbar{\mib X}^*{\mib X}}
\exp\left(+i\frac{\mib \psi}{2}\right) \ . \ \ \nonumber\\
\end{eqnarray}
With the use of the relation (\ref{6-11}) for 
$f({\hat c}^*{\hat c})=\sqrt{2t+{\hat c}^*{\hat c}}$ and 
$f({\hat d}^*{\hat d})=\sqrt{1+{\hat d}^*{\hat d}}$, 
together with the relation (\ref{6-26}) and (\ref{6-29}), 
the expectation values of $(\maru{\tau}_{\pm, 0})$ for 
$\kket{c}$ are given in the following form :
\begin{eqnarray}\label{6-30}
& &(\tau_+)_c
=\frac{1}{\sqrt{\hbar+2{\mib S}-\hbar{\mib X}^*{\mib X}}}\cdot
\sqrt{\hbar}{\mib X}^*\sqrt{2T+\hbar{\mib X}^*{\mib X}}\cdot
\sqrt{2{\mib S}-\hbar{\mib X}^*{\mib X}} \ , \nonumber\\
& &(\tau_-)_c
=\sqrt{2{\mib S}-\hbar{\mib X}^*{\mib X}}\cdot 
\sqrt{2T+\hbar{\mib X}^*{\mib X}}\cdot
\sqrt{\hbar}{\mib X}
\frac{1}{\sqrt{\hbar+2{\mib S}-\hbar{\mib X}^*{\mib X}}}
\ , \nonumber\\
& &(\tau_0)_c=\hbar{\mib X}^*{\mib X}+T \ .
\end{eqnarray}
We can prove that $(\tau_{\pm,0})_c$ obey the relations
\begin{eqnarray}
& &(\tau_-)_c^*=(\tau_+)_c \ , \qquad
(\tau_0)_c^*=(\tau_0)_c \ , \label{6-31}\\
& &[ (\tau_+)_c , (\tau_-)_c ]_{\rm P} = (-i)
\left\{-2\hbar(\tau_0)_c
+[(2T+\hbar c^* c)
(\hbar+\hbar c^* c D)
+(1-\epsilon)\hbar\cdot\hbar c^* c]D \right\}\ , 
\nonumber\\
& &[ (\tau_0)_c , (\tau_{\pm})_c ]_{\rm P}
=
(-i)(\pm\hbar (\tau_{\pm})_c) \ , \label{6-32}\\
& &(\tau_0)_c^2-\frac{1}{2}((\tau_-)_c (\tau_+)_c
+ (\tau_+)_c (\tau_-)_c) \nonumber\\
& &\ \ = {\mib T}({\mib T}-\epsilon\hbar)\nonumber\\
& &\ \ \ +\frac{1}{2}[
(2{\mib T}+\hbar c^* c)(\hbar+\hbar c^* c)
-(1-\epsilon)(2{\mib T}+\hbar c^* c)
(\hbar-\hbar c^* c)] D \ .
\label{6-33}
\end{eqnarray}
Here, $\epsilon$ and $D$ are given as 
\begin{eqnarray}
& &\epsilon = 0 \ , \label{6-34}\\
& &D=(1+d^* d)^{-1}
=1-d^*\frac{1}{1+d^*d} d \ .
\label{6-35}
\end{eqnarray}
It is interesting to see that the relations (\ref{6-31}) 
$\sim$ (\ref{6-35}) correspond to the relations (\ref{6-6}) 
$\sim$ (\ref{6-10}) completely. The quantity $\epsilon$ appears 
from the non-commutability of various operators, i.e., it is 
a kind of quantum effect. From the above correspondence, we obtain a 
conclusion that the expectation values $(\tau_{\pm,0})_c$ 
are the classical counterparts of ($\maru{\tau}_{\pm,0}$).

With the help of the relation (\ref{6-3}), together with the 
form (\ref{2-17}), the Holstein-Primakoff boson representation 
for the set $(\maru{\tau}_{\pm,0})$ is obtained easily :
\begin{eqnarray}\label{6-36}
& &{\tilde \tau}_+
=\frac{1}{\sqrt{\hbar+2S-\hbar{\hat X}^*{\hat X}}}\cdot
\sqrt{\hbar}{\hat X}^*\sqrt{2T+\hbar{\hat X}^*{\hat X}}\cdot
\sqrt{2S-\hbar{\hat X}^*{\hat X}} \ , \nonumber\\
& &{\tilde \tau}_-
=\sqrt{2S-\hbar{\hat X}^*{\hat X}}\cdot 
\sqrt{2T+\hbar{\hat X}^*{\hat X}}\cdot
\sqrt{\hbar}{\hat X}
\frac{1}{\sqrt{\hbar+2S-\hbar{\hat X}^*{\hat X}}}
\ , \nonumber\\
& &{\tilde \tau}_0=\hbar{\hat X}^*{\hat X}+T \ .
\end{eqnarray}
Here, $\maru{S}_{\pm,0}$ and $\maru{S}$ in Eq. (\ref{6-3}) 
are replaced with ${\widetilde S}_{\pm,0}$ in Eq. (\ref{2-17}) 
and $S$, respectively. We can see that under the replacements 
$({\hat X}, {\hat X}^*)\rightarrow ({\mib X}, {\mib X}^*)$ 
and $S\rightarrow {\mib S}$, the form (\ref{6-36}) is 
reduced to the form (\ref{6-30}).

\section{Description of a many-boson system interacting 
with an external harmonic oscillator}

The aim of this section was already mentioned in the beginning
part of the previous section. Corresponding the Hamiltonian (\ref{5-18}), 
we treat the following Hamiltonian for a many-boson 
system\cite{KPTY4}:
\begin{equation}\label{7-1}
\maru{K}_c=\hbar e^0\cdot{\hat c}^*{\hat c}+\hbar^2 g^0\cdot
{\hat c}^{*2}{\hat c}^2 \ .
\end{equation}
It may be self-evident why we treat the Hamiltonian (\ref{7-1}) 
in the present case. The above boson system interacts with a 
harmonic oscillator, the Hamiltonian of which is given by
\begin{equation}\label{7-2}
\maru{K}_d=\hbar\omega^0\cdot {\hat d}^*{\hat d} \ .
\end{equation}
As the interaction between the two systems, we are interested 
in the following form :
\begin{eqnarray}
& &\maru{V}_{cd}=\maru{V}_{cd}(1)+\maru{V}_{cd}(2) \ , 
\label{7-3}\\
& &\maru{V}_{cd}(1)=\hbar^2 f^0 \cdot
({\hat c}^{*}{\hat d}\cdot{\hat d}^*{\hat c}+{\hat d}^*{\hat c}\cdot
{\hat c}^*{\hat d}) \ , \label{7-4}\\
& &\maru{V}_{cd}(2)=-\hbar\gamma^0\cdot
i\biggl[\left(\sqrt{\hbar+\hbar{\hat d}^*{\hat d}}\right)^{-1}\cdot
{\hat c}^*{\hat d}\cdot\sqrt{2T+\hbar{\hat c}^*{\hat c}}
\nonumber\\
& &\qquad\qquad\qquad\qquad\ \ 
-\sqrt{2T+\hbar{\hat c}^*{\hat c}}\cdot {\hat d}^*{\hat c}\cdot
\left(\sqrt{\hbar+\hbar{\hat d}^*{\hat d}}\right)^{-1}\biggl] \ .
\label{7-5}
\end{eqnarray}
Of course, $e^0$, $g^0$, $\omega^0$, $f^0$, $\gamma^0$ and $T$ 
denote real parameters. 
In particular, we regard $T$ as the parameter obeying the condition 
$T=\hbar t$, $(t=1/2, 1, 3/2, 2, \cdots )$. Thus, the Hamiltonian 
of the entire system is expressed as 
\begin{equation}\label{7-6}
\maru{H}_{su(2)}=\maru{K}_c+\maru{K}_d+\maru{V}_{cd} \ .
\end{equation}
The meaning of the ``$su(2)$" in $\maru{H}_{su(2)}$ is interpreted 
as follows : We are now describing the system obeying the 
$su(2)$-algebra and $\maru{H}_{su(2)}$ can be expressed in 
terms of the operators $(\maru{S}_{\pm, 0})$ and $\maru{S}$ 
shown in Eqs. (\ref{6-4}) and (\ref{6-5}). The Hamiltonian 
(\ref{7-6}) can be rewritten in the form 
\begin{eqnarray}
\maru{H}_{su(2)}&=&\maru{U}_{su(2)}+\maru{K}_{su(2)}
+\maru{V}_{su(2)} \ , \label{7-7}\\
\maru{U}_{su(2)}&=&
2\left[\omega^0-2f^0\cdot\left(T-\frac{\hbar}{2}\right)\right]\cdot\maru{S}
\nonumber\\
& &-[(e^0-\omega^0)-\hbar g^0]\cdot T+(g^0-2f^0)\cdot T^2 \ ,
\label{7-8}\\
\maru{K}_{su(2)}&=&
[(e^0-\omega^0)-\hbar g^0+4f^0\maru{S}-2(g^0-2f^0)T]\maru{\tau}_0
%\nonumber\\
%& &
+(g^0-2f^0)\cdot\maru{\tau}_0^2 \ , \qquad
\label{7-9}\\
\maru{V}_{su(2)}&=&-\gamma^0 \cdot i(\maru{\tau}_+-\maru{\tau}_-) \ .
\label{7-10}
\end{eqnarray}
The operator $\maru{S}$ commutes with $\maru{H}_{su(2)}$, and 
$\maru{U}_{su(2)}$ contains only $\maru{S}$. Therefore, the dynamics 
of our present system is described in the framework of 
$(\maru{K}_{su(2)}+\maru{V}_{su(2)})$ under a given eigenvalue 
$S$ of $\maru{S}$.

In associating with the $su(2)$-spin system introduced in the above, 
we will recapitulate a possible form of the $su(1,1)$-spin 
system discussed in \S 4. Our interest is in the following 
Hamiltonian : 
\begin{eqnarray}
& &{\hat H}_{su(1,1)}={\hat K}_b-{\hat K}_a+{\hat V}_{ba} \ ,
\label{7-6b}\\
& &{\hat K}_b=\hbar e\cdot {\hat b}^*{\hat b}
+\hbar^2 g\cdot {\hat b}^{*2}{\hat b}^2 \ ,
\label{7-7b}\\
& &{\hat K}_a=\hbar e\cdot{\hat a}^*{\hat a}
+\hbar^2 g\cdot {\hat a}^{*2}{\hat a}^2 \ , 
\label{7-8b}\\
& &{\hat V}_{ba}=-\hbar\gamma\cdot i({\hat a}^*{\hat b}^*
-{\hat b}{\hat a}) \ .
\label{7-9b}
\end{eqnarray}
The Hamiltonian ${\hat K}_b$ is of the same form as that 
given in Eq. ({\ref{5-18}). The Hamiltonian (\ref{7-6b}) can be 
rewritten in the form
\begin{eqnarray}
& &{\hat H}_{su(1,1)}
={\hat U}_{su(1,1)}+{\hat K}_{su(1,1)}+{\hat V}_{su(1,1)} \ ,
\label{7-10b}\\
& &{\hat U}_{su(1,1)}=2(e-\hbar g)\left({\hat T}-\frac{\hbar}{2}
\right) \ , \label{7-11}\\
& &{\hat K}_{su(1,1)}
=4g\cdot \left({\hat T}-\frac{\hbar}{2}\right){\hat T}_0 \ ,
\label{7-12}\\
& &{\hat V}_{su(1,1)}=-\gamma\cdot i({\hat T}_+ -{\hat T}_-) \ .
\label{7-13}
\end{eqnarray}
In a sense similar to the case of the $su(2)$-spin system, 
the operator ${\hat T}$ commutes with ${\hat H}_{su(1,1)}$, 
and ${\hat U}_{su(1,1)}$ contains only ${\hat T}$. Therefore, 
the dynamics of the present system is described in the 
framework of $({\hat K}_{su(1,1)}+{\hat V}_{su(1,1)})$ 
under a given eigenvalue $T$ of ${\hat T}$. As was already 
mentioned, the Hamiltonian ${\hat H}_{su(1,1)}$ gives 
us various interesting aspects of thermal properties and dissipation 
for boson systems. In this sense, the form (\ref{7-10b}) is quite 
interesting. However, it does not represent the total energy 
of the system. On the other hand, $\maru{H}_{su(2)}$ in 
Eq. (\ref{7-6}) represents the total energy of the system. 
Therefore, it is interesting to investigate the relation 
between both systems.

At the present stage, we have two spin systems. One is the 
$su(2)$- and the other is the $su(1,1)$-spin system. 
For a given value of $S$ $(=\hbar s ; s=0, 1/2, 1, 3/2, \cdots)$, 
the $su(2)$-spin system is described by the complete set 
$\{ \kket{n;2s-n} \}$, which is expressed as 
\begin{equation}\label{7-14}
\kket{n;2s-n}=\frac{1}{\sqrt{n!(2s-n)!}}({\hat c}^*)^n
({\hat d}^*)^{2s-n} \kket{0} \ .
\end{equation}
Here, $n$ runs in the region 
\begin{equation}\label{7-15}
n=0,\ 1,\ 2,\ \cdots , \ 2s \ .
\end{equation}
For a given value of $T$ $(=\hbar t , t=1/2, 1, 3/2, 2, \cdots )$, 
the $su(1,1)$-spin system can be treated by the complete 
set $\{ \ket{n;2t-1+n} \}$, which is given by 
\begin{equation}\label{7-16}
\ket{n;2t-1+n}=\frac{1}{\sqrt{n!(2t-1+n)!}}
({\hat a}^*)^n ({\hat b}^*)^{2t-1+n} \ket{0} \ .
\end{equation}
Here, $n$ runs in the region 
\begin{equation}\label{7-17}
n=0,\ 1,\ 2,\ \cdots,\ 2s,\ 2s+1, \cdots \ .
\end{equation}
We refer to the spaces spanned by the complete set 
(\ref{7-14}) and (\ref{7-16}) as the $su(2)$-space 
with $s$ and the $su(1,1)$-space with $t$, respectively.

Let us transcribe the $su(2)$-spin system of the $su(2)$-space 
with $s$ into that in the $su(1,1)$-space with $t$. 
For this aim, first, we set up the following correspondence : 
\begin{eqnarray}
\kket{n;2s-n} &\sim& \ket{n;2t-1+n} \ , \label{7-18}\\
n&=& 0, \ 1, 2, \cdots, 2s \ .\label{7-19}
\end{eqnarray}
Following the basic idea of the boson mapping, we call the space spanned 
by the set $\{ \ket{n;2t-1+n} ; n=0,1,2,\cdots,2s \}$ 
as the physical space and define 
the operator ${\hat U}_{st}$ 
\begin{equation}\label{7-20}
{\hat U}_{st}=\sum_{n=0}^{2s} \ket{n;2t-1+n}\bbra{n;2s-n}\ .
\end{equation}
The operator ${\hat U}_{st}$ satisfies 
\begin{eqnarray}
& &{\hat U}_{st}^\dagger {\hat U}_{st}
=\sum_{n=0}^{2s} \kket{n;2s-n}\bbra{n;2s-n}=1 \ , \label{7-21}\\
& &{\hat U}_{st}{\hat U}_{st}^{\dagger}
=\sum_{n=0}^{2s}\ket{n;2t-1+n}\bra{n;2t-1+n}
=P_{st} \ . \label{7-22}
\end{eqnarray}
Here, $P_{st}$ plays the role of the projection operator 
to the physical space 
$(P_{st}^{*}=P_{st}, \ P_{st}^2=P_{st})$. 
The operators ${\hat U}_{st}$ and ${\hat U}_{st}^{\dagger}$ 
give us 
\begin{eqnarray}
& &{\hat U}_{st}\kket{n;2s-n}=\ket{n;2t-1+n} \ , \label{7-23}\\
& &{\hat U}_{st}^{\dagger}\ket{n;2t-1+n}
=\cases{ \kket{n;2s-n}\ , \quad \ (n=0,1,2,\cdots 2s) \cr
         0 \ . \qquad\qquad\quad\quad (n=2s+1, 2s+2,\cdots) }
         \label{7-24}
\end{eqnarray}
The operator ${\hat U}_{st}$ permits us to transform 
$\maru{\tau}_{\pm,0}$, $\maru{S}$ and $T$ as 
\begin{eqnarray}\label{7-25}
& &{\hat U}_{st}\cdot \maru{\tau}_{\pm,0}\cdot {\hat U}_{st}^{\dagger}
=P_{st}\cdot {\hat T}_{\pm,0}\cdot P_{st} \ , \nonumber\\
& &{\hat U}_{st}\cdot \maru{S}\cdot {\hat U}_{st}^{\dagger}
={\hat U}_{st}\cdot S \cdot{\hat U}_{st}^{\dagger}
=P_{st}\cdot {S}\cdot P_{st} \ , \nonumber\\
& &{\hat U}_{st}\cdot {T}\cdot {\hat U}_{st}^{\dagger}
={P}_{st}\cdot T \cdot{P}_{st}
=P_{st}\cdot {\hat T}\cdot P_{st} \ .
\end{eqnarray}
Here, it should be noted that $S$ and $T$ are $c$-numbers.
With the aid of $U_{st}$ and the relation (\ref{7-25}), we are 
able to transcribe $\maru{H}_{su(2)}$ in the $su(1,1)$-space 
with $t$ : 
\begin{eqnarray}
& &{\hat U}_{st}\cdot\maru{H}_{su(2)}\cdot{\hat U}_{st}^{\dagger}
=P_{st}\cdot{\hat H}_{su(2)}\cdot P_{st} \ , \label{7-26}\\
& &{\hat H}_{su(2)}={\hat U}_{su(2)}+{\hat K}_{su(2)}
+{\hat V}_{su(2)} \ , \label{7-27}\\
& &{\hat U}_{su(2)}=
2\left[\omega^0-2f^0\cdot\left({\hat T}-\frac{\hbar}{2}\right)\right]
\cdot S
\nonumber\\
& &\qquad\qquad
-[(e^0-\omega^0)-\hbar g^0]\cdot {\hat T}+(g^0-2f^0)\cdot 
{\hat T}^2 \ ,
\label{7-28}\\
& &{\hat K}_{su(2)}=
[(e^0-\omega^0)-\hbar g^0+4f^0{S}-2(g^0-2f^0){\hat T}]{\hat T}_0
%\nonumber\\
%& &
+(g^0-2f^0)\cdot{\hat T}_0^2 \ , \quad\quad
\label{7-28b}\\
& &{\hat V}_{su(2)}=
-\gamma^0 \cdot i({\hat T}_+ -{\hat T}_-) \ .
\label{7-29}
\end{eqnarray}
We can regard ${\hat H}_{su(2)}$ as the Hamiltonian in the 
$su(1,1)$-space which has a connection with the original 
$\maru{H}_{su(2)}$ in the $su(2)$-space. In the same meaning 
as that mentioned for the Hamiltonian ${\hat H}_{su(1,1)}$, 
we are interested in the part 
$({\hat K}_{su(2)}+{\hat V}_{su(2)})$. 
Then, we require the following condition for the parameters 
$(e^0, \omega^0, g^0, f^0, \gamma^0, g, \gamma)$ : 
\begin{eqnarray}\label{7-30}
& &e^0-\omega^0+2g^0\left(S-\frac{\hbar}{2}\right)
=4g\left(T-\frac{\hbar}{2}\right) \ , \nonumber\\
& &g^0=2f^0 \ , \qquad 
\gamma^0=\gamma \ . 
\end{eqnarray}
Under the condition (\ref{7-30}), we obtain 
\begin{equation}\label{7-31}
P_{st}\cdot ({\hat K}_{su(2)}+{\hat V}_{su(2)})\cdot P_{st}
=P_{st}\cdot({\hat K}_{su(1,1)}+{\hat V}_{su(1,1)})\cdot P_{st} \ .
\end{equation}
This means that, under the condition (\ref{7-30}), 
${\hat H}_{su(2)}$ is expected to display various results 
similar to those given by ${\hat H}_{su(1,1)}$. 
A possible reexpression of the condition (\ref{7-30}) 
is given as follows : 
\begin{eqnarray}\label{7-32}
& &e^0=\omega^0 \ , \qquad 
f^0=\frac{g_0}{2}\cdot\left(T-\frac{\hbar}{2}\right) \ , \qquad
g^0=g_0\cdot\left(T-\frac{\hbar}{2}\right) \ , \nonumber\\
& &\gamma^0=\gamma \ , \qquad 
g=\frac{g_0}{2}\cdot\left(S-\frac{\hbar}{2}\right) \ . \quad
(g_0 : {\rm parameter})
\end{eqnarray}
Under the condition (\ref{7-32}), 
$(\maru{K}_{su(2)} , \maru{V}_{su(2)})$ and 
$({\hat K}_{su(1,1)} , {\hat V}_{su(1,1)} )$ can be expressed in 
the form 
\begin{eqnarray}
& &\maru{K}_{su(2)}
=2g_0 \left(T-\frac{\hbar}{2}\right)\left(\maru{S}-\frac{\hbar}{2}
\right)\cdot \maru{\tau}_0 \ , \nonumber\\
& &\maru{V}_{su(2)}=-\gamma\cdot i(\maru{\tau}_+ - \maru{\tau}_-) \ ,
\label{7-33}\\
& &{\hat K}_{su(1,1)}=2g_0\left({\hat T}-\frac{\hbar}{2}\right)
\left(S-\frac{\hbar}{2}\right)\cdot {\hat T}_0 \ ,\nonumber\\
& &{\hat V}_{su(1,1)}=-\gamma\cdot i({\hat T}_+ -{\hat T}_-) \ .
\label{7-34}
\end{eqnarray}
Thus, in the $su(2)$-space with $S$ and the $su(1,1)$-space 
with $T$, we have the following correspondence 
\begin{equation}\label{7-35}
\maru{K}_{su(2)} \sim {\hat K}_{su(1,1)} \ , 
\qquad
\maru{V}_{su(2)}\sim {\hat V}_{su(1,1)} \ .
\end{equation}
The above correspondence results from 
\begin{equation}\label{7-36}
\maru{\tau}_{\pm,0} \sim {\hat T}_{\pm,0} \ , 
\qquad
T\sim {\hat T}\ , \qquad
\maru{S}\sim S \ .
\end{equation}
Of course, $\maru{U}_{su(2)}$ and ${\hat U}_{su(1,1)}$ do not 
correspond in forms. However, it should be noted that there exists a 
difference between $\maru{H}_{su(2)}$ and ${\hat H}_{su(1,1)}$ 
in addition to $\maru{U}_{su(2)}$ and ${\hat U}_{su(1,1)}$. 
The Holstein-Primakoff representations shown in Eqs. (\ref{3-16}) 
and (\ref{6-36}) teach us the difference. In the case of 
$({\widetilde T}_{\pm,0})$ in Eq. (\ref{3-16}), the integer $n$ in the 
boson state $(1/\sqrt{n!})\cdot({\hat X}^*)^n\rket{0}$ is 
permissible up to the infinity. But, in the case of 
$({\tilde \tau}_{\pm,0})$ in Eq. (\ref{6-36}), $n$ is 
permissible up to $2s$. This can be seen in the factor 
$\sqrt{2S-\hbar{\hat X}^*{\hat X}}$. 
It is quite natural, because the $su(2)$-algebra is compact, 
and the $su(1,1)$-algebra noncompact. If $s$ is sufficiently 
large compared with $n$, the two algebras give almost the same 
results.

Next, we investigate the correspondence between the two systems 
in the classical counterpart. The expectation values of 
$\maru{H}_{su(2)}$ and ${\hat H}_{su(1,1)}$ with respect to 
the coherent states (\ref{6-25}) and (\ref{3-23}), respectively, 
give us the following results :
\begin{eqnarray}
& &H_{su(2)}=U_{su(2)}+K_{su(2)}+V_{su(2)} \ , \label{7-37}\\
& &U_{su(2)}=
2\left[\omega^0-2f^0\cdot\left(T-\frac{\hbar}{2}\right)\right]
\cdot {\mib S}
\nonumber\\
& &\qquad\qquad
-[(e^0-\omega^0)-\hbar g^0]\cdot T
+\left[g^0\left(1-\frac{\hbar}{2T}\right)-2f^0\right]\cdot 
T^2 \ ,
\label{7-38}\\
& &K_{su(2)}=
[e^0-\omega^0-\hbar g^0+4f^0{\mib S}-2(g^0-2f^0)T]({\tau}_0)_c
\nonumber\\
& &\qquad\qquad
+\left[g^0\left(1+\frac{\hbar}{2T}\right)-2f^0\right]\cdot
({\tau}_0)_c^2 \ , \quad\quad
\label{7-39}\\
& &V_{su(2)}=
-\gamma^0 \cdot i(({\tau}_+)_c -({\tau}_-)_c) \ ,
\label{7-40}\\
& &H_{su(1,1)}
=U_{su(1,1)}+K_{su(1,1)}+V_{su(1,1)} \ ,
\label{7-41}\\
& &U_{su(1,1)}=2(e-2\hbar g)\left({\mib T}-\frac{\hbar}{2}
\right) \ , \label{7-42}\\
& &K_{su(1,1)}
=4g\cdot \left({\mib T}-\frac{\hbar}{2}\right)
\left(1+\frac{\hbar}{2{\mib T}}\right)(T_0)_c \ ,
\label{7-43}\\
& &V_{su(1,1)}=-\gamma\cdot i((T_+)_c -(T_-)_c) \ .
\label{7-44}
\end{eqnarray}
Here, $(\tau_{\pm,0})_c$ and $(T_{\pm,0})_c$ are given in the 
relations (\ref{6-30}) and (\ref{3-32}), respectively. 
In the same meaning as that in the quantum version, 
$U_{su(2)}$ and $U_{su(1,1)}$ do not relate with the 
dynamics. For the Hamiltonian $(K_{su(2)}+V_{su(2)})$ and 
$(K_{su(1,1)}+V_{su(1,1)})$, we set up the following condition :
\begin{eqnarray}\label{7-45}
& &e^0=\omega^0 \ , \qquad 
f^0=\frac{g_0}{2}\cdot\left(T-\frac{\hbar}{2}\right)
\left(1+\frac{\hbar}{2T}\right) \ , \nonumber\\
& &g^0=g_0\cdot\left(T-\frac{\hbar}{2}\right) \ , \qquad
\gamma^0=\gamma \ , \qquad 
g=\frac{g_0}{2}\cdot S \ . 
\end{eqnarray}
The relation (\ref{7-45}) corresponds to Eq. (\ref{7-32}). 
Then, $(K_{su(2)}, V_{su(2)})$ and $(K_{su(1,1)}, V_{su(1,1)})$ 
can be expressed in the form 
\begin{eqnarray}
& &K_{su(2)}
=2g_0 \left(T-\frac{\hbar}{2}\right)
\left(1+\frac{\hbar}{2T}\right)
{\mib S}(\tau_0)_c \ , \label{7-46}\\
& &{V}_{su(2)}=-\gamma\cdot i(({\tau}_+)_c - ({\tau}_-)_c) \ ,
\label{7-47}\\
& &K_{su(1,1)}=2g_0\left({\mib T}-\frac{\hbar}{2}\right)
\left(1+\frac{\hbar}{2{\mib T}}\right)
S (T_0)_c \ , \label{7-48}\\
& &V_{su(1,1)}=-\gamma\cdot i((T_+)_c -(T_-)_c) \ .
\label{7-49}
\end{eqnarray}
It should be noted that $(S, T)$ are the parameters and 
$({\mib S}, \ {\mib T})$ denote the dynamical variables, which 
are constants of motion in our present system. 
Under the condition $({\mib S}=S, {\mib T}=T)$ and 
the correspondence $(\tau_{\pm,0})_c\sim (T_{\pm,0})_c$, 
we have 
\begin{equation}\label{7-50}
K_{su(2)}\sim K_{su(1,1)} \ , \qquad
V_{su(2)}\sim V_{su(1,1)} \ .
\end{equation}
However, in the case of the $su(2)$-spin system, we have the 
restriction $\hbar |{\mib X}|^2 \le 2S$ and in the case of 
the $su(1,1)$-spin system, these does not exist such a restriction. 
Then, if $S$ is sufficiently large compared with 
$\hbar{\mib X}^*{\mib X}$, the results of the two systems are 
almost the same as each other.

Let us investigate the difference between $(\tau_{\pm})_c$ 
in Eq. (\ref{6-30}) and $(T_{\pm})_c$ in Eq. (\ref{3-32}). 
Under ${\mib T}=T$, both are related with each other 
through the relation 
\begin{eqnarray}
& &(\tau_{\pm})_c
=(T_{\pm})_c\sqrt{1-\theta(\hbar{\mib X}^*{\mib X})} \ ,
\label{7-51}\\
& &\theta(\hbar{\mib X}^*{\mib X}) 
=\frac{\hbar}{2S+\hbar-\hbar{\mib X}^*{\mib X}} \ .
\label{7-52}
\end{eqnarray}
The behavior of the function $\theta(\hbar{\mib X}^*{\mib X})$ 
is summarized as follows :
\begin{equation}\label{7-53}
\theta(\hbar{\mib X}^*{\mib X})=
\cases{ O(\hbar^{+1}) \ , \quad (\hbar{\mib X}^*{\mib X} \ll 2S) \cr
O(\hbar^{+\frac{1}{2}}) \ , \quad 
(\hbar{\mib X}^*{\mib X} \approx 2S-O(\hbar^{+\frac{1}{2}})) \cr
O(\hbar^{0}) \ , \quad \ \ 
(\hbar{\mib X}^*{\mib X} \approx 2S\pm O(\hbar^{+1})) \cr
O(\hbar^{-\frac{1}{2}}) \ , \quad 
(\hbar{\mib X}^*{\mib X} \approx 2S+\hbar-O(\hbar^{+\frac{3}{2}})) \cr
O(\hbar^{-1}) \ . \quad \ 
(\hbar{\mib X}^*{\mib X} \approx 2S+\hbar-O(\hbar^{+2}))}
\end{equation}
Here, $O(\hbar^k)$ denotes the term of order $\hbar^k$. 
Therefore, under the condition $\hbar\rightarrow 0$, 
$\theta(\hbar{\mib X}^*{\mib X})$ can be approximated 
in the form 
\begin{equation}\label{7-54}
\theta(\hbar{\mib X}^*{\mib X})
=
\cases{0 \ , \quad \ (\hbar{\mib X}^*{\mib X}< 2S) \cr
1 \ , \quad \ (\hbar{\mib X}^*{\mib X}= 2S) \cr
\infty \ . \quad (\hbar{\mib X}^*{\mib X} > 2S)}
\end{equation}
The behavior of $\theta(\hbar{\mib X}^*{\mib X})$ given in 
the above tells us that $\theta(\hbar{\mib X}^*{\mib X})$ is 
equal to zero in the region $\hbar{\mib X}^*{\mib X} < 2S$ 
and at the point $\hbar{\mib X}^*{\mib X}=2S$, 
$\theta(\hbar{\mib X}^*{\mib X})$ behaves like an 
infinite wall. The existence of the wall suggests us that 
the quantity $\hbar{\mib X}^*{\mib X}$ shows a certain periodic 
behavior with the period $\sim \ln S$. Then, if $S$ is 
sufficiently large, we can expect that the solution 
for the pseudo $su(1,1)$-algebra is well approximated 
as that for the $su(1,1)$-algebra. The details are 
discussed in Ref. \citen{KPTY4}.

\section{The $su(2,1)$-algebra and its modified form in terms of 
three kinds of boson operators}

In the previous section, we presented a possible description 
of a boson system interacting with an external harmonic 
oscillator. Of course, two kinds of boson operators 
$({\hat c}, {\hat c}^*)$ and $({\hat d}, {\hat d}^*)$ 
are introduced and a coherent state for a possible 
time-evolution of the system is expressed in terms of these two 
kinds of boson operators, which describe the physical objects. 
Therefore, these bosons are not related to the phase space 
doubling such as mentioned in \S 5. In this sense, with the help 
of the coherent state adopted in \S 7, it may be impossible 
to treat statistically mixed state. In this section, we will 
give a theoretical framework which enables us to describe the 
mixed state of a boson system interacting with the external 
harmonic oscillator.\cite{KPTY6}

In addition to the boson operators $({\hat c}, {\hat c}^*)$ and 
$({\hat d}, {\hat d}^*)$, we introduce the boson 
$({\hat a}, {\hat a}^*)$. With the use of these bosons, the 
following bi-linear form can be defined :
\begin{eqnarray}
& &\maru{T}_+=\hbar{\hat a}^*{\hat c}^* \ , \qquad \ 
\maru{T}_-=\hbar{\hat c}{\hat a} \ , \qquad
\maru{T}_0=\frac{\hbar}{2}({\hat a}^*{\hat a}
+{\hat c}^*{\hat c}) +\frac{\hbar}{2} \ , 
\label{8-1}\\
& &\maru{R}_+=\hbar{\hat a}^*{\hat d}^* \ , \qquad
\maru{R}_-=\hbar{\hat d}{\hat a} \ , \qquad
\maru{R}_0=\frac{\hbar}{2}({\hat a}^*{\hat a}
+{\hat d}^*{\hat d}) +\frac{\hbar}{2} \ , 
\label{8-2}\\
& &\maru{\sigma}_+=\hbar{\hat c}^*{\hat d} \ , \qquad\ \ 
\maru{\sigma}_-=\hbar{\hat d}^*{\hat c} \ , \qquad
\maru{\sigma}_0=\frac{\hbar}{2}({\hat c}^*{\hat c}
-{\hat d}^*{\hat d}) \ , 
\label{8-3}
\end{eqnarray}
The set $(\maru{T}_{\pm,0})$ and $(\maru{R}_{\pm,0})$ obey 
the $su(1,1)$-algebra, respectively, and $(\maru{\sigma}_{\pm,0})$ 
the $su(2)$-algebra. The operator $\maru{\sigma}_0$ is 
related with 
\begin{equation}\label{8-4}
\maru{\sigma}_0=\maru{T}_0-\maru{R}_0 \ .
\end{equation}
We can prove that the set $(\maru{T}_{\pm,0}, \maru{R}_{\pm,0}, 
\maru{\sigma}_{\pm})$ composes the $su(2,1)$-algebra. 
In the above set, there exists an operator which commutes with 
$\maru{T}_{\pm,0}$, $\maru{R}_{\pm,0}$ and $\maru{\sigma}_{\pm,0}$ 
: 
\begin{eqnarray}
& &\maru{K}=\frac{\hbar}{2}({\hat c}^*{\hat c}+{\hat d}^*{\hat d}
-{\hat a}^*{\hat a})+\hbar \ , 
\label{8-5}\\
& &[ \maru{K} , \maru{T}_{\pm,0} ]=[ \maru{K} , \maru{R}_{\pm,0}]
= [ \maru{K} , \maru{\sigma}_{\pm,0} ]=0 \ .
\label{8-6}
\end{eqnarray}

As was shown in Ref.\citen{KPTY6}, there exist many cases for 
constructing the orthogonal set of the $su(2,1)$-algebra. 
In this paper, we show one example.
For this aim, we introduce the operator $\maru{T}$ in the form 
\begin{equation}\label{8-7}
\maru{T}=\frac{\hbar}{2}({\hat c}^*{\hat c}-{\hat a}^*{\hat a})
+\frac{\hbar}{2} \ .
\end{equation}
The operator $\maru{T}$ commutes with $\maru{K}$ and $\maru{T}_0$ 
and, then, we set up the eigenvalue equation 
\begin{eqnarray}\label{8-8}
& &\maru{K}\kket{K,T,T_0}=K\kket{K,T,T_0}\ , \quad \ \ 
K=\hbar k \ , \nonumber\\
& &\maru{T}\kket{K,T,T_0}=T\kket{K,T,T_0}\ , \quad \ \ \ 
T=\hbar t \ , \nonumber\\
& &\maru{T}_0\kket{K,T,T_0}=T_0\kket{K,T,T_0}\ , \quad \ 
T_0=\hbar t_0 \ .
\end{eqnarray}
A possible solution of the eigenvalue equation (\ref{8-8}) 
is given as follows : 
\begin{eqnarray}
\kket{K,T,T_0}
&=&
\frac{1}{\sqrt{(t_0+t-1)!(2k-2t-1)!(t_0-t)!}}\nonumber\\
& &\times ({\hat c}^*)^{t_0+t-1}({\hat d}^*)^{2k-2t-1}({\hat a}^*)^{t_0-t}
\kket{0} \nonumber\\
&=&\frac{1}{(2k-2)!}\sqrt{\frac{(2k-2t-1)!}{(t_0+t-1)!(t_0-t)!}}
\nonumber\\
& &\times
\left(\frac{\maru{T}_+}{\hbar}\right)^{t_0-t}
\left(\frac{\maru{S}_+}{\hbar}\right)^{2t-1}
({\hat d}^*)^{2k-2}\kket{0} \ . 
\label{8-9}\\
& &\ k=1,\ \frac{3}{2},\ 2, \cdots , \nonumber\\
& &\ t=\frac{1}{2}, \ 1,\ \frac{3}{2}, \cdots , \ \frac{k-1}{2} \ ,
\nonumber\\
& &\ t_0=t,\ t+1,\ t+2, \cdots . 
\label{8-10}
\end{eqnarray}
The classical counterpart can be derived by the following 
coherent state : 
\begin{equation}\label{8-11}
\kket{c}=M_c \exp\left(\frac{V}{U}\cdot\frac{\maru{T}_+}{\hbar}
\right)
\exp\left(\frac{w}{vU}\cdot\frac{\maru{S}_+}{\hbar}\right)
\exp(v{\hat d}^*)\kket{0} \ .
\end{equation}
Here, $V$, $w$ and $v$ are complex parameters and 
$U=\sqrt{1+|V|^2}$. 
Instead of the above parameters, $({\mib \theta}, {\mib K})$, 
$({\mib X}, {\mib X}^*)$ and $({\mib Y}, {\mib Y}^*)$ are 
convenient for our discussion. They are required to satisfy the 
relation 
\begin{eqnarray}
& &\bbra{c}i\hbar\partial_{\mib \theta}\kket{c}
={\mib K}-\hbar \ , \qquad
\bbra{c}i\hbar\partial_{\mib K}\kket{c}
=0 \ , \label{8-12}\\
& &\bbra{c}\partial_{\mib X}\kket{c}
=+\frac{{\mib X}^*}{2} \ , \qquad
\bbra{c}\partial_{{\mib X}^*}\kket{c}
=-\frac{\mib X}{2} \ , \label{8-13}\\
& &\bbra{c}\partial_{\mib Y}\kket{c}
=+\frac{{\mib Y}^*}{2} \ , \qquad
\bbra{c}\partial_{{\mib Y}^*}\kket{c}
=-\frac{\mib Y}{2} \ . \label{8-14}
\end{eqnarray}
Further, ${\mib K}$ obeys the condition 
\begin{equation}\label{8-15}
\bbra{c} \maru{K} \kket{c} = {\mib K} \ . 
\end{equation}
The relations (\ref{8-12}) $\sim$ (\ref{8-15}) give us 
the form 
\begin{eqnarray}\label{8-16}
& &w=\exp\left(-i\frac{{\mib \theta}}{2}\right) {\mib X} \ , 
\nonumber\\
& &v=\exp\left(-i\frac{{\mib \theta}}{2}\right) 
\sqrt{\frac{2({\mib K}-\hbar)}{\hbar}-{\mib X}^*{\mib X}} \ , 
\nonumber\\
& &V={\mib Y}\frac{1}{\sqrt{1+{\mib X}^*{\mib X}}} \ .
\end{eqnarray}
Then, the expectation values of $\maru{T}_{\pm,0}$ and 
$\maru{R}_{\pm,0}$ 
%and $\maru{\sigma}_{\pm,0}$ 
are given by 
\begin{eqnarray}\label{8-17}
& &(T_+)_c=\sqrt{\hbar}{\mib Y}^*\cdot
\sqrt{\hbar{\mib X}^*{\mib X}+\hbar+\hbar{\mib Y}^*{\mib Y}} \ , 
\nonumber\\
& &(T_-)_c=
\sqrt{\hbar{\mib X}^*{\mib X}+\hbar+\hbar{\mib Y}^*{\mib Y}}
\cdot\sqrt{\hbar}{\mib Y} \ , 
\nonumber\\
& &(T_0)_c=\hbar{\mib Y}^*{\mib Y}+\frac{1}{2}
(\hbar{\mib X}^*{\mib X}+\hbar) \ , \nonumber\\
& &(R_+)_c=\sqrt{\hbar}{\mib Y}^*\cdot
\sqrt{2({\mib K}-\hbar)-\hbar{\mib X}^*{\mib X}}\cdot
\frac{1}{\sqrt{\hbar{\mib X}^*{\mib X}+\hbar}}\cdot
\sqrt{\hbar}{\mib X} \ , 
\nonumber\\
& &(R_-)_c=\sqrt{\hbar}{\mib X}^*\cdot
\frac{1}{\sqrt{\hbar{\mib X}^*{\mib X}+\hbar}}\cdot
\sqrt{2({\mib K}-\hbar)-\hbar{\mib X}^*{\mib X}}\cdot
\sqrt{\hbar}{\mib Y} \ , 
\nonumber\\
& &(R_0)_c={\mib K}-\frac{\hbar}{2}{\mib X}^*{\mib X}
+\hbar{\mib Y}^*{\mib Y}-\frac{\hbar}{2} \ . 
\end{eqnarray}
In Ref.\citen{KPTY6}, it was discussed in detail that 
the form (\ref{8-17}) can be regarded as the classical 
counterpart of the $su(2,1)$-algebra in terms of three kinds 
of boson operators.

In order to describe a many-boson system interacting with an 
external harmonic oscillator, we modified the $su(1,1)$-algebra, 
which was called the pseudo $su(1,1)$-algebra. Then, in the present 
case, we must modify the $su(2,1)$-algebra. For this aim, 
first, we replace $\maru{\sigma}_{\pm,0}$ in 
Eq. (\ref{8-3}) with $\maru{\tau}_{\pm,0}$ defined in 
the relation (\ref{6-1}). Then, by extending the coherent 
states shown in Eqs. (\ref{6-20}) and (\ref{8-11}), 
respectively, we adopt the following form :
\begin{equation}\label{8-18}
\kket{c}=M_c \exp\left(\frac{V}{U}\cdot\frac{\maru{T}_+}{\hbar}
\right) \exp\left(\frac{\xi}{\eta U}\cdot\frac{\maru{\tau}_+}{\hbar}
\right) \exp\left(\eta\maru{\Delta}^*\right) \kket{0} \ .
\end{equation}
Here, $V$, $\xi$ and $\eta$ are complex parameters and 
$U=\sqrt{1+|V|^2}$. The operator $\maru{\Delta}^*$ is 
defined in Eq.(\ref{6-22}). In the same way as in the previous 
case, we introduce $({\mib \theta}, {\mib K})$, 
$({\mib X}, {\mib X}^*)$ and $({\mib Y}, {\mib Y}^*)$ 
which obey the conditions (\ref{8-12}) $\sim$ (\ref{8-15}). 
Then, the parameters $\xi$, $\eta$ and $V$ are expressed in 
the form 
\begin{eqnarray}\label{8-19}
& &\xi=\exp\left(-i\frac{{\mib \theta}}{2}\right)
{\mib X} \frac{1}{\sqrt{2t+{\mib X}^*{\mib X}}} \ , \nonumber\\
& &\eta=\exp\left(-i\frac{{\mib \theta}}{2}\right)
\sqrt{2({\mib K}-\hbar)-\hbar{\mib X}^*{\mib X}}
\frac{1}{\sqrt{2({\mib K}-\hbar)+\hbar-\hbar{\mib X}^*{\mib X}}} 
\ , \nonumber\\
& &V={\mib Y}\frac{1}{\sqrt{1+{\mib X}^*{\mib X}}} \ .
\end{eqnarray}
On the basis of the form presented in the above, we will 
treat statistically mixed state of a boson system interacting 
with an external harmonic oscillator. The state $\kket{c}$ in 
Eq. (\ref{8-18}) can be factorized in the form 
\begin{eqnarray}
& &\kket{c}=\kket{\Phi_{ca}}\otimes
\kket{\Psi_{d}} \ , \label{8-20}\\
& &
\kket{\Phi_{ca}}=U^{-1}(1-|\xi|^2)^t 
\exp\left(\frac{V}{U}{\hat a}^*{\hat c}^*\right)\cdot
\exp\left(\frac{\xi}{U}{\hat c}^*\sqrt{2t+{\hat c}^*{\hat c}}
\right) \kket{0}_{ca} \ , \quad\label{8-21}\\
& &
\kket{\Psi_{d}}=(1-|\eta|^2)^{\frac{1}{2}}
\exp\left(\eta{\hat d}^*\sqrt{1+{\hat d}^*{\hat d}}\right)
\kket{0}_d \ .
\label{8-22}
\end{eqnarray}
The above factorization will be used in the next section.

\section{Description of statistically mixed state of a boson 
system interacting with an external harmonic oscillator}

Before entering the concrete treatment for the present problem, 
we will contact with the basic idea of our treatment in rather 
general form. Let us investigate a many-body system consisting 
of two parts. We refer to the first and the second part as 
the $c$-part and the $d$-part. The Hamiltonian consists of 
\begin{equation}\label{9-1}
\maru{H}=\maru{K}_c+\maru{K}_d+\maru{V}_{cd} \ .
\end{equation}
Here, $\maru{K}_c$ and $\maru{K}_d$ denote the Hamiltonian of the 
$c$- and the $d$-part, respectively, and the two parts interact 
with each other through the term $\maru{V}_{cd}$. 
Therefore, we must introduce two spaces, which we call the $c$- 
and the $d$-space, and the present system is treated in the 
product of the two spaces. The operators $\maru{K}_c$ and 
$\maru{K}_d$ belong to the $c$- and the $d$-space, respectively, 
and $\maru{V}_{cd}$ connects the two spaces.

For the Hamiltonian (\ref{9-1}), we set up the time-dependent 
Schr\"odinger equation 
\begin{equation}\label{9-2}
i\hbar\partial_t \kket{c} = \maru{H} \kket{c} \ , \qquad
\bbra{c}c \rangle\rangle = 1  \ .
\end{equation}
As a possible solution of Eq. (\ref{9-2}), let us assume 
the form
\begin{eqnarray}\label{9-3}
& &\kket{c}=\kket{\Phi_c}\otimes \kket{\Psi_d} \ , \nonumber\\
& &\bbra{\Phi_c}\Phi_c\rangle\rangle=
\bbra{\Psi_d}\Psi_d\rangle\rangle=1 \ .
\end{eqnarray}
Here, $\kket{\Phi_c}$ and $\kket{\Psi_d}$ are given in the 
$c$- and the $d$-space, respectively. For the state (\ref{9-3}), 
we impose two conditions. The first is the following : 
For any state $\kket{k}$ in the $d$-space which is orthogonal 
to $\kket{\Psi_d}$, $\kket{\Psi_d}$ is governed by 
\begin{eqnarray}
& &\bbra{k} \maru{K}_d+\maru{V}_{cd}-i\hbar\partial_t 
\kket{\Psi_d}= 0 \ , \label{9-4}\\
& &\langle\langle k\kket{\Psi_d}=0 \ .\label{9-5}
\end{eqnarray}
The second condition is the following : 
For the state $\kket{\Psi_d}$ satisfying the condition 
(\ref{9-4}), $\kket{\Phi_c}$ obeys 
\begin{eqnarray}
& &i\hbar\partial_t \kket{\Phi_c}=\maru{H}_c\kket{\Phi_c} \ ,
\label{9-6}\\
& &\maru{H}_c=\maru{K}_c+
\bbra{\Psi_d} \maru{K}_d+\maru{V}_{cd}-i\hbar\partial_t 
\kket{\Psi_d} \ . \label{9-7}
\end{eqnarray}
Here, it should be noted that the left-hand side of 
Eq. (\ref{9-4}) and the second term on the right-hand side 
of Eq. (\ref{9-7}) are calculated only for operators in the 
$d$-space and, then, naturally, they are operators in the 
$c$-space. Under the conditions (\ref{9-4}) and (\ref{9-6}), 
we can prove that the state $\kket{c}$ given in Eq. (\ref{9-3}) 
is an exact solution of the Schr\"odinger equation (\ref{9-2}). 
As is clear from the forms (\ref{9-4}) $\sim$ (\ref{9-7}), 
the above-mentioned two conditions are too strong to get 
the exact solution. However, it may be permissible 
as an approximation. In general, the time-dependent 
variational procedure gives us a possible approximate 
solution of the Schr\"odinger equation. Therefore, under the 
procedure with the trial function $\kket{c}$ in the form 
(\ref{9-3}), we obtain an approximate solution of 
Eq. (\ref{9-2}).

Our chief concern is related to a possible description of 
statistically mixed state realized in the $c$-part. Then, with 
an idea similar to that used in thermo field dynamics formalism, 
we perform the phase space doubling for the $c$-part. 
This procedure is done through the introduction of a 
new space which is conjugate to the $c$-space. We call this space 
the $a$-space. Let a possible solution of the Schr\"odinger equation 
in a form analogous to $\kket{c}$ given in Eq. (\ref{9-3}) 
be obtained :
\begin{equation}\label{9-8}
\kket{c}=\kket{\Phi_{ca}}\otimes \kket{\Psi_d} \ , \qquad
\langle\langle c\kket{c}=1  \ . 
\end{equation}
The part $\kket{\Psi_d}$ obeys the conditions (\ref{9-4}) and 
(\ref{9-5}) and $\kket{\Phi_{ca}}$ is given by solving 
Eq. (\ref{9-6}) in the $ca$-space :
\begin{equation}\label{9-9}
i\hbar\partial_t \kket{\Phi_{ca}}=\maru{H}_c \kket{\Phi_{ca}} \ , \qquad
\langle\langle \Phi_{ca}\kket{\Phi_{ca}}=1  \ . 
\end{equation}
Here, $\maru{H}_c$ is given in Eq.(\ref{9-7}). With the use 
of the state $\kket{c}$ obtained in the above process, we are 
able to describe the mixed state in a manner similar to the standard 
thermo field dynamics formalism. In our present system, 
let us introduce the operator $\maru{\rho}_c$ in the form 
\begin{equation}\label{9-9b}
\maru{\rho}_c=\sum_n (n\kket{\Phi_{ca}}\bbra{\Phi_{ca}} n) \ .
\end{equation}
Here, $\{ \rket{n} \} $ denotes an orthonormal set in the 
$a$-space. Then, $\maru{\rho}_c$ is an operator given in 
the $c$-space. We can prove that $\maru{\rho}_c$ satisfies the 
conditions which the density matrix obeys : 
\begin{equation}\label{9-10}
{\rm Tr}\ \maru{\rho}_c = 1 \ , \qquad
i\hbar \partial_t \maru{\rho}_c = [ \maru{H}_c , \maru{\rho}_c ] 
\ . 
\end{equation}
Here, $\maru{H}_c$ is given in Eq.(\ref{9-7}). The ensemble 
average of the operator $\maru{O}_c$ in the $c$-space is 
expressed as 
\begin{equation}\label{9-11}
{\rm Tr}\ (\maru{O}_c \maru{\rho}_c) =
\bbra{\Phi_{ca}} \maru{O}_c \kket{\Phi_{ca}}
=\bbra{c} \maru{O}_c \kket{c} \ . 
\end{equation}
Further, the entropy $\maru{S}_c$ in the $c$-space can be 
defined as 
\begin{equation}\label{9-12}
\maru{S}_c=-\sum_n \bbra{\Phi_{ca}} n)(n \kket{\Phi_{ca}} 
\cdot \ln \bbra{\Phi_{ca}} n)(n \kket{\Phi_{ca}} \ .
\end{equation}
We can discriminate the pure state in terms of the index $n$.

On the basis of the basic idea mentioned in the above, let us 
investigate the Hamiltonian (\ref{7-6}) in the framework 
of the modification of the $su(2,1)$-algebra given in \S 8. 
The expectation values of $\maru{K}_c$, $\maru{K}_d$ and 
$\maru{V}_{cd}(1)$ shown in Eqs. (\ref{7-1}) $\sim$ 
(\ref{7-3}), respectively, are calculated as follows :
\begin{eqnarray}
& &(K_c)_c=e^0\cdot \hbar({\mib X}^*{\mib X}+{\mib Y}^*{\mib Y})
\nonumber\\
& &\qquad\quad
+g^0\cdot\left\{2[\hbar({\mib X}^*{\mib X}+{\mib Y}^*{\mib Y})]^2
-\left(1-\frac{\hbar}{2T}\right)(\hbar{\mib X}^*{\mib X})^2
[1+\hbar{\mib Y}^*{\mib Y}(\hbar+\hbar{\mib X}^*{\mib X})^2]
\right\} \ , \nonumber\\
& &\label{9-13}\\
& &(K_d)_c=\omega^0\cdot [2({\mib K}-\hbar)
-\hbar{\mib X}^*{\mib X}] \ , 
\label{9-14}\\
& &(V_{cd}(1))_c=
2f^0\cdot\hbar({\mib X}^*{\mib X}+{\mib Y}^*{\mib Y})
[2({\mib K}-\hbar)-\hbar{\mib X}^*{\mib X}]
+\hbar f^0\cdot[2({\mib K}-\hbar)+\hbar{\mib Y}^*{\mib Y}]
\ . \nonumber\\
& &\label{9-15}
\end{eqnarray}
The above three are exact. But, in the case of $(V_{cd}(2))_c$, 
the approximation form is given as 
\begin{equation}\label{9-16}
(V_{cd}(2))_c=-\gamma^0\cdot i\sqrt{\hbar}
({\mib X}^*-{\mib X})\sqrt{2\tau+\hbar{\mib X}^*{\mib X}}
\sqrt{\frac{2({\mib K}-\hbar)-\hbar{\mib X}^*{\mib X}}
{2({\mib K}-\hbar)+\hbar-\hbar{\mib X}^*{\mib X}}} \ .
\end{equation}
Here, $\tau$ is defined by 
\begin{equation}\label{9-17}
\tau=T+\hbar{\mib Y}^*{\mib Y} \ , \qquad T=\hbar t \ .
\end{equation}
Concerning the derivation of the relation (\ref{9-16}), 
the following approximate form plays a central role : 
\begin{equation}\label{9-18}
\bbra{c}\sqrt{\hbar}{\hat c}^*\sqrt{2T+\hbar{\hat c}^*{\hat c}}
\kket{c}
=
\exp\left(i\frac{\mib \theta}{2}\right)\cdot\sqrt{\hbar}{\mib X}^*
\sqrt{2\tau+\hbar{\mib X}^*{\mib X}} \ .
\end{equation}
The detail is given in Ref.\citen{KPTY7}. Further, we introduce 
the quantities $\tau_{\pm,0}$ as follows :
\begin{eqnarray}
& &\tau_+=\sqrt{\hbar}{\mib X}^*
\sqrt{2\tau+\hbar{\mib X}^*{\mib X}}
\sqrt{\frac{2({\mib K}-\hbar)-\hbar{\mib X}^*{\mib X}}
{2({\mib K}-\hbar)+\hbar-\hbar{\mib X}^*{\mib X}}} \ , \nonumber\\
& &\tau_-=\sqrt{\hbar}{\mib X}
\sqrt{2\tau+\hbar{\mib X}^*{\mib X}}
\sqrt{\frac{2({\mib K}-\hbar)-\hbar{\mib X}^*{\mib X}}
{2({\mib K}-\hbar)+\hbar-\hbar{\mib X}^*{\mib X}}} \ , \nonumber\\
& &\tau_0=\tau+\hbar{\mib X}^*{\mib X} \ , 
\label{9-19}\\
& &S={\mib K}-\hbar+\frac{\hbar}{2}{\mib Y}^*{\mib Y} \ .
\label{9-20}
\end{eqnarray}
The quantity $S$ is the expectation value of $\maru{S}$ 
$(=\hbar({\hat c}^*{\hat c}+{\hat d}^*{\hat d})/2)$ for the state 
$\kket{c}$. Then, we have 
\begin{eqnarray}
& &H_{su(2,1)}=U_{su(2,1)}+K_{su(2,1)}+V_{su(2,1)} \ , \label{9-21}\\
& &U_{su(2,1)}=
2\left[\omega^0-2f^0\cdot\left(T-\frac{\hbar}{2}\right)\right]
S
\nonumber\\
& &\qquad\qquad
-[(e^0-\omega^0)-\hbar g^0]\cdot T
+\left[g^0\left(1-\frac{\hbar}{2T}\right)-2f^0\right] 
T^2 \ ,
\label{9-22}\\
& &K_{su(2,1)}=
[(e^0-\omega^0)-\hbar g^0+4f^0 S-2(g^0-2f^0)T]{\tau}_0
\nonumber\\
& &\qquad\qquad
+\left[g^0\left(1+\frac{\hbar}{2T}\right)-2f^0\right]
\tau_0^2 \ , \quad\quad
\label{9-23}\\
& &V_{su(2,1)}=
-\gamma^0 \cdot i(\tau_+ -\tau_-) \ .
\label{9-24}
\end{eqnarray}
We can see that the Hamiltonian (\ref{9-21}) with the forms 
(\ref{9-22}) $\sim$ (\ref{9-24}) is of the same form as that 
shown in Eq. (\ref{7-37}) with the forms 
(\ref{7-38}) $\sim$ (\ref{7-40}) if $\tau$ and 
$({\mib K}-\hbar)$ in $\tau_{\pm,0}$ appearing in 
Eq. (\ref{9-19}) are replaced with $T$ and ${\mib S}$ in 
$(\tau_{\pm,0})_c$ appearing in Eq. (\ref{6-30}), 
respectively. However, as was discussed in Ref.\citen{KPTY7}, 
the meaning of $H_{su(2,1)}$ is different from that of 
$H_{su(2)}$. Since $H_{su(2,1)}$ is of the same form as $H_{su(2)}$, 
we can treat $H_{su(2,1)}$ under the same method as that adopted 
in \S\S 4 and 7. Then, we will not contact with this method 
in detail.

In order to make the problem simpler, we consider the case of 
$e^0=\omega^0$ and $g^0=f^0=0$, which corresponds 
to the harmonic oscillator. Further, $({\mib K}-\hbar)$ is 
sufficiently large, i.e., $2({\mib K}-\hbar) \ge \hbar
{\mib X}^*{\mib X}$. In this case, $\hbar{\mib X}^*{\mib X}$ shows a 
certain long periodic behavior and, with the help of 
Eqs. (\ref{4-44}), (\ref{4-47}) and (\ref{4-56}), 
we have 
\begin{equation}\label{9-25}
\hbar{\mib X}^*{\mib X}=\sqrt{{\mib I}^2+{\mib J}^2}
\cosh (2\gamma t+\chi^0)-{\mib I}  \ ,
\end{equation}
i.e., 
\begin{equation}\label{9-26}
\bbra{c}\hbar{\hat c}^*{\hat c} \kket{c}
=\sqrt{{\mib I}^2+{\mib J}^2}
\cosh (2\gamma t+\chi^0)-T  \ .
\end{equation}
Since ${\mib \theta}=2\omega t+\theta^0$, to order $\gamma^1$, 
the expectation value $\bbra{c}\hbar{\hat c}^*{\hat c}\kket{c}$ 
can be expressed as 
\begin{eqnarray}\label{9-27}
\bbra{c}\hbar{\hat c}^*{\hat c}\kket{c} 
&=& \bbra{c}\hbar{\hat c}^*{\hat c}\kket{c}_{t=0} \nonumber\\
& & -\gamma t\biggl[\bbra{c}\sqrt{\hbar}{\hat c}^*\cdot\sqrt{2T+\hbar
{\hat c}^*{\hat c}}\cdot\exp\left(-i\frac{\theta^0}{2}\right) 
\nonumber\\
& &\qquad\qquad
+\sqrt{2T+\hbar{\hat c}^*{\hat c}}\cdot\sqrt{\hbar}{\hat c}\cdot
\exp\left(i\frac{\theta^0}{2}\right) \kket{c}\biggl]_{t=0}\ .
\end{eqnarray}
Here, the symbol $t$ denotes time.

Now let us investigate our system in the framework mentioned 
in the beginning part of this section. 
With the use of $\kket{\Phi_{ca}}$ and $\kket{\Psi_d}$ in 
Eqs. ({\ref{8-21}}) and (\ref{8-22}) and the form $\eta$ 
in Eq. (\ref{8-19}), $\maru{H}_c$ defined in the relation 
(\ref{9-7}) can be expressed as 
\begin{eqnarray}
& &{\maru H}_c={\maru H}_0+{\maru H}_1+{\maru H}_2 \ , \label{9-28}\\
& &{\maru H}_0=\hbar\omega{\hat c}^*{\hat c} \ , \label{9-29}\\
& &{\maru H}_1=-\gamma\cdot i[
\sqrt{\hbar}{\hat c}^*\cdot\sqrt{2T+\hbar{\hat c}^*{\hat c}}
\cdot\eta-\sqrt{2T+\hbar{\hat c}^*{\hat c}}\cdot\sqrt{\hbar}
{\hat c}\cdot\eta^*] \ , \label{9-30}\\
& &{\maru H}_2=-\frac{i\hbar}{2}\frac{1}{1-|\eta|^2}
[({\dot \eta}+i\omega\eta)\eta^*-({\dot \eta}^*-i\omega\eta^*)\eta]
\ .
\label{9-31}
\end{eqnarray}
Substituting ${\mib \theta}=2\omega t+\theta^0$ into the form 
(\ref{8-19}), $\eta$ can be expressed as 
\begin{equation}\label{9-32}
\eta=\exp\left(-i\frac{\theta^0}{2}\right)
\exp(-i\omega t)
\sqrt{1-\frac{\hbar}{2({\mib K}-\hbar)+\hbar-\hbar{\mib X}^*{\mib X}}}
\ . 
\end{equation}
Then, we can prove that, at order $\hbar^0$, the part ${\maru H}_2$ 
vanishes and $\eta=\exp(-i\theta^0/2)$\break
$\times\exp(-i\omega t)$. 
Thus, at this order, ${\maru H}_c$ can be approximated as 
\begin{equation}\label{9-33}
{\maru H}_c={\maru H}_0+{\maru H}_1(t) \ .
\end{equation}
Here, ${\maru H}_0$ is given in Eq. (\ref{9-29}) and 
${\maru H}_1(t)$ is written in the form 
\begin{eqnarray}\label{9-34}
{\maru H}_1(t)&=&
-\gamma\cdot i\Bigl[
\sqrt{\hbar}{\hat c}^*\cdot
\sqrt{2T+{\hat c}^*{\hat c}}\cdot\exp(-i\theta^0/2)\cdot
\exp(-i\omega t) \nonumber\\
& &\qquad\qquad
-\sqrt{2T+\hbar{\hat c}^*{\hat c}}\cdot\sqrt{\hbar}{\hat c}
\cdot\exp(+i\theta^0/2)\cdot\exp(+i\omega t) \Bigl]
\ .
\end{eqnarray}
The Hamiltonian ${\maru H}_c$ indicates that the present system is 
a harmonic oscillator (or many-boson system) in a classical 
external field with the frequency $\omega$. The interaction with this 
is ${\maru H}_1(t)$. We will apply linear response theory of 
non-equilibrium statistical mechanics. For the Hamiltonian (\ref{9-28}), 
this method leads us to the following form in the case of 
$\hbar{\hat c}^*{\hat c}$ : 
\begin{eqnarray}\label{9-35}
{\rm Tr}(\hbar{\hat c}^*{\hat c}\maru{\rho}(t))
&=& {\rm Tr}(\hbar{\hat c}^*{\hat c}\maru{\rho}(0)) \nonumber\\
& & -\gamma t\cdot{\rm Tr}\biggl[\Biggl(\sqrt{\hbar}{\hat c}^*\cdot
\sqrt{2T+\hbar{\hat c}^*{\hat c}}\cdot\exp\left(-i\frac{\theta^0}{2}\right) 
\nonumber\\
& &\qquad\qquad
+\sqrt{2T+\hbar{\hat c}^*{\hat c}}\cdot\sqrt{\hbar}{\hat c}\cdot
\exp\left(+i\frac{\theta^0}{2}\right)\Biggl)\maru{\rho}(0)
\biggl]\ . \ \ \ \ \ 
\end{eqnarray}
Here, $\maru{\rho}(t)$ obeys the von Neumann equation 
\begin{equation}\label{9-36}
i\hbar\partial_t \maru{\rho}(t)
=[ \maru{H}_c , \maru{\rho}(t) ] \ .
\end{equation}
A possible solution of Eq. (\ref{9-36}) is given in the form 
$\maru{\rho}(t)$ shown in Eq. (\ref{9-9b}) and the relation 
(\ref{9-11}) tells us that, for any operator ${\hat O}_c$, we have 
\begin{equation}\label{9-37}
{\rm Tr}({\hat O}_c\maru{\rho}(0))=\bbra{c}{\hat O}_c \kket{c}_{t=0} 
\ . 
\end{equation}
Then, the results (\ref{9-27}) and (\ref{9-35}) coincide with 
each other under the correspondence (\ref{9-37}).  
In our present method, we can calculate ${\rm Tr}(\hbar{\hat c}^*
{\hat c}\maru{\rho}(t))$ for any order of $\gamma$ in the 
specific order. Also, it may be interesting to see 
that our approach is closely related to linear response theory 
of non-equilibrium statistical mechanics.\cite{Kubo}

In our present system, $\hbar{\mib Y}^*{\mib Y}$ does not 
depend on $t$, while $\hbar{\mib X}^*{\mib X}$ depends on $t$. 
Therefore, in the present system, the minimum value of 
$\hbar{\mib X}^*{\mib X}$ for $t$ can be regarded as the 
equilibrium value, that is, the equilibrium state appears. 
This state realizes at $2\gamma t + \chi^0=0$ in the relation 
(\ref{9-25}). Then, we have 
\begin{eqnarray}\label{9-38}
(\hbar{\mib X}^*{\mib X})_{\rm eq}&=&
\sqrt{{\mib I}^2+{\mib J}^2}-{\mib I} \nonumber\\
&=&\sqrt{(T+\hbar{\mib Y}^*{\mib Y})^2+{\mib J}^2}
-(T+\hbar{\mib Y}^*{\mib Y}) \ .
\end{eqnarray}
In this case, the entropy $S_c$ is derived from the relation 
(\ref{9-12}) in the following approximate form :
\begin{equation}\label{9-39}
S_c=(1+{\mib Y}^*{\mib Y})\ln (1+{\mib Y}^*{\mib Y})
-{\mib Y}^*{\mib Y}\ln ({\mib Y}^*{\mib Y})\ .
\end{equation}
At the equilibrium point, we define the free energy $F_c$ 
in the form 
\begin{eqnarray}
F_c&=&E_c-\frac{S_c}{\beta} \ , \label{9-40}\\
E_c&=&\bbra{c}\hbar\omega{\hat c}^*{\hat c}\kket{c} \nonumber\\
&=&\omega \left(\sqrt{(T+\hbar{\mib Y}^*{\mib Y})^2+{\mib J}^2}
-T\right) \ , 
\label{9-41}\\
\beta&=&(k_{\rm B}T_{\rm eq})^{-1} \ .\label{9-42}
\end{eqnarray}
By minimizing $F_c$ with respect to ${\mib Y}^*{\mib Y}$, 
we have the following relation : 
\begin{equation}\label{9-43}
{\mib I}=T+\hbar\frac{1}{
\exp\left(\hbar\omega\beta\cdot\frac{\mib I}{\sqrt{{\mib I}^2
+{\mib J}^2}}\right)-1} \ .
\end{equation}
From the above relation, we can calculate the energy $E_c$ 
as a function of $\hbar\omega\beta$. Two extreme cases are rather 
easily derived : 

\vspace{0.3cm}
\noindent
(1) The low temperature limit :
\begin{eqnarray}
& &E_c=\frac{\omega{\mib J}^2}{T+\sqrt{T^2+{\mib J}^2}}
+\hbar\omega_c\frac{1}{\exp(\hbar\omega_c\beta)-1}\nonumber\\
& &\qquad\quad
-\frac{\hbar}{2}\frac{\hbar\omega{\mib J}^2}{(\sqrt{T^2+{\mib J}^2})^3}
\cdot\left(\frac{1}{\exp(\hbar\omega_c\beta)-1}\right)^2+\cdots \ , 
\label{9-44}\\
& &\omega_c=\omega\frac{T}{\sqrt{T^2+{\mib J}^2}} \ .
\label{9-45}
\end{eqnarray}

\vspace{0.3cm}
\noindent
(2) The high temperature limit :
\begin{equation}\label{9-46}
E_c=k_{\rm B}T_{\rm eq}+\omega^2\sqrt{T^2+{\mib J}^2}\cdot
(k_{\rm B}T_{\rm eq})^{-1}+\cdots \ .
\end{equation}
We see that in the relation (\ref{9-44}), the first term 
denotes the energy of $T_{\rm eq}=0$, and 
the second represents the energy coming from the thermal 
fluctuations and the distribution function of free bose 
particle. The frequency $\omega$ is changed 
effectively to $\omega_c$, which may be the result 
of interaction with the external oscillator. 
On the other hand, the high temperature limit 
reveals that $E_c \sim k_{\rm B}T_{\rm eq}$ and we can 
see that present system obeys the well-known law of 
equi-partition of energy. The details can be found 
in Ref.\citen{KPTY7}.

\section{Time-dependent variational 
approach with squeezed state to quantum mechanical systems}

In this and succeeding sections, we introduce another 
trial state in the time-dependent 
variational approach for quantum many-particle systems 
with the aim of investigating the classical motion 
including quantum fluctuations in the quantized systems. 
The new state, that is called the (one-mode) squeezed state, 
is introduced in this section, 
which gives a possible classical motion 
in quantum mechanical systems including quantum fluctuations beyond 
the order of $\hbar$.\cite{TFKY} The similar variational 
approach is found in Ref.\citen{JK}, while our treatment is 
developed widely in the systematic way. 
We here present instructively our formalism based on 
the time-dependent variational principle with the squeezed state 
in quantum mechanical systems. 

The squeezed state is defined as 
\begin{equation}\label{10-1}
\ket{\psi(\alpha,\beta)}=(1-\beta^*\beta)^{\frac{1}{4}}
\exp\left(\frac{\beta}{2}{\hat b}^{*2}\right)\cdot
\exp\left(-\frac{1}{2}\alpha^*\alpha\right)\exp(\alpha{\hat a}^*)
\ket{0} \ .
\end{equation}
Here, ${\hat a}^*$ is a boson creation operator and 
$\ket{0}$ is a vacuum state for the boson annihilation 
operator ${\hat a}$ : ${\hat a}\ket{0}=0$. 
The operators ${\hat b}^*$ and ${\hat b}$ 
are defined as 
\begin{equation}\label{10-2}
{\hat b}^*={\hat a}^*-\alpha^* \ , \qquad
{\hat b}={\hat a}-\alpha \ .
\end{equation}
The operator ${\hat b}$ is nothing but the annihilation 
operator for the usual coherent state : 
${\hat b}
\exp\left(-\frac{1}{2}\alpha^*\alpha\right)\exp(\alpha{\hat a}^*)
\ket{0}=0$. 
Introducing the operators which correspond to the coordinate and 
momentum operators, we have another expression of the squeezed state 
(\ref{10-1}) :
\begin{eqnarray}
\ket{\psi(\alpha,\beta)}
&=&e^{i\varphi}
(2G)^{-1/4}\exp\left(\frac{i}{\hbar}(p{\hat Q}-q{\hat P})\right)
\exp\left\{\frac{1}{2\hbar}\left(
1-\frac{1}{2G}+i2\Pi\right){\hat Q}^2\right\}\ket{0} \ , 
\nonumber\\
& & \label{10-3}\\
& &{\hat Q}=\sqrt{\frac{\hbar}{2}}({\hat a}^*+{\hat a}) \ , \qquad
\qquad
{\hat P}=i\sqrt{\frac{\hbar}{2}}({\hat a}^*-{\hat a}) \ , 
\label{10-4}\\
& &{q}=\sqrt{\frac{\hbar}{2}}({\alpha}^*+{\alpha}) \ , \qquad
\qquad
{p}=i\sqrt{\frac{\hbar}{2}}({\alpha}^*-{\alpha}) \ , 
\nonumber\\
& &G=\left| \sqrt{\frac{1}{2}+|y|^2}+y \right|^2 \ , 
\qquad
\Pi=\frac{i}{2}(y^*-y)\sqrt{\frac{1}{2}+|y|^2}\ G^{-1} 
\ , \label{10-5}\\
& &e^{-i2\varphi}
=\frac{1}{\sqrt{G}}\left(\sqrt{\frac{1}{2}+|y|^2}+y\right)
\ , \nonumber
\end{eqnarray}
where $y$ is related to $\beta$ as 
$y=\beta/\sqrt{2(1-|\beta|^2)}$. The reason why we have introduced 
the new variables $y$ and $y^*$ instead of $\beta$ and $\beta^*$ 
is shown later.

The time-evolution of this quantum state is governed by 
the time-dependent variational principle :
\begin{equation}\label{10-6}
\delta\int_{t_0}^{t_1}dt \bra{\psi(\alpha,\beta)}
i\hbar\partial_t - {\hat H} \ket{\psi(\alpha,\beta)} = 0 \ .
\end{equation}
In the present method, one can deal with any Hamiltonian 
which is a function ${\hat Q}$ and ${\hat P}$. However, 
we restrict ourselves in this section on the Hamiltonian of the 
form 
\begin{equation}\label{10-7}
{\hat H}=\frac{1}{2}{\hat P}^2 + V({\hat Q}) \ .
\end{equation}
Further, the following canonicity conditions 
\begin{eqnarray}\label{10-8}
& &\bra{\psi(\alpha,\beta)}\partial_x\ket{\psi(\alpha,\beta)}
=\frac{1}{2}x^* \ , \qquad
\bra{\psi(\alpha,\beta)}\partial_{x^*}\ket{\psi(\alpha,\beta)}
=-\frac{1}{2}x \ , \nonumber\\
& &\bra{\psi(\alpha,\beta)}\partial_y\ket{\psi(\alpha,\beta)}
=\frac{1}{2}y^* \ , \qquad
\bra{\psi(\alpha,\beta)}\partial_{y^*}\ket{\psi(\alpha,\beta)}
=-\frac{1}{2}y \ 
\end{eqnarray}
give us the sets of canonical variables :
\begin{equation}\label{10-9}
x=\alpha \ , \quad x^*=\alpha^* \ ; \quad
y=\frac{\beta}{\sqrt{2(1-\beta^*\beta)}} \ , \quad
y^*=\frac{\beta^*}{\sqrt{2(1-\beta^*\beta)}} \ .
\end{equation}
This is the reason why we have already expressed $G$ and $\Pi$ 
in Eq.(\ref{10-5}) in terms of $y$ and $y^*$ instead of 
$\beta$ and $\beta^*$. 
Thus, the variational principle (\ref{10-6}) leads us 
to the following equations of motion : 
\begin{equation}\label{10-10}
{\dot q}=\partial_p {\cal H} \ , \quad
{\dot p}=-\partial_q {\cal H} \ , \quad
i\hbar{\dot y}=\partial_{y^*} {\cal H} \ , \quad
i\hbar{\dot y}^*=-\partial_{y} {\cal H} \ ,
\end{equation}
where ${\cal H}$ is the expectation value of the Hamiltonian 
with respect to the squeezed state. 
Also, the variables $G$ and $\Pi$ obey the similar canonical 
equations of motion instead of the variables $y$ and $y^*$ :
$\hbar {\dot G}=\partial_{\Pi}{\cal H}$ and $\hbar {\dot \Pi}
=-\partial_G{\cal H}$. 
Here, we summarize the 
expectation values for the various operators :
\begin{eqnarray}
& &
\bra{\psi(\alpha,\beta)}{\hat Q}\ket{\psi(\alpha,\beta)}
=q \ , \qquad\qquad\quad
\bra{\psi(\alpha,\beta)}{\hat P}\ket{\psi(\alpha,\beta)}
=p \ , \nonumber\\
& &\bra{\psi(\alpha,\beta)}{\hat Q}^2\ket{\psi(\alpha,\beta)}
=q^2+\hbar \xi \ , \qquad
\bra{\psi(\alpha,\beta)}{\hat P}^2\ket{\psi(\alpha,\beta)}
=p^2+\hbar \eta \ , \quad \ \ \ \ \label{10-11}\\
& &\bra{\psi(\alpha,\beta)}{\hat H}\ket{\psi(\alpha,\beta)}
=\frac{1}{2}p^2+\frac{1}{2}\hbar\eta
+\exp\left\{\frac{1}{2}\hbar\xi\left(\frac{\partial}{\partial q}\right)^2
\right\} V(q)\ , \label{10-12}
\end{eqnarray}
where the real quantities $\xi\ (>0)$ and $\eta\ (>0)$ are 
defined as 
\begin{equation}\label{10-13}
\xi=\left|\sqrt{\frac{1}{2}+y^*y}+y\right|^2
=G \ , \qquad
\eta=\left|\sqrt{\frac{1}{2}+y^*y}-y\right|^2
=\frac{1}{4G}+G\Pi^2 \ .
\end{equation}
The squares of the standard deviations for ${\hat Q}$ and ${\hat P}$ 
are expressed as 
$\bra{\psi(\alpha,\beta)}({\hat Q}-q)^2\ket{\psi(\alpha,\beta)}
=\hbar\xi$ and 
$\bra{\psi(\alpha,\beta)}({\hat P}-p)^2\ket{\psi(\alpha,\beta)}
=\hbar\eta$. 
Thus, the uncertainty
relation is expressed in terms of $\xi$ and $\eta$ 
as 
$\bra{\psi(\alpha,\beta)}({\hat Q}-q)^2\ket{\psi(\alpha,\beta)}
\bra{\psi(\alpha,\beta)}({\hat P}-p)^2\ket{\psi(\alpha,\beta)}
=\hbar^2\xi\eta=\hbar^2(1/4+G^2\Pi^2)$. 
It can be seen from this uncertainty relation that, 
if one direction of the uncertainty is relaxed, 
the other can be squeezed. 
Also, one can see from (\ref{10-12}) that the quantum effects beyond 
the order of $\hbar$ are included in this formalism. 
This novel feature is originated from the degree of freedom of 
the squeezing.

We have to give the initial conditions for solving the 
Hamilton equations of motion derived in Eq. (\ref{10-10}). 
Among four initial values 
for $q_0$, $p_0$, $y_0$ and $y_0^*$ 
which lead to a unique set of solutions of Eq.(\ref{10-10}), 
those for $q_0$ and $p_0$ can be selected arbitrary 
as is usually done in the ordinary TDHF 
theory. Thus, our task is to determine the initial conditions 
for $y$ and $y^*$. Since we are interested in classical motions in 
quantal systems, it may be natural to require that, at least, 
the initial state has the least quantal effects. 
Thus, the initial state satisfies the minimal uncertainty 
relation. Also, this means that the absolute value of 
the energy due to quantum effect at the very beginning 
is as small as possible. These two conditions give us 
the initial values of $y_0$ and $y_0^*$ :
\begin{equation}\label{10-14}
\xi_0\eta_0=\left|\frac{1}{2}+y_0^*y_0-y_0^2\right|^2=\frac{1}{4}
\ , \qquad
|{\cal H}(q_0,p_0,y_0,y_0^*)-{\cal H}_c(q_0,p_0)| : {\rm minimal}
\ .
\end{equation}
Here, each variable with suffix 0 denotes the initial value 
of each dynamical variable, and ${\cal H}_c(q,p)$ denotes 
the classical value of ${\cal H}$ defined as 
\begin{equation}\label{10-15}
H_c(q,p)=\frac{1}{2}p^2+V(q) \ .
\end{equation}
The first condition in Eq.(\ref{10-14}) gives us $y_0^*=y_0\ (=\rho)$, 
which is identical with $\Pi_0=0$. 
The second condition in Eq. (\ref{10-14}) together with $y_0^*=y_0$ 
is reduced to 
\begin{equation}\label{10-16}
\left|\frac{1}{8G_0}+\frac{1}{\hbar}\left[\exp\left\{
\frac{1}{2}\hbar G_0\left(\frac{\partial}{\partial q_0}\right)^2
\right\}-1\right]V(q_0)\right| \ : \ {\rm minimal}.
\end{equation}
Here, $G_0$ has already been defined in the relation (\ref{10-5}).

In the order of $\hbar^0$ and $\hbar^1$, the condition 
(\ref{10-16}) is further reduced to 
\begin{equation}\label{10-17}
\left|\frac{1}{G_0}+4vG_0+\hbar w G_0^2 \right| \ : \ {\rm minimal}
\ , \qquad
v=\frac{\partial^2}{\partial q_0^2}V(q_0) \ , \quad
w=\frac{\partial^4}{\partial q_0^4}V(q_0) \ .
\end{equation}
Approximate solutions of Eq. (\ref{10-17}) and the corresponding 
energies are given as :
\begin{eqnarray}
& &{\rm (i)}\ v>0 ; \ \qquad\qquad\quad
G_0=\frac{1}{2\sqrt{v}}-\frac{\hbar w}{16v^2} \ , \quad
\ \ \ 
E=E_{c,0}+\frac{\hbar}{2}\sqrt{v}+\frac{\hbar^2}{32}\frac{w}{v} 
 \ , \label{10-18}\\
& &{\rm (ii)}\ v<0 ; \qquad\qquad\quad
G_0=\frac{1}{2\sqrt{-v}}+\frac{\hbar w}{32v^2} \ , \quad
\ E=E_{c,0} \ , \label{10-19}\\
& &{\rm (iii)}\ v=0\ , w>0 ; \ \quad
G_0=\hbar^{-1/3}(2w)^{-1/3} \ , \quad
\ E=E_{c,0}+\hbar^{4/3}\frac{3}{16}(2w)^{1/3} \ , \qquad
\ \ \ \ \ \label{10-20}\\
& &{\rm (iv)}\ v=0\ , w<0 ; \ \quad
G_0=\hbar^{-1/3}(-w)^{-1/3} \ , \quad
E=E_{c,0} \ . \label{10-21}
\end{eqnarray}
Here, the quantity $E_{c,0}$ is defined by 
$E_{c,0}={\cal H}_c(q_0,p_0)$. 
The quantity $\rho\ (=y_0^*=y_0)$ is 
obtained from $G_0$ with the use of the relation 
$G_0=(\sqrt{(1/2)+\rho^2}+\rho)^2$.

For various potentials, let us investigate the energy expectation 
values under the above initial conditions : \\
(1) The harmonic oscillator potential $\omega^2{\hat Q}^2/2$ : 
In this case, $v=\omega^2$ and $w=0$ are obtained. 
We then have $\rho=(1/2\sqrt{2})(1/\sqrt{\omega}-\sqrt{\omega})$. 
Thus, the total energy $E$ is expressed as 
\begin{equation}\label{10-22}
E=\frac{1}{2}p_0^2+\frac{1}{2}\omega^2 q_0^2 +\frac{1}{2}\hbar\omega \ .
\end{equation}

\noindent
(2) The Morse potential 
$W_0[\exp(-2\mu{\hat Q})-2\exp(-\mu{\hat Q})]$ with 
$W_0>\hbar^2\mu^2/8$ : 
In this case, $v=2\mu^2 W_0[2\exp(-2\mu q_0)-\exp(-\mu q_0)]$. 
We will show the results of the order $\hbar^0$ for $G_0$. 
If $q_0<\log 2/\mu$ (case(i)), the total energy $E$ is expressed as 
\begin{equation}\label{10-23}
E=\frac{1}{2}p_0^2+W_0[\exp(-2\mu q_0)-2\exp(-\mu q_0)]
+\hbar\mu\sqrt{\frac{W_0}{2}}
[2\exp(-2\mu q_0)-\exp(-\mu q_0)]^{\frac{1}{2}} \ .
\end{equation}
If $q_0=p_0=0$, we have the minimal classical energy
\begin{equation}\label{10-24}
E_0=-W_0+\hbar\mu\sqrt{\frac{W_0}{2}} \ .
\end{equation}
This energy should be compared with the exact eigenvalue 
of this quantum Hamiltonian which has the form 
\begin{equation}\label{10-25}
E_n^{\rm exact}=-W_0\left[1-\hbar\frac{\mu}{\sqrt{2W_0}}
\left(n+\frac{1}{2}\right)\right]^2 \ . \ \ \ (n=0,1,2,\cdots)
\end{equation}
If $n=0$, in the order of $\hbar^1$, the eigenvalue (\ref{10-25}) 
is reduced to the result (\ref{10-24}). This means that, 
even in the limit of $\hbar\rightarrow 0$, our 
formalism automatically contains the zero-point energy 
of order of $\hbar^1$. 

\noindent
(3) The upside-down harmonic oscillator potential 
$-\omega^2{\hat Q}^2/2$ : 
In this case, $v=-\omega^2$ and $w=0$ (case(ii)). 
Then, $G_0$ is equal to $1/(2\omega)$. 

\noindent
(4) The double-well potential 
$\lambda({\hat Q}^2-\mu^2)^2/24$ with $\lambda>0$ : 
In this case, the cases (i), (ii) and (iii) appear depending 
on the initial value $q_0$. 
If $q_0>\mu/\sqrt{3}$ or $<-\mu/\sqrt{3}$ (case(i)), 
$G_0$ and $E$ are given as 
\begin{eqnarray}\label{10-26}
& &G_0=\sqrt{\frac{3}{2\lambda(3q_0^2-\mu^2)}}-
\hbar\cdot\frac{9}{4\lambda(3q_0^2-\mu^2)^2} \ , \nonumber\\
& &E=\frac{1}{2}p_0^2+V(q_0)+\hbar\sqrt{\frac{\lambda}{24}
(3q_0^2-\mu^2)}+\hbar^2\cdot
\frac{3}{16(3q_0^2-\mu^2)} \ .
\end{eqnarray}
If $-\mu/\sqrt{3}<q_0<\mu/\sqrt{3}$ (case(ii)), 
$G_0$ is as follows : 
\begin{equation}\label{10-27}
G_0=\sqrt{\frac{3}{2\lambda(\mu^2-3q_0^2)}}+
\hbar\cdot\frac{9}{8\lambda(\mu^2-3q_0^2)^2} \ .
\end{equation}
If $q_0=\pm \mu/\sqrt{3}$, $G_0$ and $E$ are given as 
\begin{equation}\label{10-28}
G_0=\hbar^{-\frac{1}{3}}(2\lambda)^{-\frac{1}{3}} \ , \qquad
E=\frac{1}{2}p_0^2+V(q_0)+\hbar^{\frac{4}{3}}\frac{3}{16}
(2\lambda)^{\frac{1}{3}} \ .
\end{equation}
We can conclude that the conditions in Eq.(\ref{10-14}) 
are plausible for describing the classical motion in quantal 
system in our squeezed state approach. 
The time-evolution of the dynamical variables in the 
squeezed state, which 
governs the time-evolution of the quantum state, is 
presented in Ref.\citen{TF1} in the case of 
$V({\hat Q})=0$, $V({\hat Q})=\gamma {\hat Q}$ 
and the upside-down harmonic oscillator potential.

Before closing this section, we give two comments : 
First, our squeezed state formalism for quantal systems 
automatically contains the quantum effects beyond the order of 
$\hbar$. If we take the limit of $\hbar\rightarrow 0$, the 
WKB result is reproduced. Further, we can 
show that the Maslov 
phase which appears in the Bohr-Sommerfeld quantization rule 
has the geometrical property such as the Berry phase 
in the limit of $\hbar\rightarrow 0$ in our formalism. 
Of course, we can take into account of 
the higher order contributions of $\hbar$. 
The details can be found in Ref.\citen{T}.
Secondly, we point out that our squeezed state formalism 
can be easily extended to the case of the scalar field theory.
The details can be found in Ref.\citen{TF2}.

\section{Coherent and squeezed states in the su(2)-boson model}

We extend the squeezed state approach developed 
in \S 10 to the su(2)-algebraic model. The $su(2)$-algebraic 
models such as the Lipkin model and the pairing model are 
regarded as the schematic models which describe 
many-nucleon and/or fermion system. The $su(2)$-algebra 
can be expressed in terms of two kinds of boson operators, which 
is known as the Schwinger boson realization.\cite{Schw}
In this sense, we call the $su(2)$-model treated here the 
$su(2)$-boson model. 
Focussing on the least quantal effect, in this section, 
we give an analysis of the $su(2)$-boson model based on 
the coherent and squeezed state.\cite{YKT} 

We introduce two kinds of boson operators ${\hat a}$ and ${\hat b}$. 
The generators of the $su(2)$-algebra are expressed by these 
boson operators, which have already given in Eq. (\ref{2-1}). 
The Casimir operator ${\hat \Gamma}_{su(2)}$ has also been 
defined in Eq. (\ref{2-4}) with (\ref{2-5}). 
The magnitude of the $su(2)$-spin, $S$, as a mean value is 
given by 
\begin{equation}\label{11-1}
\bra{w}{\hat S}\ket{w} = S \ ,
\end{equation}
where $\ket{w}$ is a certain state. If $\ket{w}$ is an eigenstate 
for ${\hat S}$, then the expectation value of the Casimir 
operator is obtained as 
$\bra{w}{\hat \Gamma}_{su(2)}\ket{w}=S^2+\hbar S$. 
The second term may be regarded as the quantal effect because 
the mean value of ${\hat S}$ is $S$. In general, 
this relation is not satisfied. 

First, let us adopt a coherent state as $\ket{w}$. 
The investigation is expected to be a powerful help 
to the classical description of the $su(2)$-boson model.
As a preparation, we introduce the new sets of boson operators :
\begin{equation}\label{11-2}
{\hat c}=U{\hat a}-V{\hat b} \ , \qquad
{\hat d}=V^*{\hat a}+U{\hat b} \ .
\end{equation}
Here, $U$ is real and $V$ is complex and they obey the condition 
\begin{equation}\label{11-3}
U^2+V^*V=1 \ .
\end{equation}
Then, (${\hat c},{\hat c}^*$) and (${\hat d}, {\hat d}^*$) are 
independent boson annihilation and creation operators, respectively. 
Further, we introduce the boson operators 
(${\hat C}, {\hat C}^*)$ and $({\hat D}, {\hat D}^*)$ which are defined 
as
\begin{equation}\label{11-4}
{\hat C}={\hat c} \ , \qquad
{\hat D}={\hat d}-\exp(-i\theta/2)\sqrt{\frac{2\sigma}{\hbar}} \ .
\end{equation}
The boson operator ${\hat D}$ expresses the displacement from 
the complex value $e^{-i\theta/2}\sqrt{2\sigma/\hbar}$.

The coherent state $\ket{c}$ has the following standard form :
\begin{eqnarray}\label{11-5}
& &\ket{c}=N_c\ \exp(e^{-i\chi/2}\cdot(\alpha{\hat a}^*
+\beta{\hat b}^*)) \ket{0} \ , \nonumber\\
& &N_c=\exp(-(|\alpha|^2+\beta^2)/2) \ . 
\end{eqnarray}
Here, $\alpha$ is complex and $\beta$ is real with the help of 
the introduction of the phase angle $\chi$. 
Let $\alpha$, $\beta$ and $\chi$ correspond to 
\begin{equation}\label{11-6}
\alpha=\sqrt{\frac{2\sigma}{\hbar}}\cdot V \ , \qquad
\beta=\sqrt{\frac{2\sigma}{\hbar}}\cdot U \ , \qquad
\chi=\theta \ .
\end{equation}
Here, $V$, $U$ and $\theta$ have been introduced in Eqs. (\ref{11-2}) 
and (\ref{11-4}). Then, the coherent state $\ket{c}$ is 
rewritten as 
\begin{eqnarray}\label{11-7}
& &\ket{c}=N_c\cdot \exp\left(e^{-i\theta/2}\cdot
\sqrt{\frac{2\sigma}{\hbar}}\cdot{\hat d}^*\right)\ket{0} \ , \nonumber\\
& &N_c=\exp(-\sigma/\hbar) \ .
\end{eqnarray}
The state $\ket{c}$ is also the coherent state of ${\hat c}$ 
and ${\hat d}$ with the eigenvalues 0 and $\exp(-i\theta/2)\cdot
\sqrt{2\sigma/\hbar}$, respectively. Therefore, 
${\ket{c}}$ is the vacuum for ${\hat C}$ and ${\hat D}$ : 
\begin{equation}\label{11-8}
{\hat c}\ket{c}=0 \ , \quad
{\hat d}\ket{c}=\exp(-i\theta/2)\cdot\sqrt{\frac{2\sigma}{\hbar}} 
\  ; \qquad
{\hat C}\ket{c}={\hat D}\ket{c}=0 \ .
\end{equation}
It is noted that the coherent state $\ket{c}$ is specified 
by four variables $\theta$, $\sigma$, $V$ and $V^*$. 
Let us investigate the quantal fluctuations 
in the coherent state $\ket{c}$. 
The squares of standard deviations for the quasi-spin operators 
are calculated as follows : 
\begin{equation}\label{11-9}
(\Delta S_x^2)_c=(\Delta S_y^2)_c=(\Delta S_z^2)_c=\frac{\hbar S}{2} 
\ ,
\end{equation}
where $(\Delta S_x^2)_c=\bra{c}{\hat S}_x^2\ket{c}-\bra{c}{\hat S}_x
\ket{c}^2$ etc., and 
${\hat S}_x=({\hat S}_++{\hat S}_-)/2$, 
${\hat S}_y=({\hat S}_+-{\hat S}_-)/(2i)$ and ${\hat S}_z={\hat S}_0$. 
It should be noted that the standard deviations are fixed in the 
coherent state. 
Further, it is interesting to investigate the quantal 
fluctuations by means of the expectation value of the 
Casimir operator ${\hat \Gamma}_{su(2)}$. 
We can calculate it as 
\begin{eqnarray}\label{11-10}
\bra{c}{\hat \Gamma}_{su(2)}\ket{c}
&=&\bra{c}{\hat S}^2+\hbar{\hat S}\ket{c} \nonumber\\
&=&S^2+\hbar S + \frac{\hbar S}{2}\cdot(1-W_c) \ ,
\end{eqnarray}
where $W_c$ is related to the standard deviation for 
${\hat S}$, i.e., 
$(\Delta S^2)_c
=\bra{c}{\hat S}^2\ket{c}-\bra{c}{\hat S}\ket{c}^2$ : 
\begin{equation}\label{11-11}
W_c=1-\frac{2(\Delta S^2)_c}{\hbar S} \ .
\end{equation}
Here, for the coherent state $\ket{c}$, we obtain $W_c=0$. 
In other words, $(\Delta S^2)_c$ is equal to $\hbar S/2$. 
Namely, it is of the same order as that of $\hbar S$, 
which is the exact quantal fluctuation. 
From the above fact, we have to get the conclusion that 
the coherent state cannot give a correct quantal fluctuation 
with respect to the Casimir operator ${\hat \Gamma}_{su(2)}$. 
The fictitious condition $W_c=1$ gives us the exact relation for 
the Casimir operator. Therefore, it is inevitable to 
look for the state $\ket{w}$ which makes the order of 
$(\Delta S^2)_w$ less than that of $\hbar S$.

In order to overcome the above-mentioned shortcoming of the 
coherent state, we extend the coherent state $\ket{c}$ given in
Eq.(\ref{11-5}) to the squeezed state $\ket{s}$ in the 
following form : 
\begin{eqnarray}
& &\ket{s}=N_s \exp(u{\hat C}^{*2}+v{\hat D}^{*2})
\cdot \exp\left(e^{-i\theta/2}\sqrt{\frac{2\sigma}{\hbar}}
\cdot{\hat d}^*\right)\ket{0} \ , 
\label{11-12}\\
& &u=\frac{\xi e^{-i\theta}}{\sqrt{2(1+2\xi^*\xi)}} \ , \qquad
v=\frac{\eta e^{-i\theta}}{\sqrt{2(1+2\eta^*\eta)}} \ , 
\label{11-13}\\
& &N_s=[(1+2\xi^*\xi)(1+2\eta^*\eta)]^{-1/4}\cdot \exp\left(
-\frac{\sigma}{\hbar}\right) \ .\label{11-14}
\end{eqnarray}
It should be noted that the term related to the quadratic 
for the bosons, which is characteristic to the squeezed state, is added. 
The state $\ket{s}$ contains new complex variables 
$(\xi^*, \xi)$ and $(\eta^*, \eta)$. Under the condition 
(\ref{11-1}) for $\ket{w}=\ket{s}$, $\sigma$ can be related to $S$ 
in the following form : 
\begin{equation}\label{11-15}
S=\bra{s}{\hat S}\ket{s}=\sigma+\hbar(\xi^*\xi+\eta^*\eta) \ .
\end{equation}
First, let us investigate the square of the standard deviation 
$(\Delta S^2)_s$. The quantity $(\Delta S^2)_s$ can be 
calculated as 
\begin{eqnarray}\label{11-16}
(\Delta S^2)_s
&=& \bra{s}{\hat S}^2\ket{s}-\bra{s}{\hat S}\ket{s}^2 \nonumber\\
&=& \frac{\hbar}{2}\cdot(S-\hbar\xi^*\xi)
[1+\sqrt{2}(\eta^*+\eta)\sqrt{1+2\eta^*\eta}+4\eta^*\eta] 
\nonumber\\
& &+\hbar^2\cdot
\left(2(\xi^*\xi)^2+\xi^*\xi+\frac{\eta^*\eta}{2}-
\frac{1}{\sqrt{2}}(\eta^*+\eta)\eta^*\eta\sqrt{1+2\eta^*\eta}
\right) \ .\qquad\quad
\end{eqnarray}
The differential of $(\Delta S^2)_s$ for $\eta$ and $\eta^*$ 
leads us to the following relations for the minimum value 
of $(\Delta S^2)_s$, while we cannot make $(\Delta S^2)_s$ 
vanish any choice of $\eta$ and $\eta^*$ for given $S$ and 
$\xi^*\xi$ : 
\begin{equation}\label{11-17}
\eta=\eta^*=-\frac{1}{2\sqrt{2}}(\sqrt{\zeta}
-\frac{1}{\sqrt{\zeta}}) \ ,
\end{equation}
where $\zeta$ satisfies the following equation :
\begin{equation}\label{11-18}
\zeta^3-2+\frac{1}{\zeta}=8\left(\frac{S}{\hbar}-\xi^*\xi\right) \ .
\end{equation}
Here, Eq.(\ref{11-18}) has a real solution if 
$S/\hbar+(1-4\xi^*\xi)/4 \ge 1/2\sqrt[4]{27}$. 
Substituting the solutions (\ref{11-17}) into 
the relation (\ref{11-16}), 
the deviation $(\Delta S^2)_s$ is given as 
\begin{equation}\label{11-19}
\frac{(\Delta S^2)_s}{\hbar S}
=\frac{3}{8}\sqrt[3]{\frac{\hbar}{S}}\cdot
\biggl[1+\frac{8}{3}\sqrt[3]{\left(\frac{\hbar}{S}\right)^2}
\left(-\frac{1}{8}+\xi^*\xi+2(\xi^*\xi)^2\right) 
%\nonumber\\
%& &\qquad\qquad
+\frac{1}{6}\cdot\frac{\hbar}{S}\cdot(1-4\xi^*\xi)+\cdots \biggl] 
\ , 
\end{equation}
where $S$ and $\xi^*\xi$ are regarded as the same order. 
Thus, up to the order of $\hbar^1$, $(\Delta S^2)_s$ can 
be neglected : $(\Delta S^2)_s=0$. From this result, 
we have 
\begin{equation}\label{11-20}
\bra{s}{\hat \Gamma}_{su(2)}\ket{s} = S^2+\hbar S \ .
\end{equation}
This is the desired result. As was discussed previously, 
the condition $W_s=1$ is realized, while $W_c=0$ in the coherent 
state. 
Thus, the squeezed state $\ket{s}$ 
gives us an appropriate quantal fluctuation for the Casimir 
operator in contrast to the coherent state $\ket{c}$. 

Before investigating the deviations 
$(\Delta S_x^2)_s$ etc., let us parametrize the state $\ket{s}$ 
in terms of the canonical variables. We have new variables 
$\xi^*$ and $\xi$ adding to the four variables characterizing 
the coherent state $\ket{c}$. We impose the following 
canonicity conditions :
\begin{eqnarray}
& &\bra{s}i\hbar \partial_{\theta} \ket{s}=S \ , 
\qquad
\bra{s}i\hbar \partial_S \ket{s}=0 \ , \label{11-21}\\
& &\bra{s}\partial_{X} \ket{s}= \frac{X^*}{2} \ , \qquad
\bra{s}\partial_{X^*} \ket{s}= -\frac{X}{2} \ , \label{11-22}\\
& &\bra{s}\partial_{x} \ket{s}= \frac{x^*}{2} \ , \qquad
\bra{s}\partial_{x^*} \ket{s}= -\frac{x}{2} \ , \label{11-23}
\end{eqnarray}
where $S$ is nothing but the expectation value of ${\hat S}$ 
and $\theta$ corresponds to its conjugate angle variable. 
A possible solutions of Eqs. (\ref{11-22}) and (\ref{11-23}) 
are obtained as 
\begin{equation}\label{11-24}
V=\sqrt{\frac{\hbar}{2(S-2\hbar x^* x)}}\cdot X \ , \quad
U=\sqrt{1-\frac{\hbar X^* X}{2(S-2\hbar x^*x)}} \ ; \quad
\xi=x \ , \quad \xi^*=x^* \ .
\end{equation}
Then, we can express 
the expectation values of the generators of the $su(2)$-algebra 
in terms of these canonical variables :
\begin{eqnarray}\label{11-25}
& &(S_x)_s=\bra{s}{\hat S}_x \ket{s} 
=
Q \sqrt{S-(P^2+Q^2)/4-2\hbar x^* x} \ , \nonumber\\
& &(S_y)_s=\bra{s}{\hat S}_y \ket{s} 
=
-P\sqrt{S-(P^2+Q^2)/4-2\hbar x^* x} \ , \nonumber\\
& &(S_z)_s=\bra{s}{\hat S}_z \ket{s} 
=
-S+(P^2+Q^2)/2+2\hbar x^* x \ ,
\end{eqnarray}
where $Q=\sqrt{\hbar/2}\cdot(X^*+X)$ and 
$P=i\sqrt{\hbar/2}\cdot (X^*-X)$. 
The expression (\ref{11-25}) contains the terms related to the 
quantal fluctuations, i.e., $\hbar x^* x$. The $su(2)$-generators 
are expressed in terms of the quadratic forms 
for the boson operators. In contrast to the case of the coherent 
state, the forms are not of the normal ordered product for 
the squeezed state $\ket{s}$. From this reason, 
the expressions contain the terms of the quantal fluctuations. 

Now we are ready to give the deviations 
$(\Delta S_x^2)_s$ etc., which are expressed up to the order 
of $\hbar^1$ in terms of the canonical variables : 
\begin{eqnarray}\label{11-26}
(\Delta S_x^2)_s &=& 
\frac{\hbar}{8S}\cdot [
4S^2-Q^2(4S-(P^2+Q^2))W_s] \nonumber\\ 
& &+\frac{\hbar}{4S}\cdot [4S^2-Q^2(4S+(P^2-Q^2))]
\cdot (x^*+x)\sqrt{\frac{1}{2}+x^* x} \nonumber\\
& &-\frac{\hbar}{2S}\cdot QP(2S-Q^2)\cdot i(x^*-x)\sqrt{\frac{1}{2}
+x^* x}
\nonumber\\
& &
+\frac{\hbar}{2S}\cdot[4S^2-Q^2(4S-(P^2+Q^2))]\cdot x^* x \ , 
\nonumber\\
(\Delta S_y^2)_s &=& 
\frac{\hbar}{8S}\cdot [
4S^2-P^2(4S-(P^2+Q^2))W_s] \nonumber\\ 
& &-\frac{\hbar}{4S}\cdot [4S^2-P^2(4S-(P^2-Q^2))]
\cdot (x^*+x)\sqrt{\frac{1}{2}+x^* x} \nonumber\\
& &-\frac{\hbar}{2S}\cdot QP(2S-P^2)\cdot i(x^*-x)\sqrt{\frac{1}{2}
+x^* x}\nonumber\\
& &+\frac{\hbar}{2S}\cdot[4S^2-P^2(4S-(P^2+Q^2))]\cdot x^* x \ , 
\nonumber\\
(\Delta S_z^2)_s &=& 
\frac{\hbar}{8S}\cdot [
4S^2-(2S-(P^2+Q^2))^2W_s] \nonumber\\ 
& &-\frac{\hbar}{4S}\cdot [(P^2-Q^2)(4S-(P^2+Q^2))]
\cdot (x^*+x)\sqrt{\frac{1}{2}+x^* x} \nonumber\\
& &+\frac{\hbar}{2S}\cdot QP(4S-(P^2+Q^2))\cdot i(x^*-x)\sqrt{\frac{1}{2}
+x^* x}\nonumber\\
& &+\frac{\hbar}{2S}\cdot(P^2+Q^2)[4S-(P^2+Q^2)]\cdot x^* x \ .
\end{eqnarray}
Here, $W_s$ is defined by 
\begin{equation}\label{11-27}
W_s=1-\frac{2(\Delta S^2)_s}{\hbar S} = 1 \ ,
\end{equation}
where the last equality is valid for the approximation 
up to the order of $\hbar^1$ as was seen in Eq.(\ref{11-19}).
In contrast to the case of the coherent state, the above uncertainties 
do not have fixed value $\hbar S/2$ and $W_s$ is equal to unity 
even if the variables describing 
the quantal effects, $x^*$ and $x$, are equal to zero. This fact 
comes from the inclusion of the parameters $\eta$ and $\eta^*$.

As an example, we sketch our basic idea of a possible 
application to an $su(2)$-model, that is, the Lipkin model 
whose Hamiltonian is given as 
\begin{equation}\label{11-28}
{\hat H}=2\epsilon{\hat S}_z-G({\hat S}_x^2-{\hat S}_y^2) \ .
\end{equation}
Here, $\epsilon$ and $G$ denote positive constant. 
Main interest is to investigate the time-evolution of this quantal 
system with the least quantal effects. The expectation value of 
this Hamiltonian with respect to the squeezed state $\ket{s}$ is
expressed as 
\begin{eqnarray}
& &\bra{s}{\hat H}\ket{s}+2\epsilon S = H_s = H_{s,{\rm cl}}
+H_{s,{\rm ql}} \ ,\label{11-29}\\
& &H_{s,{\rm cl}}=\epsilon(P^2+Q^2)+GS(P^2-Q^2)
-\frac{G}{4}\cdot(P^4-Q^4) \ , \label{11-30}\\
& &H_{s,{\rm ql}}=\hbar\biggl[
4\epsilon x^*x-\frac{G}{8S}\cdot(P^2-Q^2)(4S-(P^2+Q^2)) \nonumber\\
& &\qquad\qquad
-\frac{G}{4S}\cdot [8S^2-4S(P^2+Q^2)+(P^2-Q^2)^2]
(x^*+x)\sqrt{\frac{1}{2}+x^*x} \nonumber\\
& &\qquad\qquad
+\frac{G}{2S}\cdot QP(P^2-Q^2)
i(x^*-x)\sqrt{\frac{1}{2}+x^*x} 
\nonumber\\
& &\qquad\qquad
-\frac{G}{2S}\cdot(P^2-Q^2)[8S-(P^2+Q^2)]x^*x \biggl] \ . 
\label{11-31}
\end{eqnarray}
The part $H_{s,{\rm cl}}$ ($H_{s,{\rm ql}}$) 
denotes the classical (quantal) part of $H_s$. 
It is seen from (\ref{11-29}) $\sim$ (\ref{11-31}) that 
the $c$-number Hamiltonian is expressed in terms of the 
canonical variables $(\theta, S)$, $(Q, P)$ and $(x^*, x)$. 
Then, the Hamilton equations of motion are written as 
\begin{eqnarray}
& & {\dot \theta}=\partial_S H_s \ , \qquad
{\dot S}=-\partial_{\theta} H_s=0 \ , \label{11-32}\\
& & {\dot Q}=\partial_P H_s \ , \qquad
{\dot P}=-\partial_{Q} H_s \ , \label{11-33}\\
& & i\hbar{\dot x}=\partial_{x^*} H_s \ , \qquad
i\hbar{\dot x}^*=-\partial_{x} H_s \ . \label{11-34}
\end{eqnarray}
The second equation of Eq.(\ref{11-32}) 
shows us that $S$ is a constant of motion. 
By solving Eqs.(\ref{11-33}) and (\ref{11-34}) with appropriate initial 
conditions, the time-dependence of $Q$, $P$, $x$ and $x^*$ can 
be obtained. 
Here, let us analyze the Lipkin model under the adiabatic 
approximation like the adiabatic time-dependent Hartree-Fock (ATDHF) 
theory. In the ATDHF theory, the classical Hamiltonian is 
approximately expressed 
in terms of the quadratic form for the momentum. 
Following the same viewpoint as that of the ATDHF theory, 
$H_{s,{\rm cl}}$ can be expressed in terms of the variables 
$Q$ and $P$ as follows :
\begin{eqnarray}\label{11-35}
& &H_{s,{\rm cl}}=\frac{P^2}{2M}+V(Q) \ , \nonumber\\
& &M=\frac{1}{2(\epsilon+GS)} \ , \qquad
V(Q)=(\epsilon-GS)\cdot Q^2+\frac{G}{4}\cdot Q^4 \ .
\end{eqnarray}
Further, we approximate $H_{s,{\rm ql}}$ in the frame of the term with 
$P^0$ : 
\begin{eqnarray}\label{11-36}
H_{s,{\rm ql}}&=&
\hbar\biggl( \frac{G}{8S}\cdot Q^2(4S-Q^2)
+\left(4\epsilon+\frac{G}{2S}
\cdot Q^2(8S-Q^2)\right) x^* x \nonumber\\
& &\qquad
-\frac{G}{4S}\cdot (8S^2-4SQ^2+Q^4)(x^*+x)\sqrt{\frac{1}{2}
+x^* x} \biggl) \ .
\end{eqnarray}
In the same way as that in $H_{s,{\rm ql}}$, we have the following 
approximate uncertainty relations : 
\begin{eqnarray}\label{11-37}
(D_x)_s&=&\frac{4}{\hbar^2}\left[
(\Delta S_y^2)_s\cdot(\Delta S_z^2)_s - \frac{\hbar^2}{4}
\cdot (S_x)_{s, {\rm cl}}^2\right] 
=Q^2(4S-Q^2)[i(x^*-x)]^2\left(\frac{1}{2}+x^* x\right) \ , \nonumber\\
(D_y)_s&=&\frac{4}{\hbar^2}\left[
(\Delta S_z^2)_s\cdot(\Delta S_x^2)_s - \frac{\hbar^2}{4}
\cdot (S_y)_{s, {\rm cl}}^2\right] \nonumber\\
&=&\frac{1}{4S^2}\cdot Q^2(4S-Q^2)(2S-Q^2)^2 
\left(x^*+\sqrt{\frac{1}{2}+x^* x}\right)^2
\left(x+\sqrt{\frac{1}{2}+x^* x}\right)^2
\ , \nonumber\\
(D_z)_s&=&\frac{4}{\hbar^2}\left[
(\Delta S_x^2)_s\cdot(\Delta S_y^2)_s - \frac{\hbar^2}{4}
\cdot (S_z)_{s, {\rm cl}}^2\right] 
=(2S-Q^2)^2[i(x^*-x)]^2\left(\frac{1}{2}+x^* x\right) \ .
\nonumber\\
\end{eqnarray}
Here, $(S_x)_{s,{\rm cl}}=Q\sqrt{S-(P^2+Q^2)/4}$, 
$(S_y)_{s,{\rm cl}}=-P\sqrt{S-(P^2+Q^2)/4}$ and 
$(S_z)_{s,{\rm cl}}=-S+(P^2+Q^2)/2$. 
It is noted that the variable $Q$ should run in the region $Q^2 \le 4S$. 
Our task is to give the initial conditions 
in order to solve the Hamilton equations of motion (\ref{11-33}) 
and (\ref{11-34}). 
As for the initial conditions for $Q$ and $P$, they are 
taken arbitrary. 
It is necessary to give the initial conditions for $x^*$ and $x$. 
The basic idea has been already presented in the previous section, 
that is, the least quantal effects should be realized at the initial 
time. The idea for investigating the least quantal effects in the classical 
motion of quantized system is given in Ref. \citen{YK2}. 
Translating the basic idea into the present case, the following criteria 
to give the initial conditions 
should be taken into account for the squeezed state :
\begin{eqnarray}
& &|H_{s,{\rm ql}}^0| \ : \ {\rm minimal} \ , \label{11-38}\\
& &{\rm any\ two\ of\ }(D_x)_s^0 \ , \ (D_y)_s^0 \ {\rm and} \ 
(D_z)_s^0 = 0 \ ,\label{11-39}
\end{eqnarray}
where the superscript 0 denotes the initial value for each 
quantity. As can be seen in Eq.(\ref{11-37}), the criterion 
(\ref{11-39}) is satisfied for the case 
$(D_x)_s^0=(D_z)_s^0=0$ which gives us 
$x_0=x_0^*$, that is, $x_0$ is real. 
For convenience, let us introduce the quantities $g$, $z$ 
and $y$ instead of $GS/\epsilon$, $Q_0$ and $x_0$ :
\begin{equation}\label{11-40}
\frac{GS}{\epsilon}=g \ , \quad
\frac{Q_0^2}{4S}=z \ , \quad
x_0=\frac{1}{2\sqrt{2}}\cdot\left(\sqrt{y}-\frac{1}{\sqrt{y}}\right) 
\ .
\end{equation}
Then, the quantities $g$ and $z$ run from 0 to $\infty$ and 
from 0 to 1, respectively. 
From the criterion (\ref{11-38}), the initial condition 
is obtained together with the initial value of the energy 
of the quantal part : 

\noindent
(1) in the regions $0\le z \le z_1$ and $z_2 \le z \le 1$,  
\begin{eqnarray}
& &y=\frac{g\cdot 2z+1-\sqrt{g^2\cdot (8z^3-4z^2-4z+1)+g\cdot 4z(z-1)}}
{-g\cdot(4z^2-6z+1)+1} \ , \label{11-41}\\
& &H_{s,{\rm ql}}^0=0 \ , \label{11-42}
\end{eqnarray}
(2) in the region $z_1 \le z \le z_2$, 
\begin{eqnarray}
& &y=\sqrt{\frac{g\cdot (2z+1)+1}
{-g\cdot(4z^2-6z+1)+1}} \ , \label{11-43}\\
& &H_{s,{\rm ql}}^0=\hbar \epsilon
[\sqrt{[g\cdot(2z+1)+1)][-g\cdot(4z^2-6z+1)+1]}-(g\cdot 2z+1) ] \ . 
\qquad \ \ 
\label{11-44}
\end{eqnarray}
Here, $z_1$ and $z_2$ are given as 
\begin{eqnarray}\label{11-45}
& &z_1=\frac{1}{6g}\cdot\left(
g-1+2\sqrt{7g^2+4g+1}\ \cos(\phi/3+\pi/3) \right) \ , \nonumber\\
& &z_2=\frac{1}{6g}\cdot\left(
g-1+2\sqrt{7g^2+4g+1}\ \cos(\phi/3-\pi/3) \right) \ , \nonumber\\
& &\ \ 
\cos \phi =\frac{7g^3+6g^2+12g+2}{2(\sqrt{7g^2+4g+1})^3} \ .
\end{eqnarray}
The details are found in Ref.\citen{YKT}. 
It may be interesting to calculate $(D_y)_s^0$ which 
is given by $(D_y)_s^0=S^2\cdot f(z)$ where 
$f(z)=4z(1-2z)^2(1-z)y^2$. 
It can be shown that the factor $f(z)$ is rather small 
with some exceptions. The minimal uncertainty for the 
quantity $(D_y)_s^0$ is satisfied at rather high accuracy. 
The detailed numerical estimation is given in Ref.\citen{YKT}.

\section{Quasi-spin squeezed state approach to many-fermion systems}

A possible classical description of many-fermion systems such as nucleus 
is given by the TDHF theory \cite{YK87}. 
In this theory, the Slater determinant is usually 
adopted as a trial state for the variational procedure. 
The Slater determinant is essentially a coherent state which 
yields a minimal uncertainty relation where each uncertainty 
is fixed. 
Therefore, in order to take into account of appropriate quantal effects 
in many-fermion system in terms of classical mechanics, 
we need to proceed beyond the usual coherent state approach. 

In our formalism developed in the preceding two sections, 
the boson squeezed state, which is introduced 
as a trial state in the time-dependent variational approach 
in the quantum mechanical or many-boson systems, has been originally 
constructed as an extension of the familiar boson coherent state. 
In many-fermion systems, our first task is to define an appropriate 
trial state similar to the boson squeezed state.  

We formulate our idea for the $su(2)$-algebraic models, namely, 
a single-level model with pairing interaction and a 
two-level model with particle-hole 
interaction. In these models the trial state of the time-dependent 
Hartree-Fock or Hartree-Fock-Bogoliubov theory is 
reduced to the $su(2)$-coherent state. 
With the aim of going beyond this theory, 
we consider a possible extension of the $su(2)$-coherent state. 
This new state, which we call ``a quasi-spin squeezed state," 
is used as a trial state in the variational method, and 
we show in this section 
that the quasi-spin squeezed state approach is workable and 
successful.

First, we investigate a simple system: $N$ identical nucleons are 
moving in a spherical orbit with the pairing interaction. 
The single-particle state is specified by a set of quantum numbers 
$(j, m)$, where $j$ and $m$ denote 
the magnitude of angular momentum of the single-particle state 
(the half-integer) and its projection to the $z$-axis, respectively.
The annihilation and the creation operators for the nucleon 
with $m$, ${\hat c}_m$ and ${\hat c}_m^*$, satisfy the 
usual anti-commutation relations : 
$\{ {\hat c}_m , {\hat c}_{m'}^* \} = \delta_{mm'}$ 
and the other combinations are equal to 0. 
The Hamiltonian is written as 
\begin{eqnarray}\label{12-1}
{\hat H}&=&\sum_{m=-j}^{j} \hbar\epsilon {\hat c}_m^*{\hat c}_m 
-\frac{1}{4}\hbar^2 G \sum_{m=-j}^{j}(-)^{j-m}
{\hat c}_m^*{\hat c}_{-m}^* \sum_{m'=-j}^{j}(-)^{j-m'}
{\hat c}_{-m'}{\hat c}_{m'} \nonumber\\
&=&2\epsilon ({\hat S}_0+S_j)-G{\hat S}_+ {\hat S}_- \ ,
\end{eqnarray}
where $\hbar\epsilon$ and $\hbar^2 G$ denote the single-particle 
energy and the force strength, respectively. Also, $S_j=\hbar\Omega/2$ 
where $\Omega$ is the half of the degeneracy : 
$\Omega=j+1/2$. 
This model has been used in \S 4 as an example of the 
$su(2)$-algebraic models.
Here, we have defined the quasi-spin operators 
$({\hat S}_0 , {\hat S}_{\pm})$ as 
\begin{eqnarray}\label{12-2}
& &{\hat S}_+=\frac{\hbar}{2}\sum_m (-)^{j-m}{\hat c}_m^*{\hat c}_{-m}^*
\ , \qquad
{\hat S}_-=\frac{\hbar}{2}\sum_m (-)^{j-m}{\hat c}_{-m}{\hat c}_{m}
\ , \nonumber\\
& &{\hat S}_0=\frac{\hbar}{2}
(\sum_m {\hat c}_m^*{\hat c}_m-\Omega)
=\frac{\hbar}{2}({\hat N}-\Omega) \ , 
\end{eqnarray}
where ${\hat N}$ is a number operator defined by 
${\hat N}=\sum_m{\hat c}_m^*{\hat c}_m$. The set (\ref{12-2}) 
is governed by the $su(2)$-algebra :
\begin{equation}\label{12-3}
[ {\hat S}_+ , {\hat S}_- ]=2\hbar{\hat S}_0 \ , 
\qquad
[ {\hat S}_0 , {\hat S}_{\pm} ] = \pm \hbar{\hat S}_{\pm} \ .
\end{equation}

As is well-known, this model can be solved exactly. The exact 
eigenvalue $H_{\rm ql}$ in the ground state 
for this Hamiltonian is given by 
\begin{eqnarray}\label{12-4}
H_{\rm ql}=\bra{q}{\hat H}\ket{q}&=&
2\epsilon \Lambda-G\Lambda(2S_j-\Lambda)-\hbar G \Lambda \nonumber\\
&=&H_{\rm cl}-\hbar G \Lambda
\ (=E_{\rm ql}) \ .
\end{eqnarray}
Here, $\Lambda=\hbar N/2$ in which $N$ is the particle number. 
The state $\ket{q}$ is expressed in the following form :
\begin{equation}\label{12-5}
\ket{q}=\left(\frac{\Omega !(N/2)!}{(\Omega-N/2)!}\right)^{-\frac{1}{2}}
\left(\frac{{\hat S}_+}{\hbar}\right)^{\frac{N}{2}}\ket{0}
\cdot\exp\left(-\frac{i}{\hbar}E_{\rm ql}t\right) \ .
\end{equation}

Let us investigate the state $\ket{w}$ 
which gives a classical counterpart for the quasi-spin. 
The state $\ket{w}$ should satisfy the exact relation for 
the Casimir operator ${\hat \Gamma}_{su(2)}$. 
We set up the following conditions : 
\begin{eqnarray}
\Gamma_w &=& \bra{w}{\hat \Gamma}_{su(2)} \ket{w} 
=\bra{w} {\hat S}_0^2+\frac{1}{2}({\hat S}_-{\hat S}_+
+{\hat S}_+{\hat S}_-)\ket{w} \nonumber\\
&=&S_j^2+\hbar S_j \ , \label{12-5b}\\
N_w &=& \bra{w} {\hat N} \ket{w} =N \ .
\end{eqnarray}
First, we consider a quasi-spin coherent state $\ket{c}$, 
which is generalized from the boson coherent state : 
\begin{eqnarray}\label{12-6}
& &\ket{c}=C_c^{-1/2}\cdot \exp\left(\frac{V}{U}\cdot\frac{{\hat S}_+}{\hbar}
\right)\ket{0} \ , \nonumber\\
& & \ \ 
C_c=\sum_{k=0}^{\Omega}\frac{\Omega !}{k!(\Omega-k)!}\cdot
\left(\left|\frac{V}{U}\right|^2\right)^k
=(U^{-2})^\Omega \ ,
\end{eqnarray}
where $V$ is complex and $U$ is real parameter and 
the relation $U^2+|V|^2=1$ is satisfied.
This state is known as the $su(2)$-coherent state 
which corresponds to the BCS ground-state. 
It has been regarded as an approximate state for 
$\ket{q}$ in Eq.(\ref{12-5}). 
This state is a vacuum state for the Bogoliubov-transformed 
fermion operator ${\hat a}_m$ : 
\begin{equation}\label{12-7}
{\hat a}_m=U{\hat c}_m-V(-)^{j-m}{\hat c}_{-m}^* \ , \qquad
{\hat a}_m \ket{c}=0 \ .
\end{equation}
By the use of this operator, let us define new operators :
\begin{equation}\label{12-8}
{\hat B}^*=\sqrt{\frac{\hbar}{8S_j}}\sum_m
(-)^{j-m}{\hat a}_m^*{\hat a}_{-m}^* \ , \quad
{\hat B}=\sqrt{\frac{\hbar}{8S_j}}\sum_m
(-)^{j-m}{\hat a}_{-m}{\hat a}_{m} \ , \quad
{\hat M}=\sum_m {\hat a}_m^*{\hat a}_m \ .
\end{equation}
These operators satisfy the following commutation relations :
\begin{equation}\label{12-9}
[{\hat B} , {\hat B}^* ] = 1-\frac{\hbar}{2S_j}{\hat M} \ , 
\quad
[{\hat M} , {\hat B}^* ] = 2{\hat B}^*  \ , \quad
[{\hat M} , {\hat B} ] = -2{\hat B} \ .
\end{equation}
It should be noted that the state $\ket{c}$ gives the following 
expectation value for the quasi-particle number operator ${\hat M}$ :
$n=\bra{c}{\hat M}\ket{c}=0$. 
The expectation value of the Hamiltonian is expressed as 
\begin{eqnarray}\label{12-10}
& &H_{\rm ch}=\bra{c}{\hat H}\ket{c}=H_{\rm cl}-\hbar G \Lambda 
+F_{\rm ch} \ , \nonumber\\
& &F_{\rm ch}=\frac{\hbar G}{2S_j}\Lambda(2S_j-\Lambda) \ .
\end{eqnarray}
The expectation value is clearly different from the exact 
energy eigenvalue. Further, 
it is interesting to investigate the uncertainty relations. 
Let us introduce another $su(2)$-generators, 
(${\hat s}_{\pm} , {\hat s}_0$), which consist of the quasi-particle 
operators ${\hat a}_m$ and ${\hat a}_m^*$ instead of 
${\hat c}_m$ and ${\hat c}_m^*$ in Eq.(\ref{12-2}) : 
${\hat s}_+=({\hbar}/{2})\sum_m (-)^{j-m}{\hat a}_m^*{\hat a}_{-m}^*$, 
${\hat s}_-=(\hbar/2)\sum_m (-)^{j-m}{\hat a}_{-m}{\hat a}_{m}$
and 
${\hat s}_0=(\hbar/2)\cdot (\sum_m {\hat a}_m^*{\hat a}_m-\Omega)$. 
Then, the following squares of the standard deviations 
for ${\hat s}_{x,y,z}$, which are defined as 
${\hat s}_x=({\hat s}_++{\hat s}_-)/2$, ${\hat s}_y=({\hat s}_+
-{\hat s}_-)/2i$ and ${\hat s}_z={\hat s}_0$, 
are obtained :
\begin{equation}\label{12-11}
(\Delta s_x^2)_{\rm ch} = (\Delta s_y^2)_{\rm ch}
=
\frac{\hbar S_j}{2} \ , \qquad
(\Delta s_z^2)_{\rm ch}=0 \ , 
\end{equation}
where $(\Delta s_x^2)_{\rm ch}=
\bra{c}{\hat s}_x^2\ket{c}-\bra{c}{\hat s}_x\ket{c}^2$, etc., 
and 
$\bra{c}{\hat s}_z\ket{c}=-S_j$. 
The above relations mean that the state $\ket{c}$ is 
governed by the minimal uncertainty for the quasi-spin 
operators ${\hat s}_x$, ${\hat s}_y$ and ${\hat s}_z$, 
namely, 
$(\Delta s_x^2)_{\rm ch}(\Delta s_y^2)_{\rm ch}
=(\hbar^2/4)\bra{c}{\hat s}_z\ket{c}^2$, etc. 
Also, each standard deviation has a fixed value. 
We see that the above-mentioned situation is similar to 
the case of the boson coherent state.

In order to relax the expectation value, $n$, of the quasi-particle 
number operator, ${\hat M}$, and the fixed standard deviations 
for the quasi-spin operators, 
we here introduce a new state $\ket{s}$, which we call the 
quasi-spin squeezed state, as an extension of 
the quasi-spin coherent state $\ket{c}$. 
The new state is defined as 
\begin{eqnarray}\label{12-12}
& &\ket{s}=C_s^{-1/2}\cdot\exp\left(\frac{1}{2}\cdot
\frac{v}{u}\cdot {\hat B}^{*2}\right)\ket{c} \ , \nonumber\\
& & \ \ 
C_s=\sum_{k=0}^{[\Omega/2]}\frac{(2k-1)!!}{(2k)!!}\cdot
\prod_{p=0}^{2k-1}\left(1-\frac{p}{\Omega}\right)\cdot 
z^k \ . \quad (z=(|v|/u)^2)
\end{eqnarray}
For the state $\ket{s}$, we have 
\begin{equation}\label{12-13}
n=\bra{s}{\hat M}\ket{s}=4z\frac{C_s'}{C_s} \ ,
\end{equation}
where the prime denotes the first derivative with respect to 
$z$. 
The dynamical variables in the quasi-spin squeezed state 
are $V$, $V^*$, $v/u$ and $v^*/u$. For the canonicity 
conditions for the present variables, the following 
relations are adopted :
\begin{eqnarray}\label{12-14}
& &\bra{s}i\hbar\partial_{\Phi}\ket{s}=\Lambda \ , 
\qquad
\bra{s}i\hbar\partial_{\Lambda}\ket{s}=0 \ , \nonumber\\
& &\bra{s}i\hbar\partial_{\phi}\ket{s}=\hbar\lambda \ , 
\qquad
\bra{s}i\hbar\partial_{\lambda}\ket{s}=0 \ .
\end{eqnarray}
We then get possible solutions for the above canonicity 
conditions : 
\begin{eqnarray}
& &V=\sqrt{\frac{\Lambda-\hbar\lambda}{2(S_j-\hbar\lambda)}}
\cdot\exp(-i\Phi)  \ , \quad
U=\sqrt{1-|V|^2}=\sqrt{1-\frac{\Lambda-\hbar\lambda}{2(S_j-\hbar\lambda)}}
 \ , \qquad \quad
\label{12-15}\\
& &v=\sqrt{\lambda}\cdot\exp(-2i(\Phi+\phi)) \ , \nonumber\\
& &u=\sqrt{\frac{2C_s'}{C_s}}
= \sqrt{
(\lambda+1)\left(1-\frac{\lambda}{\Omega}\right)
\left(1-\frac{\lambda+1}{\Omega}\right)(1+\Delta)^{-1}} \ . 
\label{12-16}
\end{eqnarray}
The explicit form of $\Delta$ is found in Ref. \citen{TKY5}. 
Here, it is easily shown that 
$\lambda$ is related to the quasi-particle number : 
\begin{equation}\label{12-17}
\lambda=\frac{1}{2}\bra{s}{\hat M}\ket{s}=\frac{n}{2} \ .
\end{equation}
If $\lambda=0$, the quasi-spin squeezed state is 
reduced to the quasi-spin coherent state. 
Thus, we see that the variables ($\phi, \lambda$) 
describe the quantal fluctuation and the 
variables ($\Phi, \Lambda$) mainly describe the 
classical motion. 

The expression of the factor 
$(\lambda+1)(1-\lambda/\Omega)(1-(\lambda+1)/\Omega)$ in $u^2$ 
is interesting. If $\Omega$ is even or odd, then the maximum 
of $\lambda$ should be $\Omega$ or $\Omega-1$, respectively. 
This realizes the Pauli principle. On the other hand, 
if $\lambda \ll \Omega$, $u^2$ is approximated as 
$(\lambda+1)$. 
In this case, we have $u^2-|v|^2=1$, which leads us to the 
boson approximation for ${\hat B}$ and ${\hat B}^*$.

Hereafter, the discussion is restricted to the first order 
quantal fluctuation. In this approximation, it is enough to adopt 
the following form for $V$, $U$, $v$ and $u$ :
\begin{equation}\label{12-18}
V=\sqrt{\frac{\Lambda}{2S_j}}e^{-i\Phi} \ ,\quad
U=\sqrt{1-\frac{\Lambda}{2S_j}} \ , \quad
v=\sqrt{\lambda}e^{-2i(\Phi+\phi)} \ , \quad
u=\sqrt{1+\lambda} \ .
\end{equation}
Then, the expectation value of the Hamiltonian 
with respect to the quasi-spin squeezed state is approximately 
given as 
\begin{eqnarray}
& &H_{\rm sq}=\bra{s}{\hat H}\ket{s}
=H_{\rm cl}-\hbar G \Lambda + F_{\rm sq} \ , \label{12-19}\\
& & \ \ 
F_{\rm sq}=F_{\rm ch}\cdot f(\lambda, \phi) \ , \nonumber\\
& & \ \ 
f(\lambda,\phi)=1+2\lambda+2\sqrt{\lambda(\lambda+1)}\cdot\cos 2\phi
\ .\label{12-20}
\end{eqnarray}
Here, $F_{\rm ch}$ has been defined in Eq.(\ref{12-10}). 
In this approximation, the squares of the standard deviations 
for the quasi-spin operators ${\hat s}_{x,y,z}$ are given as 
follows :
\begin{eqnarray}\label{12-21}
& &(\Delta s_x^2)_{\rm sq}
=\frac{\hbar S_j}{2}\cdot[1+2\lambda+2\sqrt{\lambda(\lambda+1)}
\cdot \cos[2(\Phi+\phi)]] \ , \nonumber\\
& &(\Delta s_y^2)_{\rm sq}
=\frac{\hbar S_j}{2}\cdot[1+2\lambda-2\sqrt{\lambda(\lambda+1)}
\cdot \cos[2(\Phi+\phi)]] \ , \nonumber\\
& &(\Delta s_z^2)_{\rm sq}
=0 \ ,
\end{eqnarray}
where $(\Delta s_x^2)_{\rm sq}=\bra{s}{\hat s}_x^2\ket{s}
-\bra{s}{\hat s}_x\ket{s}^2$, etc., and 
$\bra{s}{\hat s}_z\ket{s}=-S_j$.

Our problem is to solve the equations of motion derived from 
the time-dependent variational principle : 
\begin{eqnarray}\label{12-22}
& &{\dot \Lambda}=-\frac{\partial H_{\rm sq}}{\partial \Phi} \ , 
\qquad
{\dot \Phi}=\frac{\partial H_{\rm sq}}{\partial \Lambda} \ , 
\nonumber\\
& &\hbar{\dot \lambda}=-\frac{\partial H_{\rm sq}}{\partial \phi} \ , 
\qquad
\hbar{\dot \phi}=\frac{\partial H_{\rm sq}}{\partial \lambda} \ .
\end{eqnarray}
The initial conditions for $\lambda$ and $\phi$ which describe 
the classical image of quantal fluctuation 
are obtained under the similar criterion to that given in the case 
of quantum mechanical system in \S 10 or the $su(2)$-boson 
model in \S 11. 
Namely, at the initial time, the minimal uncertainty should be 
realized. Further, the energy originated from quantum fluctuations 
is minimal at the initial time. 
\begin{equation}\label{12-23}
(\Delta s_x^2)_{\rm sq}^0\cdot(\Delta s_y^2)_{\rm sq}^0
=\left(\frac{\hbar}{2}\right)^2\cdot
|\bra{s}{\hat s}_z\ket{s}^0|^2 \ , \qquad
f(\lambda_0, \phi_0) : {\rm minimal} \ ,
\end{equation}
where the superscript and subscript ``0" denotes the initial value 
of corresponding variable. 
These criteria lead us to the following initial conditions :
\begin{equation}\label{12-24}
\phi_0=\frac{k\pi}{2}-\Phi_0 , \quad (k=0,\pm 1, \pm 2, \cdots) \ ,
\qquad
\lambda_0=\frac{1}{2}\left(\frac{1}{|\sin 2\Phi_0|}-1\right) 
\ .\ \ (\cos 2\phi_0 \le 0)
\end{equation}
If $\cos 2\phi_0 >0$, then $\lambda_0=0$. 
For the solution (\ref{12-24}), $H_{\rm sq}$ is given in the 
form 
\begin{equation}\label{12-25}
H_{\rm sq}=H_{\rm cl}-\hbar G \Lambda
+\frac{\hbar G}{2S_j}\cdot\Lambda(2S_j-\Lambda)\cdot
|\sin 2\Phi_0| \ .
\end{equation}
It may be interesting to see that if $|\sin 2\Phi_0|$ vanishes, 
$F_{\rm sq}=0$ and if $|\sin 2\Phi_0|$ is equal to 1, 
$F_{\rm sq}=\hbar G/2S_j\cdot\Lambda(2S_j-\Lambda)$. 
The former and the latter correspond to the situations 
given by $\ket{q}$ and $\ket{c}$, respectively. 
Therefore, by the parameter 
$|\sin 2\Phi_0|$, we can reproduce the intermediate situation. 
Concerning the minimal uncertainty, we have 
\begin{eqnarray}\label{12-26}
& &(\Delta s_x^2)_{\rm sq}^0=\frac{\hbar S_j}{2}\cdot \rho_0^2 
 \ , \qquad
 (\Delta s_y^2)_{\rm sq}^0=\frac{\hbar S_j}{2}\cdot \rho_0^{-2} 
 \ , \nonumber\\
& &\rho_0=\frac{\sqrt{1+|\sin 2\Phi_0|}\pm\sqrt{1-|\sin 2\Phi_0|}}
{\sqrt{2|\sin 2\Phi_0|}} \ .
\end{eqnarray}
The relation (\ref{12-26}) shows us that at the initial time 
the magnitudes of $(\Delta s_x^2)_{\rm sq}^0$ and 
$(\Delta s_y^2)_{\rm sq}^0$ are controlled by the quantity 
$\sin 2\Phi_0$, which is the same situation as that in the case of 
$F_{\rm sq}$. 
Under the above initial conditions, we obtain the following 
solutions for the Hamilton equations of motion :
\begin{eqnarray}\label{12-27}
& &\Lambda(t)=\Lambda \ , \nonumber\\
& &\Phi(t)=2\omega_{\rm sq} t + \Phi_0 \ , \nonumber\\
& &\lambda(t)=\left(\frac{G}{S_j}\right)\Lambda^2(2S_j-\Lambda)^2 
|\sin 2\Phi_0| t^2 \nonumber\\
& &\ \ \ \ \ \ \ \ \ +
\frac{G}{S_j}\Lambda(2S_j-\Lambda)|\cos 2\Phi_0| t
+\frac{1}{2}\left(\frac{1}{|\sin 2\Phi_0|}-1\right) \ , \nonumber\\
& &\phi(t)=\frac{1}{2}\cdot \cos^{-1}\chi(t) \ ,
\end{eqnarray}
where $\omega_{\rm sq}$ and $\chi(t)$ are defined as 
\begin{eqnarray}\label{12-28}
& &\omega_{\rm sq}
=\epsilon - G(S_j-\Lambda)
-\frac{\hbar G}{2}\cdot (1-|\sin 2\Phi_0|)
-\frac{\hbar G \Lambda}{2S_j}|\sin 2\Phi_0| \ , \nonumber\\
& &\chi(t)=\frac{|\sin 2\Phi_0|-(1+2\lambda(t))}{
2\sqrt{\lambda(t)(\lambda(t)+1)}} \ .
\end{eqnarray}

Next, let us investigate another many-fermion system 
which consists of two-energy levels with same degeneracy 
$2\Omega \ (=2j+1)$. The two-body interaction is active only 
for a particle-hole pair with coupled momentum being 0. 
This model is called the Lipkin model whose 
Hamiltonian is written as 
\begin{equation}\label{12-29}
{\hat H} =
 2\epsilon {\hat S}_0 
           - \frac{1}{2} G ({{\hat S}_+}^2 + {{\hat S}_-}^2 ) \ , 
\end{equation}
where $2\hbar\epsilon$ is 
the level spacing between lower and upper energy levels
labeled by $g=1$ and 2, respectively. 
We have already used this model in \S\S 4 and 11 as an $su(2)$-boson model. 
However, in this section, we deal with this model as an example of 
many-fermion system. 
Thus, the quasi-spin operators consist of the fermion operators, 
which are defined as 
\begin{equation}\label{12-30}
{\hat S}_+ = \hbar\sum_{m=-j}^{j} {\hat c}_{2m}^{*} {\hat c}_{1m} 
\ , \quad
{\hat S}_- = \hbar\sum_{m=-j}^{j} {\hat c}_{1m}^{*} {\hat c}_{2m} \ , 
\quad
{\hat S}_0 = \frac{\hbar}{2}\sum_{m=-j}^{j} 
           ({\hat c}_{2m}^{*} {\hat c}_{2m} - 
           {\hat c}_{1m}^{*} {\hat c}_{1m}) \ , 
\end{equation}
where ${\hat c}_{gm}^{*}$ and ${\hat c}_{gm}$ 
are fermion creation and annihilation operators. 
Here, the above quasi-spin operators satisfy the commutation relations of 
the $su(2)$-algebra :
$[{\hat S}_+ \ , \ {\hat S}_- ] = 2\hbar{\hat S}_0$ and 
$[{\hat S}_0 \ , \ {\hat S}_\pm ] = \pm \hbar{\hat S}_\pm$. 
Hereafter, we concentrate on the case $N=2\Omega$ where $N$ 
represents particle number. 

The $su(2)$-coherent state is adopted as a trial state 
in the time-dependent Hartree-Fock theory. This state is given as
\begin{equation}\label{12-31}
\ket{c} \equiv \frac{1}{(1+|Z|^2)^\Omega} 
\exp \left(Z\frac{{\hat S}_+}{\hbar}\right) \ket{0_{\rm ph}} 
\end{equation}
with ${\hat c}_{2m}\ket{0_{\rm ph}} = {\hat c}_{1m}^{*} 
\ket{0_{\rm ph}} =0$. 
This coherent state is a vacuum state with respect to the 
unitary-transformed fermion operators as 
\begin{eqnarray}\label{12-32}
& &{\hat a}_{2m} = 
\frac{1}{(1+|Z|^2)^{1/2}} {\hat c}_{2m}
- \frac{Z}{(1+|Z|^2)^{1/2}} {\hat c}_{1m} \ , \nonumber\\
& &{\hat a}_{1m} = 
\frac{Z^*}{(1+|Z|^2)^{1/2}} {\hat c}_{2m}
+ \frac{1}{(1+|Z|^2)^{1/2}} {\hat c}_{1m}
\end{eqnarray}
with ${\hat a}_{2m}\ket{c}={\hat a}_{1m}^*\ket{c}=0$.

Our aim is to extend the coherent state to the squeezed state 
in order to go beyond the usual Hartree-Fock (HF) approximation. 
Now, let us introduce a quasi-spin squeezed state from the 
$su(2)$-coherent state in a manner similar to that for the case 
of the pairing model discussed previously. 
The new operators which consist of the unitary-transformed 
quasi-fermion operators are introduced like (\ref{12-8}) :
\begin{eqnarray}\label{12-33}
& &{\hat B}^{*} = \sqrt{\frac{\hbar}{2\Omega}}
  \sum_{m=-j}^{j} {\hat a}_{2m}^{*} {\hat a}_{1m} \ , \qquad
{\hat B} = \sqrt{\frac{\hbar}{2\Omega}}
  \sum_{m=-j}^{j} {\hat a}_{1m}^{*} {\hat a}_{2m} \ , \nonumber\\
& &{\hat M} = \sum_{m=-j}^{j} 
           ({\hat a}_{2m}^{*} {\hat a}_{2m} + {\hat a}_{1m} 
           {\hat a}_{1m}^{*})\ . 
\end{eqnarray}
The operator ${\hat M}$ is nothing but the number operator of the 
quasi-particles and holes. 
The commutation relations among these operators have the same forms 
as Eq.(\ref{12-9}) except for the replacement of $S_j$ to $\Omega$. 
The quasi-spin squeezed state in this model is defined as 
\begin{eqnarray}\label{12-34}
& &\ket{s}=C_s^{-1/2}\cdot\exp\left(\frac{1}{2}\cdot
\frac{v}{u}\cdot {\hat B}^{*2}\right)\ket{c} \ , \nonumber\\
& & \ \ 
C_s=1+\sum_{k=1}^{[\Omega]}\frac{(2k-1)!!}{(2k)!!}\cdot
\prod_{p=0}^{2k-1}\left(1-\frac{p}{2\Omega}\right)\cdot 
z^k \ . \quad (z=(|v|/u)^2)
\end{eqnarray}
Let us investigate the classical 
counterpart given by the quasi-spin squeezed state. 
We impose the canonicity conditions so as to 
transform the variables, which are contained in the $\ket{s}$, from 
$Z$, $Z^*$, $v/u$ and $v^*/u$ to $b$, $b^*$, $\beta$ and $\beta^*$ :
\begin{equation}\label{12-35}
\bra{s}\frac{\partial}{\partial b} \ket{s} = \frac{1}{2}b^* \ , \quad
\bra{s}\frac{\partial}{\partial b^*} \ket{s} = -\frac{1}{2}b \ ; \quad
\bra{s}\frac{\partial}{\partial \beta} \ket{s} = \frac{1}{2}\beta^* \ , \quad
\bra{s}\frac{\partial}{\partial \beta^*} \ket{s} = -\frac{1}{2}\beta \ .
\end{equation}
Under these conditions, the sets of variables $(b, b^*)$ and 
$(\beta , \beta^*)$ are those of canonical variables, respectively. 
Solving the above canonicity conditions, we can express the expectation 
values of the quasi-spin operators in terms of 
$b$, $b^*$, $\beta$ and $\beta^*$ :
\begin{equation}\label{12-36}
\bra{s}{\hat S}_+ \ket{s} = \sqrt{\hbar}
b^* \sqrt{2\Omega-\hbar |b|^2-4\hbar |\beta|^2} 
=\bra{s}{\hat S}_- \ket{s}^* \ , \quad
\bra{s} {\hat S}_0 \ket{s} = -\Omega + \hbar |b|^2 + 2\hbar |\beta|^2 \ .
\end{equation}
These expressions remind us of the Holstein-Primakoff boson 
representation for 
the $su(2)$-algebra. 
Further, we can calculate 
the energy expectation value with respect to 
the quasi-spin squeezed state. The result is almost identical with 
the exact eigenvalue for the ground state in the wide region 
of the coupling strength $G$. 
The details are found numerically in Refs. \citen{TAKY1} and 
\citen{TAKY2}.

Finally, let us investigate the implication to the RPA.\cite{TAKY2}
We now set the operator ${\hat B}$ as a pure boson operator 
${\hat {\cal B}}$. 
Further, under this replacement, the number operator of the 
quasi-particles and quasi-holes, ${\hat M}$, should be replaced 
so as to satisfy the original commutation relation (\ref{12-9}). 
Namely, 
\begin{equation}\label{12-37}
{\hat B} \longrightarrow {\hat {\cal B}} \ , \quad
{\hat B}^{*} \longrightarrow {\hat {\cal B}}^{*} \ , \qquad
[ {\hat {\cal B}} \ , \ {\hat {\cal B}}^{*} ] = 1 \ ; \qquad
 {\hat M} \longrightarrow {\hat{\cal M}}
 \equiv 2{\hat {\cal B}}^{*}{\hat {\cal B}} \ . 
\end{equation}
Also, the quasi-spin squeezed state $\ket{s}$ has to be 
replaced into the usual boson squeezed state $\ket{b}$ : 
\begin{equation}\label{12-38}
\ket{s}\longrightarrow 
\ket{b}\equiv \Gamma_b^{-1/2} 
              \exp\left(\frac{1}{2}\cdot \frac{v}{u}
               {\hat{\cal B}}^{*2} \right) \ket{c} \ , 
\end{equation}
where the relation $u^2-|v|^2 =1$ is satisfied. 
As is well known, the boson squeezed state is a vacuum with respect to 
the unitary-transformed boson operator : 
\begin{equation}\label{12-39}
{\hat b} \ket{b} = 0 \ , \quad 
{\hat b} = u{\hat {\cal B}}-v{\hat {\cal B}}^{*} \ , \quad
[{\hat b} \ , \ {\hat b}^{*} ]=1 \ . 
\end{equation}
Let us consider to diagonalize the model Hamiltonian with respect to the 
boson operators ${\hat b}$ and ${\hat b}^{*}$. 
The Hamiltonian (\ref{12-29}) is expressed in terms of ${\hat B}$, 
${\hat B}^{*}$ and ${\hat M}$ as
\begin{equation}\label{12-40}
{\hat H} = E_{\rm ch} + E_{11}{\hat M} + E_{20}{\hat B}^{*} 
                                       + E_{20}^*{\hat B}
           + E_{MM}{\hat M}^2
           + E_{22}{\hat B}^{*} {\hat B} 
           +(E_{40}{\hat B}^{*2} + E_{60}{\hat B}{\hat M} + h.c.) \  , 
\end{equation}
where $E_{\rm ch}\equiv \bra{c}{\hat H}\ket{c}$, $E_{ij}$ denote $c$-number 
coefficients and $h.c.$ means the Hermitian conjugate. 
In the HF approximation, $E_{20}=E_{20}^*=0$ are demanded because 
${\hat B}^{*}$ is a particle-hole pair-creation operator. 
Now, let us take into account of the higher order terms, which are neglected 
in the HF approximation. 
Following the replacement (\ref{12-37}), neglecting the terms of 
third and fourth order with respect to the boson operators 
${\hat {\cal B}}$ and ${\hat {\cal B}}^{*}$, and transforming 
${\hat {\cal B}}$ 
to ${\hat b}$ in (\ref{12-39}), we obtain the Hamiltonian as 
\begin{eqnarray}\label{12-41}
{\hat H}_b &\simeq& 
  E_{\rm ch} + \alpha |v|^2 + \frac{\beta}{2}(uv+uv^*)
  + \left[ \left( \alpha u v + \frac{\beta}{2} (u^2+v^2)\right) 
                 {\hat b}^{*2} + h.c. \right] \nonumber\\
  & & + \left[ \alpha(u^2 + |v|^2) + \beta (uv+uv^*) \right] 
         {\hat b}^{*}{\hat b} \ . 
\end{eqnarray}
Here, if we define $\chi=\hbar^2 G\Omega(1-1/2\Omega)/(\hbar\epsilon)$, 
then 
$\alpha$ and $\beta$ are given as 
$\alpha = 2\hbar\epsilon$ for $\chi \leq 1$ or 
$\hbar\epsilon(3\chi^2 -1)/\chi$ 
for $\chi \geq 1$ and 
$\beta = -2\hbar\epsilon \chi$ for $\chi \leq 1$ or 
$-\hbar\epsilon(\chi^2+1)/\chi$ for $\chi \geq 1$, respectively. 
We demand that 
the terms which is proportional to ${\hat b}^{*2}$ 
and ${\hat b}^2$ should be zero. 
As a result, in the limit of $2\Omega \gg 1$, $u$ and $v$ are determined as 
\begin{equation}\label{12-42}
u^2 
 \simeq \frac{1}{2} 
          \left(1+\frac{\alpha}{\sqrt{\alpha^2-\beta^2}}\right) \ , 
\qquad
|v|^2 
 \simeq -\frac{1}{2}
          \left(1-\frac{\alpha}{\sqrt{\alpha^2-\beta^2}}\right) \ . 
\end{equation}
The Hamiltonian is finally expressed as 
\begin{eqnarray}\label{12-43}
{\hat H}_b &\simeq& 
  E_{\rm RPA} + \hbar\omega_b {\hat b}^{*}{\hat b} \ , \nonumber\\
  & & E_{\rm RPA} 
       \equiv E_{\rm ch} 
              + \frac{1}{2}(\sqrt{\alpha^2-\beta^2}-\alpha) \ , \qquad
      \hbar\omega_b = \sqrt{\alpha^2-\beta^2} \ .
\end{eqnarray}
Here, $E_{\rm RPA}$ is identical to the ground state energy of 
the RPA calculation. 
Namely, the ground state correlation is included 
automatically in our approach. 
The boson operator ${\hat b}$ corresponds to the phonon operator. 
Thus, our quasi-spin squeezed state approach includes the RPA 
in a certain limit.

\section{Concluding remarks}

On the basis of the celebrated boson expansion theory\cite{MYT} 
proposed by Marumori, together with 
Yamamura (one of the present authors) and Tokunaga, 
we have developed the idea of the boson mapping 
method, which is contained in the Marumori-Yamamura-Tokunaga 
boson expansion theory, to describe the thermal states 
in quantum many-body systems. 
The original boson mapping method in which the fermion space 
is mapped to the physical boson-space is not restricted 
to the many-fermion systems such as nucleus. We have used this method 
in \S\S 2 and 3 to give the classical counterparts of the 
$su(2)$- and the $su(1,1)$-algebras realized by the Schwinger boson 
representation. As a result, it has been shown that the classical 
counterparts of these algebras realized in terms of the 
Schwinger bosons 
could be obtained in the same forms as those of 
the Holstein-Primakoff representations of the $su(2)$- 
and the $su(1,1)$-algebras, respectively. 
With the aim of describing 
the thermal behaviors of the quantum many-body systems, 
we have learned in \S\S 4 and 5 that the thermal properties 
appear with the $su(1,1)$-algebraic features in the 
``damped and amplified oscillator" model. 
With this insight, in \S 6, 
we have considered the $su(2)$-spin 
system and our idea has been applied to this model. 
The $su(2)$-spin model is often regarded as a schematic 
model of nucleus. 
It has been indicated that the $su(2)$-spin 
model shows the $su(1,1)$-like behavior with the help of a 
certain kind of the coherent state giving a classical counterpart. 
Further, we have treated the $su(2)$-spin system interacting 
with an external harmonic oscillator. With the help of the MYT 
boson mapping method again, we have shown that the 
$su(1,1)$-like behavior is realized in the physical space 
spanned by the boson mapping from the original space. 
Thus, it becomes possible to describe the thermal behavior 
in the system composed of the $su(2)$-spin interacting 
with the harmonic oscillator. It is seen in \S 7 $\sim$ \S9 that 
our formalism is workable to describe the thermal behaviors 
in this model. 

Another powerful method to describe the quantum many-body 
systems in the microscopic way is the time-dependent 
Hartree-Fock theory. To investigate the small amplitude 
nuclear collective vibrational motion, the idea of the TDHF 
theory was proposed by Marumori.\cite{M} 
In the TDHF theory, 
an approximation of 
the small amplitude fluctuation around a static Hartree-Fock field 
results in the random phase approximation. 
This approximation is not enough to investigate the large amplitude 
nuclear collective motion. To go beyond the RPA, the TDHF theory 
devised by the canonical form was proposed by Marumori,\cite{MMSK} 
together 
with Maskawa, Sakata and Kuriyama (one of the present authors), 
in which the collective coordinate was extracted self-consistently 
from many-nucleon degrees of freedom. 
The time-dependent variational approach with the canonical form 
developed by Marumori et al. is not restricted to treating the 
many-nucleon or -fermion systems. Also, it is not restricted 
to using the Slater determinant, which is essentially a 
coherent state, as a trial state for the time-dependent variational 
procedure. We have extended this formulation to 
quantum mechanical systems and quantum many-boson systems 
in order to describe the time-evolution of these systems in 
terms of the classical mechanics. 
The key idea is to use squeezed states as trial states in the 
time-dependent variational procedure. As a result, it has been 
shown in \S\S 10 and 11 that the higher order 
quantal effects than the order of $\hbar^1$ have been 
automatically included in our squeezed state formalism. 
Further, our idea is applied to the quantum many-fermion 
systems governed by the $su(2)$-algebra. We have defined and 
introduced the quasi-spin squeezed state to go beyond the 
usual TDHF theory in \S 12. 
It has been shown that, 
in the single orbit shell model with pairing interaction, 
the BCS result and the exact one for the ground state 
have been connected by the use of this new state. 
Also, in the Lipkin mode, we have shown that the 
quasi-spin squeezed state approach could reproduce 
the result obtained in the RPA under a certain limit 
and, further, 
seemed to give us a possible way to go beyond the RPA. 

The boson expansion theory and the TDHF theory in the canonical form 
are powerful theories to investigate diverse quantum many-body 
systems in various situations. It is expected that these 
theories will be developed in and successfully applied 
to quantum physics in addition to the microscopic 
studies of nuclear collective motions. 
For example, we can find such studies in Ref. \citen{KPPY}.

\section*{Acknowledgements}

For publishing this review, the authors should acknowledge 
to Professor\break
T. Marumori for his continuous guide to 
the study of microscopic theory of collective motion. 
One of the authors (M. Y.) is indebted to Professor 
J. da Provid\^encia (the co-author of this review) for giving 
him many chances to stay in Coimbra and 
for suggesting various subjects which are discussed in this 
review. 

Dr. Y. Fujiwara should be acknowledged. At the early stage 
of the studies on the squeezed states, he collaborated with 
the authors (A. K., Y. T. and M. Y.). 

Some parts reported in this paper are based on the works 
supported by the Kansai University Grand-in-Aid for the 
Faculty Joint Research Program, 2000.
%\appendix
%\section{First Appendix} %Empty argument \section{} yields `Appendix'. 
%
%\section{Second Appendix}
%


\begin{thebibliography}{99}
%%%%%%%%%%%%%%%%%%%%%%%%%%%%%%%%%%%%%%%%%%%%%%%%%%%%%%%%%%%%%
% Some macros are available for the bibliography:
%   o for general use
%      \JL : general journals          \andvol : Vol (Year) Page
%   o for individual journal 
%      \PR  : Phys. Rev.               \PRL : Phys. Rev. Lett.
%      \NP  : Nucl. Phys.              \PL  : Phys. Lett.
%      \JMP : J. Math. Phys.           \CMP : Commun. Math. Phys.
%      \PTP : Prog. Theor. Phys.       \JPSJ: J. Phys. Soc. Jpn.
%      \JP  : J. of Phys.              \NC  : Nouvo Cim.
%      \IJMP: Int. J. Mod. Phys.       \ANN : Ann. of Phys.
% Usage:
%   \PR{D45,1990,345}            ==> Phys.~Rev.\ {\bf D45} (1990), 345
%   \JL{Phys.~Lett.,A30,1981,56} ==> Phys.~Lett.\ {\bf A30} (1981), 56
%   \andvol{B123,1995,1020}      ==> {\bf B123} (1995), 1020
%%%%%%%%%%%%%%%%%%%%%%%%%%%%%%%%%%%%%%%%%%%%%%%%%%%%%%%%%%%%%
\bibitem{M}
T. Marumori, 
        Prog.~Theor.~Phys.\ {\bf 24} (1960), 331.   %End with period .(not ;)
\bibitem{AVB}
R. Arvieu and M. Veneroni, 
        Compt. Rend.\ {\bf 250} (1960), 992.\\   %End with period .(not ;)
M. Baranger, Phys. Rev. {\bf 120} (1960), 957.
\bibitem{BZ}
S. T. Belyaev and V. G. Zelevinsky, 
        Nucl. Phys.\ {\bf 39} (1962), 582.   %End with period .(not ;)
\bibitem{MYT}
T. Marumori, M. Yamamura and A. Tokunaga, 
        Prog. Theor. Phys.\ {\bf 31} (1964), 1009.
\bibitem{P}
J. da Provid\^encia, 
        Nucl. Phys.\ {\bf A108} (1968), 589,\\
J. da Provid\^encia and J. Weneser, 
Phys. Rev. {\bf C1} (1970), 825.
\bibitem{Mar}
E. R. Marshalek, Nucl. Phys. {\bf A161} (1971), 401.
\bibitem{KM}
A. Klein and E. R. Marshalek, 
        Rev. Mod. Phys.\ {\bf 63} (1991), 375.
\bibitem{N}
M. Nogami, 
        Soryushiron Kenkyu (Kyoto),\ {\bf 10} (1955-1956), 600. 
        (in Japanese)
\bibitem{BV}
M. Baranger and M Veneroni, 
        Ann. of Phys.\ {\bf 112} (1978), 123. \\
D. M. Brink, M. J. Giannoni and M. Veneroni, 
        Nucl. Phys. {\bf A258} (1976), 237.\\
F. Villars, Nucl. Phys. {\bf A285} (1977), 269.\\
K. Goeke and P. G. Reinhard, Ann. of Phys. {\bf 112} (1978), 328.\\
A. K. Mukherjee and M. K. Pal, Nucl. Phys. {\bf A373} (1982), 289.\\
G. Holzwarth and T. Yukawa, Nucl. Phys. {\bf A219} (1974), 125.\\
D. J. Rowe and R. Bassermann, Can. J. Phys. {\bf 54} (1976), 1941.
\bibitem{MMSK}
T. Marumori, T. Maskawa, F. Sakata and A. Kuriyama, 
        Prog. Theor. Phys.\ {\bf 64} (1980), 1294.
\bibitem{MH}
E. R. Marshalek and G. Holzwarth, Nucl. Phys. {\bf A191} 
(1972), 438. 
\bibitem{YK87}
M. Yamamura and A. Kuriyama, 
        Prog.~Theor.~Phys.~Suppl. No. 93 (1987), 1.
\bibitem{YPKF}
M. Yamamura, J. da Provid\^encia, A. Kuriyama and C. Fiolhais, 
Prog. Theor. Phys. {\bf 81} (1989), 1198 : {\bf 83} (1990), 749.\\
M. Yamamura, J da Provid\^encia and A. Kuriyama, 
Nucl. Phys. {\bf A514} (1990), 461.\\
A. Kuriyama, J. da Provid\^encia and M. Yamamura, 
Prog. Theor. Phys. {\bf 84} (1990), 452, 115 : 
{\bf 85} (1991), 939 : {\bf 86} (1991), 419 : {\bf 87} (1992), 
135, 911, 1377.\\
A. Kuriyama, M. Yamamura, J. da Provid\^encia and C. Provid\^encia, 
Phys. Rev. {\bf C45} (1992), 2196.
\bibitem{TU}
Y. Takahashi and H. Umezawa, Collect. Phenom. {\bf 2} (1975), 55.\\
H. Umezawa, H. Matsumoto and M. Tachiki, 
{\it Thermo Field Dynamics and Condensed States} (North Holland, 
Amsterdam, 1982).
\bibitem{Schw}
J. Schwinger, in {\it Quantum Theory of Angular Momentum}, ed. L. Biedenharn 
and H. Van Dam (Academic Press, New York, 1955), p.229.
\bibitem{Vitiello}
E. Celeghini, M. Rasetti, M. Tarlini and G. Vitiello, 
Mod. Phys. Lett. {\bf B3} (1989), 1213.\\
E. Celeghini, M. Rasetti and G. Vitiello, Ann. of Phys. 
{\bf 215} (1992), 156. 
\bibitem{TKY1}
Y.~Tsue, A. Kuriyama and M.~Yamamura, 
        Prog.~Theor.~Phys.\ {\bf 91} (1994), 49.   %End with period .(not ;)
\bibitem{TKY2}
Y.~Tsue, A. Kuriyama and M.~Yamamura, 
        Prog.~Theor.~Phys.\ {\bf 91} (1994), 469.   %End with period .(not ;)
\bibitem{YKT-p}
M. Yamamura, A. Kuriyama and Y. Tsue, 
{\it Proceedings of the International Conference on 
Many-Body Physics}, Coimbra, Portugal, 1993 (World Scientific, 
1994), p. 311.
\bibitem{TKY3}
Y.~Tsue, A. Kuriyama and M.~Yamamura, 
        Prog.~Theor.~Phys.\ {\bf 93} (1995), 541.   %End with period .(not ;)
\bibitem{TKY4}
Y.~Tsue, A. Kuriyama and M.~Yamamura, 
        Prog.~Theor.~Phys.\ {\bf 93} (1995), 1037.   %End with period .(not ;)
\bibitem{TKY-p}
Y. Tsue, A. Kuriyama and M. Yamamura, 
{\it Thermal Field Theories and Their Applications}, Forth 
International Workshop on Thermal Field Theories 
and Their Applications, 1995, Dalian, China (World Scientific, 1996), 
p. 373. 
\bibitem{KPTY1}
A. Kuriyama, J. da Provid\^encia, Y. Tsue and M.~Yamamura, 
        Prog.~Theor.~Phys.\ {\bf 95} (1996), 79.  
\bibitem{KPTY2}
A. Kuriyama, J. da Provid\^encia, Y. Tsue and M.~Yamamura, 
        Prog.~Theor.~Phys.\ {\bf 96} (1996), 387.  
\bibitem{KPTY3}
A. Kuriyama, J. da Provid\^encia, Y. Tsue and M.~Yamamura, 
        Prog.~Theor.~Phys.\ {\bf 98} (1997), 381.  
\bibitem{KPTY4}
A. Kuriyama, J. da Provid\^encia, Y. Tsue and M.~Yamamura, 
        Prog.~Theor.~Phys.\ {\bf 98} (1997), 893.   %End with period .(not ;)
\bibitem{KPTY5}
A. Kuriyama, J. da Provid\^encia, Y. Tsue and M.~Yamamura, 
        Prog.~Theor.~Phys.\ {\bf 99} (1998), 819.   %End with period .(not ;)
\bibitem{KPTY6}
A. Kuriyama, J. da Provid\^encia, Y. Tsue and M.~Yamamura, 
        Prog.~Theor.~Phys.\ {\bf 100} (1998), 993. 
\bibitem{KPTY7}
A. Kuriyama, J. da Provid\^encia, Y. Tsue and M.~Yamamura, 
        Prog.~Theor.~Phys.\ {\bf 100} (1998), 1181. 
\bibitem{PYK}
J. da Provid\^encia, M. Yamamura and A. Kuriyama, 
Prog. Theor. Phys. {\bf 101} (1999), 139. 
\bibitem{TFKY}
Y.~Tsue, Y. Fujiwara, A. Kuriyama and M.~Yamamura, 
        Prog.~Theor.~Phys.\ {\bf 85} (1991), 693.   %End with period .(not ;)
\bibitem{TF1}
Y.~Tsue and Y. Fujiwara 
        Prog.~Theor.~Phys.\ {\bf 86} (1991), 443.  
\bibitem{TF2}
Y.~Tsue and Y. Fujiwara, 
        Prog.~Theor.~Phys.\ {\bf 86} (1991), 469.  
\bibitem{YK2}
M. Yamamura and A. Kuriyama, 
        Prog. Theor. Phys. {\bf 88} (1992), 711.
\bibitem{YKT}
M. Yamamura, A. Kuriyama and Y.~Tsue, 
       Prog.~Theor.~Phys.\ {\bf 88} (1992), 719.  
\bibitem{T}
Y.~Tsue, 
        Prog.~Theor.~Phys.\ {\bf 88} (1992), 911. \\
Y. Tsue, 
{\it Third International Workshop on Squeezed States and 
Uncertainty Relations}, Baltimore, Maryland, USA, 1993 
(NASA Conference Publication 3270, 1994), p.413.
\bibitem{TKY5}
Y.~Tsue, A. Kuriyama and M.~Yamamura, 
        Prog.~Theor.~Phys.\ {\bf 92} (1994), 545. 
\bibitem{TAKY1}
Y.~Tsue, N. Azuma, A. Kuriyama and M.~Yamamura, 
        Prog.~Theor.~Phys.\ {\bf 96} (1996), 729.  
\bibitem{TAKY2}
Y.~Tsue, N. Azuma, A. Kuriyama and M.~Yamamura, 
{\it NASA/CP-1998-206855} 
(NASA Conference Publication, 1998), p.~221.
\bibitem{HP}
T. Holstein and H. Primakoff, Phys. Rev. {\bf 58} (1940), 
1098.\\
S. C. Pang, A. Klein and R. M. Dreizler, 
Ann. of Phys. {\bf 49} (1968), 477.
\bibitem{Kubo}
R. Kubo, J. Phys. Soc. Jpn {\bf 12} (1957), 570.
%\\
\bibitem{JK}
R.~Jackiw and A.~Kerman, 
        Phys.~Lett.\ {\bf 71A} (1979), 158.
\bibitem{KPPY}
A. Kuriyama, C. Provid\^encia, J. da Provid\^encia and M.~Yamamura, 
        Prog.~Theor.~Phys.\ {\bf 103} (2000), 733. 
\end{thebibliography}
\end{document}